\documentclass[10pt,amsmath,amssymb,aps,prx,twocolumn,nobibnotes]{revtex4-2}

\usepackage{graphicx}
\usepackage{bm}
\usepackage{color}
\usepackage{braket}
\usepackage{siunitx}
\usepackage{tabularx}
\usepackage[dvipsnames]{xcolor}
\usepackage{mathtools}
\usepackage{footmisc}

\usepackage{blindtext}

\usepackage{hyperref}
\usepackage{soul}

\usepackage[dvipsnames]{xcolor}

\newcommand{\beq}{\begin{equation}}
\newcommand{\eeq}{\end{equation}}
\newcommand{\bea}{\begin{eqnarray}}
\newcommand{\eea}{\end{eqnarray}}
\newcommand{\bec}{\begin{center}}
\newcommand{\enc}{\end{center}}
\newcommand{\bfr}{\begin{flushright}}
\newcommand{\efr}{\end{flushright}}
\newcommand{\tblue}{\textcolor{blue}}

\newcommand{\di}{\textrm{d}}

\newcommand{\bin}{\hat{b}_{\mathrm{in}}}
\newcommand{\bout}{\hat{b}_{\mathrm{out}}}
\newcommand{\cin}{\hat{c}_{\mathrm{in}}}
\newcommand{\cout}{\hat{c}_{\mathrm{out}}}
\newcommand{\din}{\hat{d}_{\mathrm{in}}}
\newcommand{\vin}{\hat{v}_{\mathrm{in}}}
\newcommand{\uin}{\hat{u}_{\mathrm{in}}}
\newcommand{\ain}{\hat{a}_{\mathrm{in}}}
\newcommand{\aout}{\hat{a}_{\mathrm{out}}}

\newcommand{\eint}{\eta_0}
\newcommand{\eext}{\eta}
\newcommand{\kint}{\kappa_0}
\newcommand{\kext}{\kappa}

\DeclareSIUnit{\belmilliwatt}{Bm}
\DeclareSIUnit{\dBm}{\deci\belmilliwatt}

\makeatletter
\renewcommand\frontmatter@affiliationfont{\normalfont\small\vspace{6pt}}
\makeatother

\DefineFNsymbols*{lamport}[math]{*AB\S\P\|{**}{AA}{BB}}
\setfnsymbol{lamport}

\begin{document}

\title{High-gain and large-bandwidth Josephson parametric amplifier \\ influenced by Fabry--P\'erot interference}

\author{S. Kono$^{1,2\dag}$}
\thanks{These authors contributed equally to this work.\\
$\dag$ shingo.kono@nbi.ku.dk and $\ddag$ jesper.ilves@qipe.t.u-tokyo.ac.jp\\}

\author{J. Ilves$^{3\ddag}$}
\thanks{These authors contributed equally to this work.\\
$\dag$ shingo.kono@nbi.ku.dk and $\ddag$ jesper.ilves@qipe.t.u-tokyo.ac.jp\\}

\author{A. F. van Loo$^{1,3}$}
\thanks{Present address: Alice \& Bob, France}

\author{Y. Sunada$^{3}$}
\author{C. W. Sandbo Chang$^{1}$}
\author{Y. Takeda$^{3}$}
\author{K. Yuki$^{3}$}
\author{T. Miyamura$^{3}$}
\author{K. Matsuura$^{3}$}
\author{K. Koshino$^{4}$}
\thanks{Present address: Quantum Laboratory, Fujitsu Research, Japan}
\author{Y. Nakamura$^{1,3}$}%

\affiliation{$^{1}$RIKEN Center for Quantum Computing, 2-1 Hirosawa, Wako-shi, Saitama 351-0198, Japan}
\affiliation{$^{2}$NNF Quantum Computing Programme, Niels Bohr Institute, University of Copenhagen, 2100 Copenhagen, Denmark}
\affiliation{$^{3}$Department of Applied Physics, Graduate School of Engineering, The University of Tokyo, 7-3-1 Hongo, Bunkyo-ku, Tokyo 113-8656, Japan}
\affiliation{$^{4}$Institute for Liberal Arts, Institute of Science Tokyo, 2-8-30 Konodai, Ichikawa-shi, Chiba 272-0827, Japan}


\date{Apr 15, 2026}

\begin{abstract}

Quantum-limited parametric amplifiers are essential components for many quantum technologies operating in the microwave domain. Achieving both high gain and broad bandwidth, however, remains challenging due to trade-offs between gain and bandwidth, pump efficiency, and dynamic range. Moreover, high-gain broadband amplifiers become increasingly sensitive to their external electromagnetic environment, which can distort their gain spectra and hinder reliable operation. Here, we present an accurate theoretical model and a systematic design methodology for a flux-driven, lumped-element Josephson parametric amplifier based on a SQUID array. Our device achieves near-quantum-limited, phase-preserving amplification with a net gain of 20 (maximally 44)~dB and a 3-dB bandwidth of $\sim$50 ($\lesssim 0.2$)~MHz. We further show that the gain spectra exhibit pronounced sensitivity to weak reflections in the input-output waveguide caused by impedance mismatches in the microwave environment. By incorporating Fabry--P\'erot-type interference into a quantum input-output model, we analytically reproduce these complex spectral features and identify how they depend on the physical parameters of the environment. More generally, our results provide a practical framework for separating the intrinsic dynamics of parametric amplifiers from environmental effects. This approach enables reliable characterization and optimization of amplifier performance while providing a systematic strategy for diagnosing microwave reflections and engineering environmental interference to shape amplifier gain spectra, thereby offering a pathway toward robust, reproducible, and truly quantum-limited microwave amplification.

\end{abstract}

\maketitle

\section{Introduction}

Quantum-limited amplifiers enable the precise measurement of weak signals at the quantum level~\cite{caves1982quantum, bib:clerk2010introduction}. 
They are therefore essential for microwave quantum technologies~\cite{bib:krantz2019quantum, bib:blais2021circuit}, where efficient photon detectors remain difficult to realize despite several recent experimental demonstrations~\cite{inomata2016single, bib:kono2018quantum, bib:besse2020parity, bib:kokkoniemi2020bolometer}. 
The Josephson nonlinearity, when combined with a strong microwave pump field, enables efficient parametric processes at the quantum limit~\cite{bib:synthesis}, allowing quantum-limited amplification~\cite{bib:castellanos2008amplification, bib:fluxJPA1}, vacuum squeezing~\cite{bib:castellanos2008amplification, bib:zhong2013squeezing, kono2017nonclassical}, and the generation of entangled microwave fields~\cite{bib:eichler2011observation, bib:menzel2012path, flurin2012generating, bib:jpateleport}. In particular, quantum-limited microwave amplifiers have enabled fast and high-fidelity readout of superconducting qubits~\cite{vijay2011observation, bib:optdispersive, sunada2022fast}. With the rapid scaling of superconducting quantum processors, exemplified by recent large-scale demonstrations~\cite{bib:arute2019quantum}, the demand for high-performance quantum-limited amplifiers has increased substantially~\cite{heinsoo2018rapid, bib:googlero}.

The ideal performance of such amplifiers would combine large and broadband gain, quantum-limited added noise, and a wide dynamic range while requiring minimal pump power. Achieving these goals, however, is challenging due to the fundamental gain--bandwidth constraint of resonator-based parametric amplifiers~\cite{bib:clerk2010introduction}. Moreover, strong parametric interactions generally introduce residual nonlinearities that limit operation at high gain and restrict the achievable dynamic range~\cite{abdo2013nondegenerate, bib:kochetov2015higher, bib:liu2017josephson, bib:sqarray2}.

To overcome these challenges, two main strategies have been pursued. The first is to employ traveling waves in a nonlinear transmission line, where a signal is amplified by a parametric process induced by a co-propagating pump~\cite{bib:esposito2021perspective}. This approach can achieve very large bandwidths~\cite{bib:macklin2015near, bib:white2015traveling, bib:sandbotwpa}, but requires careful phase matching among propagating modes~\cite{o2014resonant}, complicating design~\cite{bib:planat2020photonic, bib:ranadive2022kerr, bib:wang2025high} and fabrication~\cite{bib:macklin2015near}, and potentially limiting reproducibility. The second strategy is to engineer the external microwave environment of resonator-based amplifiers~\cite{aumentado2020superconducting} to modify and broaden the gain spectrum with impedance-matching networks~\cite{bib:impa1,bib:impa2}, for example, based on taper structures~\cite{bib:impa1,bib:googlero,bib:impa3} and impedance transformers~\cite{bib:impa2,bib:duan2021broadband,bib:duan2021broadband,bib:grebel2021flux,bib:ranzani2022wideband,bib:ezenkova2022broadband,bib:googleimpa2, bib:hung2025broadband}. 
In parallel, nonlinear engineering techniques have been developed to improve pump efficiency, suppress residual self-Kerr effects, and maximize dynamic range~\cite{bib:eichler2014controlling,bib:zhou2014high,bib:frattini2018optimizing}.

A key observation underlying both approaches is that high-gain, broadband parametric amplifiers are intrinsically sensitive to their external microwave environment. In superconducting qubit readout systems, for example, ferrite-based circulators or isolators are typically inserted between the qubit device and the amplifier~\cite{bib:optdispersive, sunada2022fast}. These components inevitably introduce small impedance mismatches and reflections in the measurement chain. Although such imperfections are often weak, the large gain and bandwidth of parametric amplifiers can make them highly sensitive to these reflections, leading to distortions of the gain spectrum and reduced reproducibility even for otherwise well-engineered devices~\cite{bib:impa1, bib:impa3, bib:ljpa2, bib:hung2025broadband}. Quantitatively understanding and controlling these environmental effects is therefore essential for the reliable operation of high-performance parametric amplifiers.

In this work, we develop resonator-based Josephson parametric amplifiers (JPAs) designed to achieve high gain, broad bandwidth, and a large dynamic range. In particular, we focus on a flux-driven JPA~\cite{bib:fluxJPA1} with a SQUID array~\cite{bib:zhou2014high, bib:esposito2019development}, which naturally separates the pump from the signal path~\cite{bib:ljpa2} and enables independent optimization of pump efficiency and dynamic range~\cite{bib:kaufman2025simple}. Guided by an accurate theoretical model and a systematic design optimization methodology, we demonstrate near-quantum-limited, phase-preserving amplification with net gains of 20~dB (up to 44~dB) and 3-dB bandwidths exceeding 50~MHz ($\lesssim 0.2$~MHz).

Furthermore, we show that the operation of these amplifiers is highly sensitive to weak impedance mismatches introduced by the surrounding microwave environment. 
Rather than employing a full circuit-theory treatment~\cite{bib:impa1}, we extend the quantum input--output formalism to incorporate impedance mismatches as Fabry--P\'erot-type interference effects~\cite{bib:fabry1899theorie}. This minimal theoretical model, involving only a small number of parameters, quantitatively reproduces the complex gain spectra observed in experiments. We also qualitatively describe the dependence of the gain spectra on the model parameters. This framework provides a clear physical picture of how reflections in the measurement chain modify the effective amplifier response. More importantly, it enables the separation of the intrinsic amplifier dynamics from the external environment, allowing independent characterization and optimization of both. As a result, our approach provides a practical tool for diagnosing reflections in cryogenic microwave wiring, identifying weak points in the measurement chain, and intentionally shaping amplifier gain spectra through controlled environmental engineering. These capabilities offer a general strategy for achieving robust and reproducible quantum-limited microwave amplification in realistic experimental settings.

This paper is organized as follows. In Sec.~II, we present the JPA design methodology. In Sec.~III, we develop the theoretical model describing the environmental interference effects. In Sec.~IV, we apply the theoretical model to experimentally characterize the amplifier performance. Finally, in Sec.~V, we summarize the results and discuss future prospects.

\section{Device modeling}

\subsection{Circuit description}
\begin{figure}[t]
\centering
  \includegraphics[width=78.864mm]{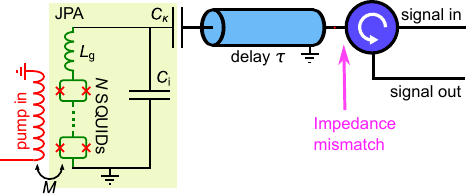} 
\caption{
Circuit diagram of a flux-driven JPA coupled to a waveguide containing a circulator with an impedance mismatch, forming an effective Fabry--P\'erot cavity between the JPA and the circulator.
The JPA consists of a capacitor in parallel with a SQUID array.
} 
  \label{fig:f1_1st} 
\end{figure}

\begin{figure}[t]
\centering
  \includegraphics[width=60mm]{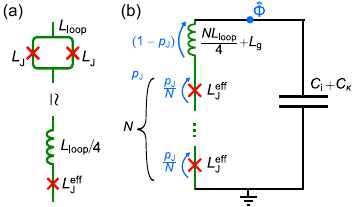} 
\caption{
Effective circuit modeling for
(a)~a symmetric SQUID with a finite loop inductance, and
(b)~a JPA incorporating a SQUID array.
} 
  \label{fig:f1-2_1st} 
\end{figure}

We consider a flux-driven lumped-element JPA incorporating a SQUID array. The circuit diagram of the JPA is shown in Fig.~\ref{fig:f1_1st}. The JPA consists of a parallel combination of a capacitor ($C_\mathrm{i}$) and an inductive element composed of a series connection of $N$ SQUIDs and a geometric inductor $L_\mathrm{g}$. 
Each SQUID is assumed to be symmetric and consists of a pair of Josephson junctions, each characterized by a Josephson inductance \(L_\mathrm{J}\), as shown in Fig.~\ref{fig:f1-2_1st}(a).
Additionally, it is capacitively coupled to a signal line via a coupling capacitor $C_\kappa$, while each SQUID is inductively coupled to a pump line with mutual inductance $M$.

To accurately derive the JPA Hamiltonian and optimize the circuit parameters, the loop inductance of each SQUID, characterized by \(L_\mathrm{loop}\), must be taken into account. However, the nonlinearity of the Josephson junctions precludes an analytical expression for the inductive energy of the SQUID in the presence of the finite loop inductance and its external-flux dependence~\cite{rymarz2023consistent}.

As detailed in Appendix~\ref{app:squidtheory}, we employ a perturbative approach to analytically derive the charge--flux relation of the SQUID with an external magnetic flux by retaining only the leading nonlinear terms. 
As schematically shown in Fig.~\ref{fig:f1-2_1st}(a), this analysis shows that each SQUID can be approximately reduced to a series combination of a linear inductor $L_{\mathrm{loop}}/4$ and a Josephson junction with a tunable Josephson inductance
$L_{\mathrm{J}}^{\mathrm{eff}} = L_{\mathrm{J}}/(2|\!\cos\!\phi_\mathrm{ex}^\mathrm{eff}|)$,
where $\phi_\mathrm{ex}^\mathrm{eff}$ is the effective external flux through the SQUID loop, which slightly differs from the applied external flux due to the finite loop inductance.
Note that the Josephson capacitance can be safely neglected in our circuit model within the Born--Oppenheimer approximation~\cite{kafri2017tunable}.
Furthermore, we show that the total energy of a series combination of linear inductors and Josephson junctions can be written as the sum of the energies of the individual components, with the flux distribution across each element determined by the linear parts of inductance via Kirchhoff’s laws.
This result is consistent with analyses based on energy-participation ratios~\cite{minev2021energy} and Josephson harmonics~\cite{willsch2024observation, kim2025emergent}.

It is therefore convenient to define the participation ratio of the sum of the flux-tunable Josephson inductance to the total inductance as
\begin{equation}
p_\mathrm{J} = \frac{N L_\mathrm{J}^\mathrm{eff}}{L_\mathrm{tot}},
\end{equation}
where the total inductance of the SQUID array is given by $L_\mathrm{tot}=N \left(L_\mathrm{loop}/4 + L_\mathrm{J}^\mathrm{eff} \right) + L_\mathrm{g}$. For convenience, we also define the total linear inductance as $L_l=N (L_\mathrm{loop}/4) + L_\mathrm{g}$, yielding its participation ratio as $1-p_\mathrm{J}$.

As a result, as schematically shown in Fig.~\ref{fig:f1-2_1st}(b), the flux distribution across the individual inductive elements can be expressed in terms of the JPA flux operator \(\hat{\Phi}\), yielding \((1-p_\mathrm{J})\hat{\Phi}\) for the linear inductance and \(p_\mathrm{J}\hat{\Phi}/N\) for each tunable Josephson junction.
Thus, the total inductive energy of the present JPA is given by
\begin{equation}\label{eq:squidenergy}
    \hat{U} = E_{L_l} \!\Bigg(\! (1-p_\mathrm{J}) \frac{\hat{\Phi}}{\Phi_0} \!\Bigg)^2\! + NE_{L_\mathrm{J}^\mathrm{eff}} \: (-2)\!\cos\!\left( \frac{p_\mathrm{J}}{N}\frac{\hat{\Phi}}{\Phi_0}\right),
\end{equation}
where 
$E_{L_l}  = \Phi_0^2/(2L_l)$
and 
$E_{L_\mathrm{J}^\mathrm{eff}} = \Phi_0^2/(2L_\mathrm{J}^\mathrm{eff})$
are the inductive energies associated with a single reduced quantum of magnetic flux (\( \Phi_0 = \hbar/(2e) \)) for the total linear inductance and for each flux-tunable Josephson junction, respectively.

For circuit quantization, it is useful to define the inductive energy associated with a single flux quantum for the total inductance as $E_{L_\mathrm{tot}} = \Phi_0^2/(2L_\mathrm{tot})$.
Using
$E_{L_l} = E_{L_\mathrm{tot}}/(1-p_\mathrm{J})$
and
$E_{L_\mathrm{J}^\mathrm{eff}} = N E_{L_\mathrm{tot}}/p_\mathrm{J}$,
the total inductive energy can be expressed in terms of $E_{L_\mathrm{tot}}$ as
\begin{equation}
\label{eq:squidenergy2}
\hat{U}
= E_{L_\mathrm{tot}}
\left[
\left(\!\frac{\hat{\Phi}}{\Phi_0}\!\right)^2
- \frac{p_\mathrm{J}^3}{12 N^2}\!
\left(\!\frac{\hat{\Phi}}{\Phi_0}\!\right)^4
+ \cdots
\right],
\end{equation}
where the cosine potential has been expanded to fourth order, which is crucial to calculate the self-Kerr nonlinearity for the next section.

On the other hand, the capacitive energy of the JPA is simply obtained as
\begin{equation}\label{eq:capenergy}
    \hat{T}=E_{C_\mathrm{tot}} \left(\frac{\hat{Q}}{2e}\right)^2,
\end{equation}
where $E_{C_\mathrm{tot}}=(2e)^2/(2C_\mathrm{tot})$ is the charging energy of a single Cooper pair ($2e$) with the total capacitance of $C_\mathrm{tot}=C_\mathrm{i} + C_\kappa$, and $\hat{Q}$ is the charge operator for the JPA.

\subsection{JPA Hamiltonian}
By combining the inductive energy in Eq.~\eqref{eq:squidenergy} with the capacitive energy in Eq.~\eqref{eq:capenergy}, and retaining only the leading nonlinear term, the Hamiltonian of the flux-pumped JPA can be expressed in the Fock basis as 
\begin{equation} \label{eq:JPAhamiltonian_main}
\hat{H} = \hbar\omega_\mathrm{a} \: \hat{a}^\dag \hat{a} + \frac{\hbar K}{2}\hat{a}^{\dag^2}{\hat{a}}^2 + \frac{\hbar \Omega_\mathrm{p}}{4}\!\left(\hat{a}^{\dag^2}e^{-i\omega_\mathrm{p}t} + \mathrm{H.c.}\right),
\end{equation}
where $\hat{a}$ is the annihilation operator, $\omega_\mathrm{a}$ is the resonance frequency, and $K$ is the self-Kerr coefficient.
As detailed in Appendix~\ref{app:fluxpumping}, the two-photon drive term originates from the external-flux derivative of the inductive energy, with \(\omega_\mathrm{p}\) and \(\Omega_\mathrm{p}\) (assumed real for simplicity) denoting the frequency and amplitude of the two-photon drive, respectively.
See Appendix~\ref{app:jpasquidham} for a full derivation of the system parameters.

Instead of directly optimizing the JPA based on circuit parameters, we introduce a set of dimensionless parameters to facilitate the analysis. 
Here, the system parameters in the Hamiltonian can be expressed in a dimensionless form using the JPA resonance frequency $\omega_\mathrm{a} = 1/\sqrt{L_\mathrm{tot}C_\mathrm{tot}} = 2\sqrt{E_{L_\mathrm{tot}}E_{C_\mathrm{tot}}}/\hbar$, and are subsequently rewritten in terms of the following characteristic dimensionless quantities.

The effective fine-structure constant~\cite{kuzmin2019quantum} of the JPA is defined as 
\begin{equation}
\label{eq:alphaa}
\alpha_\mathrm{a} = \frac{Z}{Z_\mathrm{Q}} = \sqrt{\frac{E_{C_\mathrm{tot}}}{E_{L_\mathrm{tot}}}},
\end{equation}
where $Z = \sqrt{L_\mathrm{tot}/C_\mathrm{tot}}$ is the characteristic impedance of the JPA, and $Z_\mathrm{Q} = \Phi_0/(2e)$ is the quantum of resistance. 
Similarly, the fine-structure constant of the input/output waveguide with characteristic impedance $Z_0$ is defined as $\alpha_0 = Z_0/Z_\mathrm{Q}$.
In addition to the participation ratio $p_\mathrm{J}$, the ratio of the effective Josephson inductance to the loop inductance in the reduced circuit model of each SQUID is defined as 
\begin{equation}
\label{eq:pSQ}
p_\mathrm{SQ} = \frac{L_\mathrm{J}^\mathrm{eff}}{L_\mathrm{loop}/4 + L_\mathrm{J}^\mathrm{eff}}.
\end{equation}
This gives the achievable maximum $p_\mathrm{J}$, namely, $p_\mathrm{J} \leq p_\mathrm{SQ} $ with equality if and only if $L_\mathrm{g} = 0$~H.
Moreover, the participation ratio of the coupling capacitor to the total capacitance is defined as
\begin{equation}
\label{eq:pkappa}
p_\kappa = C_\kappa/C_\mathrm{tot}.
\end{equation}

Using the above dimensionless quantities, we derive that the dimensionless self-Kerr coefficient takes the form
\begin{equation}
\label{eq:barK}
\bar{K} = \frac{K}{\omega_\mathrm{a}} = -\frac{\alpha_\mathrm{a} \, p_\mathrm{J}^3}{8N^2},
\end{equation}
and the dimensionless two-photon drive amplitude is
\begin{equation}
\label{eq:barOmegap}
\bar{\Omega}_\mathrm{p} = \frac{\Omega_\mathrm{p}}{\omega_\mathrm{a}} = \frac{p_\mathrm{J}\: p_\mathrm{SQ} \tan\!\phi_\mathrm{ex}^\mathrm{eff}}{4} \, \phi_\mathrm{p},
\end{equation}
where \(\phi_\mathrm{p}\) denotes the amplitude of the pump flux, expressed in units of the reduced flux quantum, threading the SQUID loop.
The dimensionless external coupling rate, whose inverse corresponds to the quality factor, is given by
\begin{equation}
\label{eq:barkappa}
\bar{\kappa} = \frac{\kappa}{\omega_\mathrm{a}} = \frac{\alpha_0}{\alpha_\mathrm{a}}\, p_\kappa^2.
\end{equation}

In this dimensionless framework, once the parameters $\alpha_\mathrm{a}$, $p_\kappa$, $p_\mathrm{J}$, and $p_\mathrm{SQ}$, along with the resonance frequency $\omega_\mathrm{a}$, are specified to achieve target values of $\bar{K}$, $\bar{\Omega}_\mathrm{p}$, and $\bar{\kappa}$, all circuit parameters are uniquely determined.

\subsection{Design methodology}
The gain spectra of JPAs can be primarily characterized by three key metrics: gain $G$, gain bandwidth $B$, and dynamic range, often quantified by the 1-dB compression point.
Given the gain--bandwidth relationship for single-pole devices, $B \approx \kappa/(2\sqrt{G})$~\cite{bib:clerk2010introduction}, the gain bandwidth can be increased by enhancing the external-coupling rate $\kappa$. 
To achieve sufficient gain, however, the JPA must be pumped such that the two-photon drive amplitude approaches the external coupling rate, i.e., $\Omega_\mathrm{p} \approx \kappa$. 
Consequently, the gain bandwidth is often limited by the achievable two-photon drive amplitude $\Omega_\mathrm{p}$, indicating that a larger participation ratio $p_\mathrm{J}$ is desirable to enhance the pump efficiency, thereby increasing the gain bandwidth, as suggested by Eq.~(\ref{eq:barOmegap}).
On the other hand, the compression point is known to be inversely proportional to the self-Kerr coefficient $K$~\cite{bib:sqarray2, bib:frattini2018optimizing} (also see Appendix~\ref{app:gainsat}), implying that the dynamic range improves as $K$ decreases, i.e., as $p_\mathrm{J}$ decreases, according to Eq.~(\ref{eq:barK}). 
This reveals a fundamental trade-off between pump efficiency and dynamic range in the JPA optimization.



In our design, the target operating frequency of the JPA is set to $\omega_\mathrm{a}/2\pi = 9.5~\mathrm{GHz}$.
Due to design and fabrication constraints, the loop inductance is fixed at $L_\mathrm{loop}=20~\mathrm{pH}$, corresponding to a loop area of approximately
$10\times10~\mu\mathrm{m}^2$.
This yields a mutual inductance between each SQUID loop and the pump line of approximately \(M \approx 5~\mathrm{pH}\).
To enhance the pump efficiency, namely \(p_\mathrm{J}\) in Eq.~\eqref{eq:barOmegap}, we aim to minimize the geometric inductance \(L_\mathrm{g}\), thereby approaching the limit imposed by the finite loop inductance, i.e., \(p_\mathrm{J}\rightarrow p_\mathrm{SQ}\). Moreover, to achieve a larger external coupling rate in Eq.~\eqref{eq:barkappa} and a smaller self-Kerr coefficient in Eq.~\eqref{eq:barK}, we target a smaller fine-structure constant in Eq.~\eqref{eq:alphaa}, which further motivates minimizing \(L_\mathrm{g}\). These considerations lead to a minimal residual geometric inductance in our circuit geometry, namely \(L_\mathrm{g}=80~\mathrm{pH}\).
Additionally, the Josephson participation ratio $p_\mathrm{J}$ must be chosen to balance the tradeoff between the self-Kerr nonlinearity in Eq.~\eqref{eq:barK} and the pump efficiency in Eq.~\eqref{eq:barOmegap}.
As a preliminary design choice, we set \(p_\mathrm{J}=0.8\) and suppress the nonlinearity by increasing the number of SQUIDs to \(N=5\)~\cite{bib:zhou2014high, bib:esposito2019development}.
Finally, the normalized external coupling rate to the 50-\(\Omega\) waveguide is preliminarily set to \(\bar{\kappa}=0.04\), guided by the amplification condition \(\bar{\Omega}_\mathrm{p}\approx\bar{\kappa}\).

These design choices fully determine the remaining dimensionless design parameters of the JPA, yielding \(\alpha_\mathrm{a}=0.03\), \(p_\mathrm{SQ}=0.94\), and \(p_\kappa=0.15\).
As a result, we obtain a dimensionless Kerr coefficient of $\bar{K}=-7.5\times10^{-5}$.
Finally, these dimensionless parameters determine the remaining circuit parameters, yielding
$C_\mathrm{i}=470~\mathrm{fF}$,
$C_\kappa=80~\mathrm{fF}$, and
$L_\mathrm{J}^\mathrm{eff}=80~\mathrm{pH}$, where the value of $L_\mathrm{J}^\mathrm{eff}=L_{\mathrm{J}}/(2|\!\cos\!\phi_\mathrm{ex}^\mathrm{eff}|)$ is targeted at the bias point $\phi_\mathrm{ex}^\mathrm{eff}=\pi/3$.

\section{Theoretical descriptions}
The ideal performance characteristics of JPAs, particularly high gain and large bandwidth, make them highly sensitive to their external electromagnetic environment, namely the waveguide that interfaces their input--output fields. In practice, when a JPA is deployed in a measurement chain, a circulator is required to separate the incoming and outgoing signals (see Fig.~\ref{fig:f1_1st}). However, the circulator often introduces a finite, unavoidable impedance mismatch, resulting in small reflections of the input--output fields. These reflections give rise to Fabry--Pérot-like interference effects that distort the gain spectra.

In this section, we present an analytical, quantum-mechanical investigation of how such imperfections affect the gain spectra of the JPA within the input--output formalism.

\subsection{Heisenberg equation and input--output relation}

\begin{figure}[t]
\centering
  \includegraphics[width=85.241mm]{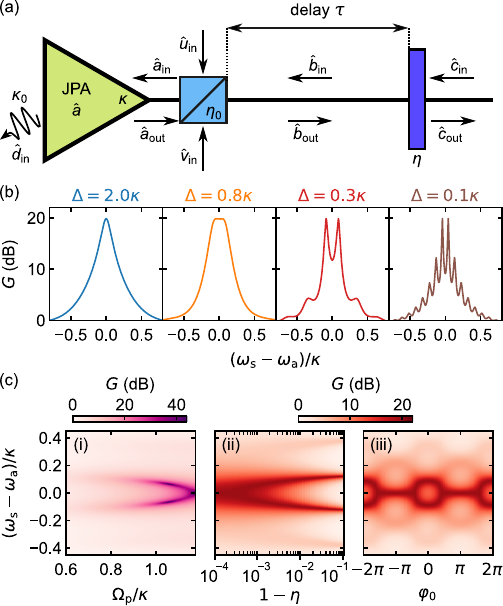} 
\caption{
(a)~Theoretical model of a JPA coupled to a waveguide through a Fabry--Pérot cavity based on the input--output formalism.
(b)~Calculated gain spectra based on the theoretical model as a function of input signal detuning from the JPA frequency for four different Fabry--P\'erot cavity free spectral ranges $\Delta$. 
(c)~Gain spectra dependence on 
(i)~JPA pump amplitude $\Omega_\mathrm{p}$, 
(ii)~Fabry--P\'erot-cavity mirror-reflection coefficient $1 - \eext$, and 
(iii)~round-trip phase $\varphi_0$. 
When not explicitly varied in the plots, we fix the parameters as $\Delta = 0.3\kappa$, $\eext = 0.995$, $\eint = 0.9$, and $\varphi_0 = 0$.
The JPA pump amplitudes are chosen such that the maximum gain is \SI{20}{\decibel} in all panels except (c)(i).
} 
  \label{fig:f1_1st_b}
\end{figure}

Here, we consider a general model in which the field operator \(\hat{a}\) of a quantum system is coupled to its input and output fields. 
Specifically, as shown in Fig.~\ref{fig:f1_1st_b}(a), a partially reflecting mirror is placed at a distance corresponding to a time delay $\tau$, which induces Fabry--P\'erot interference in the waveguide. 
As we demonstrate below, the Heisenberg equation and input–output relation of the quantum system can be compactly formulated by combining the individually defined input--output relations of all relevant fields, with the time delay explicitly incorporated.

Within a standard theoretical description, the Heisenberg equation for the field operator $\hat{a}$ of the quantum system, driven by input fields $\hat{a}_\mathrm{in}$ and $\hat{d}_\mathrm{in}$ from the waveguide and the intrinsic loss channel, respectively, can be written as
\begin{equation}
\label{eq:Heisenbergeq_a}
\begin{split}
\frac{\di\hat{a}(t)}{\di t} =  \frac{i}{\hbar} \left[ \hat{H}, \:\hat{a}(t) \right] &- \frac{\kext + \kint}{2}\hat{a}(t) \\ &- i \sqrt{\kext}\:\ain\!(t)  - i \sqrt{\kint}\:\din\!(t),
\end{split}
\end{equation}
while the input--output relation between the fields of $\ain$ and $\aout$ is given by
\begin{equation}
\label{eq:aoutaina}
    \aout\!(t) = \ain\!(t) - i \sqrt{\kappa} \: \hat{a}(t),
\end{equation}
where $\kext$ and $\kint$ are the external-coupling rate and the intrinsic-loss rate of the quantum system, respectively.

In our model, the input--output fields of the system are subject to propagation losses in the waveguide as well as reflections from a partially reflecting mirror that represents the impedance mismatch down the line, for example, due to a circulator. As schematically shown in Fig.~\ref{fig:f1_1st_b}(a), the propagation losses are modeled as mixing with vacuum fields of $\uin$ and $\vin$, and the corresponding field relations are given by
\begin{align}
\label{eq:beamsplitter_eta0}
    \ain\!(t) &= \sqrt{\eint} \:\bin\!(t) + \sqrt{1- \eint} \:\uin\!(t),\\
\label{eq:beamsplitter_eta0-2}
    \bout\!(t) &= \sqrt{\eint} \:\aout\!(t) + \sqrt{1- \eint} \:\vin\!(t),
\end{align}
where the propagation loss coefficients for both directions are assumed to be equivalent and are characterized by $\eint$.
Additionally, the input–output relations between the fields inside the Fabry--P\'erot cavity and the external fields at the mirror are described by
\begin{equation}
\label{eq:beamsplitter_eta}
\begin{pmatrix}
\bin\!(t+\tau) \\
\cout\!(t-\tau)
\end{pmatrix}
=
\begin{pmatrix}
T^* & R \\
-R^* & T
\end{pmatrix}
\!
\begin{pmatrix}
\cin\!(t+\tau) \\
\bout\!(t-\tau)
\end{pmatrix},
\end{equation}
where $T=|T|e^{i\phi_T}$ and $R=|R|e^{i\phi_R}$ are the amplitude transmission and reflection coefficients of the mirror, respectively. 
The transmittance can also be defined as $\eta = |T|^2 = 1 - |R|^2$.
In the formulation of the above field relations, the time associated with each field is defined relative to the moment when it first interacts with the quantum system.

By eliminating the auxiliary field degrees of freedom, namely, $\ain$, $\aout$, $\bin$ and $\bout$, using the above field relations, we can derive the Heisenberg equation for the field operator $\hat{a}$, driven by the external signal field $\cin$ and the vacuum fields $\din$, $\uin$, and $\vin$ from the individual loss channels.
However, the partial reflections at the mirror cause the incoming and outgoing fields to return to the quantum system with a round-trip time delay of $2\tau$, leading to nontrivial interference that complicates the analytical derivation of the effective Heisenberg equation and the input--output relation. 
Nevertheless, when only a specific frequency component \(\omega\) is relevant, the time delay can be treated as an effective phase shift of \(2\omega\tau\).
As explained in detail in Appendix~\ref{app:iojpafp}, it is convenient to define the coefficient accumulated during a round trip inside the Fabry--Pérot cavity as
\begin{equation}
\label{eq:r}
    \mathcal{R}_{(\omega)} = \eint \sqrt{1-\eext}\: e^{i (2 \omega \tau+\phi_R)}.
\end{equation}

By considering the infinite number of reflections occurring in the Fabry--P\'erot cavity, the explicit form of the effective Heisenberg equation can be described as
\begin{widetext}
\begin{equation}
\begin{split}\label{eq:heisenberg_main}
    \frac{\di\hat{a}(t)}{\di t} = &\frac{i}{\hbar} \left[ \hat{H}, \:\hat{a}(t) \right]  - \frac{\kext + \kint}{2}\hat{a}(t)  -\kext 
 \frac{\mathcal{R}_{(\omega)}}{1 - \mathcal{R}_{(\omega)}}\hat{a}(t) -i \frac{T^*\!\sqrt{\eint\kext}}{1 - \mathcal{R}_{(\omega)}}\cin\!(t)\\
    & -i \frac{\sqrt{(1-\eint)\kext}}{1 - \mathcal{R}_{(\omega)}}\uin\!(t) -i \sqrt{\frac{(1-\eint)\kext}{\eint}}\frac{\mathcal{R}_{(\omega)}}{1 - \mathcal{R}_{(\omega)}}\vin\!(t)  - i \sqrt{\kint}\:\din\!(t),
\end{split}
\end{equation}
while the input--output relation can also be described as
\begin{equation}
\label{eq:inout_main}
    \cout\!(t) = -i  \frac{T\sqrt{\eint \kappa}}{1 - \mathcal{R}_{(\omega)}} \hat{a}(t)
    + \left( - \frac{\mathcal{R}^*_{(\omega)}}{\eint} 
    + \frac{|T|^2 \eint}{1 - \mathcal{R}_{(\omega)}} \right) \cin\!(t)
    + \frac{T\sqrt{\eint(1 - \eint)}}{1 - \mathcal{R}_{(\omega)}} \uin\!(t) 
    + \frac{T\sqrt{1-\eint} }{1 - \mathcal{R}_{(\omega)}} \vin\!(t).
\end{equation}
We note that the Heisenberg equation and the input--output relations are valid when the dynamics are considered within a narrow frequency range around \(\omega\).
In the continuous-wave (stationary) experiments throughout this work, the frequency \(\omega\) corresponds to the signal frequency in reflection measurements (without flux pumping) or to the signal and idler frequencies of the corresponding fields in parametric gain measurements (with flux pumping).

\end{widetext}

\subsection{Reflection spectra}
When it is not flux-pumped and is weakly probed, the JPA behaves as a linear harmonic oscillator with Hamiltonian
\(
\hat{H} = \hbar\omega_\mathrm{a} \: \hat{a}^\dag \hat{a}
\).
The reflection spectrum can be obtained by applying a Fourier transform to the Heisenberg equation~\eqref{eq:heisenberg_main} together with the input--output relation~\eqref{eq:inout_main}, evaluated at the signal frequency \(\omega_\mathrm{s}\). The resulting reflection spectrum is given by
\begin{equation}
\begin{split}
\label{eq:ana_reflextion_spe}
    S_{11}(\omega_\mathrm{s}) = &- \frac{\mathcal{R}^*_{(\omega_\mathrm{s})}}{\eint} 
    + \frac{\eext \eint}{1 - \mathcal{R}_{(\omega_\mathrm{s})}} \\ &+ \frac{\eext \eint\kext}{\left(1 - \mathcal{R}_{(\omega_\mathrm{s})}\right)^2\chi^{-1}_{(\omega_\mathrm{s})}}.
\end{split}
\end{equation}
In the above equation, the response function $\chi^{-1}_{(\omega)}$ is defined as
\begin{equation}
\label{eq:chi-1omega}
\chi^{-1}_{(\omega)} = i(\omega - \omega_\mathrm{a}) - \frac{\kext + \kint}{2} - \kext 
\frac{\mathcal{R}_{(\omega)}}{1 - \mathcal{R}_{(\omega)}}.
\end{equation}
The real and imaginary parts of the last term correspond to a modification of the decay rate and a frequency shift, respectively, both of which depend on the signal frequency.

In contrast, when the device is flux pumped at approximately twice the JPA resonance frequency (\(\omega_\mathrm{p} \approx 2\omega_\mathrm{a}\)), the three-wave-mixing parametric process couples a signal field at \(\omega_\mathrm{s}\) to the corresponding idler field at \(\omega_\mathrm{i}\), satisfying the frequency relation \(\omega_\mathrm{p}=\omega_\mathrm{s}+\omega_\mathrm{i}\).
Consequently, both the signal and idler fields are included in the calculation of the gain spectrum.
As detailed in Appendix~\ref{app:effectiveJPAinput}, the coupled equations for the signal and idler components are described in the frequency domain, where the effects of Fabry--Pérot interference are explicitly evaluated at the corresponding frequencies. Neglecting the self-Kerr effect in the Hamiltonian of Eq.~\eqref{eq:JPAhamiltonian_main}, the reflection spectrum of the flux-pumped JPA (gain spectrum) can be expressed as
\begin{equation}
\label{eq:ana_gain_spe}
\begin{split}
    S_{11}(\omega_\mathrm{s}) = &- \frac{\mathcal{R}^*_{(\omega_\mathrm{s})}}{\eint} 
    + \frac{\eext \eint}{1 - \mathcal{R}_{(\omega_\mathrm{s})}} \\ &+ \frac{\eext \eint\kext \chi^{-1*}_{(\omega_\mathrm{i})}}{\left(1 - \mathcal{R}_{(\omega_\mathrm{s})}\right)^2\left(\chi^{-1}_{(\omega_\mathrm{s})}\chi^{-1*}_{(\omega_\mathrm{i})} - \frac{|\Omega_\mathrm{p}|^2}{4}\right)}.
\end{split}
\end{equation}

In our analysis of both the unpumped and flux-pumped cases, the experimentally obtained reflection spectra are divided by a reference signal to facilitate comparison with the analytical expressions. The reference signal is obtained as a reflection spectrum measured when the JPA is far detuned from the measurement frequency window using a DC flux bias.
In this regime, the input--output relation at the JPA can be approximated as
\begin{equation}
    \aout\!(t) = \ain\!(t) e^{i\phi_\mathrm{r}},
\end{equation}
where $\phi_\mathrm{r}$ is an additional phase factor introduced by the reflection from the far-detuned JPA.
The reference reflection spectrum is then given analytically by
\begin{equation}
S_{11}^\mathrm{ref}(\omega_\mathrm{s}) =- \frac{\mathcal{R}^*_{(\omega_\mathrm{s})}}{\eint} 
    + \frac{\eext \eint e^{i\phi_\mathrm{r}}}{1 - \mathcal{R}_{(\omega_\mathrm{s})}e^{i\phi_\mathrm{r}}}.
\end{equation}
Using this reference, the analytical expression for the normalized reflection spectrum is given by
\begin{equation}
\label{eq:jpagainbgsub}
\tilde{S}_{11}(\omega_\mathrm{s})
= \frac{S_{11}(\omega_\mathrm{s})}{S_{11}^\mathrm{ref}(\omega_\mathrm{s})},
\end{equation}
which enables direct comparison with the experimental results, denoted by
$\tilde{S}_{11}^\mathrm{meas}(\omega_\mathrm{s})$, obtained using the same
normalization procedure as in the analytical derivation.
While we use Eq.~\eqref{eq:jpagainbgsub} combined with Eq.~\eqref{eq:ana_reflextion_spe} to fit the experimental data of the normalized reflection spectra without flux pumping, we use the modulus squared of Eq.~\eqref{eq:jpagainbgsub} combined with Eq.~\eqref{eq:ana_gain_spe}, i.e., $\tilde{G}(\omega_\mathrm{s})=\lvert \tilde{S}_{11}(\omega_\mathrm{s}) \rvert^2$ to fit the experimentally obtained gain spectra, $\tilde{G}_{\mathrm{meas}}(\omega_\mathrm{s}) = \lvert \tilde{S}^{\mathrm{meas}}_{11}(\omega_\mathrm{s}) \rvert^2$.


Here, it is crucial to minimize the number of model parameters in order to stabilize the following fitting process, particularly for the analysis of the gain spectra of the JPA. 
To remain consistent with the experimental conditions, the pump frequency is fixed to twice the JPA resonance frequency, \(\omega_\mathrm{p}=2\omega_\mathrm{a}\).
In addition, we describe the reflection spectra in terms of the frequency offset $\delta$ from half the pump frequency, i.e., the JPA frequency. Accordingly, the signal and idler frequencies are redefined as $\omega_\mathrm{s} = \delta + \omega_\mathrm{a} $ and $\omega_\mathrm{i} = -\delta + \omega_\mathrm{a}$, respectively.
Namely, we redefine the notation, for example, \(S_{11}(\omega_\mathrm{s}) \rightarrow S_{11}(\delta)\), where \(\delta = \omega_\mathrm{s}-\omega_\mathrm{a}\).
Moreover, it is useful to define the free spectral range of the Fabry--P\'erot cavity as 
\begin{equation}
\Delta = 2\pi/(2\tau), 
\end{equation}
which corresponds to the periodicity observed in the spectra.
Thus, the reflection spectrum $S_{11}(\delta)$ can be described as a function of $\delta$ using
\begin{equation}
\label{eq:r_simple}
    \mathcal{R}_{(\omega_\mathrm{s/i})} = \eint \sqrt{1-\eext}\: e^{i (\pm 2\pi \delta/\Delta+\varphi_0)}
\end{equation}
and 
\begin{equation}
\label{eq:chi-1omega_simple}
\chi^{-1}_{(\omega_\mathrm{s/i})} = \pm i\delta - \frac{\kext + \kint}{2} - \kext 
\frac{\mathcal{R}_{(\omega_\mathrm{s/i})}}{1 - \mathcal{R}_{(\omega_\mathrm{s/i})}},
\end{equation}
where the round-trip phase $\varphi_0$ at half the pump frequency, corresponding to the JPA resonance frequency in our setting, is defined as 
\begin{equation}
\varphi_0 = \omega_\mathrm{a}\cdot (2\pi/\Delta) + \phi_\mathrm{R}. 
\end{equation}
Due to $\phi_\mathrm{R}$, the round-trip phase \(\varphi_0\) is an independent model parameter of $\Delta$ and plays a central role, as it quantifies the frequency detuning between the JPA and the Fabry--Pérot cavity modes and thus determines how interference effects modify the reflection spectra.

With these definitions, the analytical reflection spectrum \(S_{11}(\delta)\) is fully specified by the following independent fitting parameters: the external-coupling and intrinsic-loss rates of the JPA, \(\kext\) and \(\kint\); the transmittances of the external and internal mirrors of the Fabry--P\'erot cavity, \(\eext\) and \(\eint\); the free spectral range of the Fabry--P\'erot cavity, \(\Delta\); the round-trip phase, \(\varphi_0\); the reference phase offset, \(\phi_\mathrm{r}\); and, for the gain spectra, the pump amplitude, \(\Omega_\mathrm{p}\).

In Figs.~\ref{fig:f1_1st_b}(b) and (c), we qualitatively study the effect of the Fabry--Pérot interference on the JPA gain spectra by using Eq.~\eqref{eq:ana_gain_spe} to calculate analytical gain spectra for different combined system parameters. In Fig.~\ref{fig:f1_1st_b}(b), we vary the free spectral range $\Delta$, observing that for $\Delta > \kappa$, the gain spectrum has an approximately Lorentzian shape, while for $\Delta < \kappa$ the gain spectrum contains ripples with a spacing corresponding to $\Delta$. In the regime $\Delta \approx \kappa$, the ripples blend in together to form a flat-top gain spectrum. Thus, the characteristic shape of the gain spectrum depends strongly on the free spectral range, or equivalently on the length, of the Fabry--P'erot cavity.

In Fig.~\ref{fig:f1_1st_b}(c)(i), we show the dependence of the gain spectra on the JPA pump amplitude $\Omega_\mathrm{p}$. We observe that the spacing between the gain peaks becomes narrower as pump amplitude is increased. This can be explained by the gain bandwidth of the JPA becoming narrow compared to the free spectral range of the Fabry--Pérot cavity, reducing the impact of the interference on the gain spectrum.

As shown in Fig.~\ref{fig:f1_1st_b}(c)(ii), the Fabry--Pérot cavity mirror-reflection coefficient also affects the gain spectra considerably. 
For very low amounts of reflection ($1 - \eext \lesssim 0.001$), the effect of the interference is minimized since most of the signal is directly emitted to the waveguide, resulting in a single-peaked Lorentzian gain spectrum. 
For higher amounts of reflection ($1 - \eext \gtrsim 0.01$), the signal propagating inside the Fabry--Pérot cavity becomes stronger, eventually leading to strong coupling between the two systems, resulting in a well-separated double-peaked gain spectrum.

Finally, we consider how the accumulated round-trip phase, \(\varphi_0\), affects the gain spectra in Fig.~\ref{fig:f1_1st_b}(c)(iii). When the JPA is resonant with one of the Fabry--P\'erot cavity modes (\(\varphi_0 = 0\)), the Fabry--P\'erot interference causes the gain spectrum to split into two peaks. In contrast, when the JPA lies midway between adjacent Fabry--P\'erot resonances and is therefore detuned from all cavity modes (\(\varphi_0 = \pm \pi/2\)), the two peaks merge into a single peak at the JPA frequency, indicating a reduced influence of the Fabry--P\'erot cavity, although weak side peaks remain visible.


\subsection{Combined Fabry--Pérot-cavity-JPA system}
Because the JPA is strongly affected by imperfections in the circulator, which induce Fabry--Pérot interference, we analyze the gain spectra and added noise by treating the combined system of the JPA and the Fabry--Pérot cavity as the amplifier under test, hereafter referred to as the FPJPA.
This approach provides direct performance metrics for the practical use of reflection-type amplifiers.

In our case, the reflection at the Fabry--Pérot mirror (the circulator) is negligible when the JPA is not flux-pumped, i.e., $\eta \approx 1$, such that the reference spectrum can be approximated as $S_{11}^\mathrm{ref}(\delta) \approx \eta_0$.
By obtaining $\eta_0$ from the fitting analysis for the gain spectrum normalized by the reference, $\tilde{G}_{\mathrm{meas}}(\delta)$, and using the relation in Eq.~\eqref{eq:jpagainbgsub}, the net gain spectrum of the FPJPA system can be obtained as
\begin{equation}
\begin{split}
G_{\mathrm{meas}}(\delta) &\coloneqq \lvert S_{11}^\mathrm{ref}(\delta) \rvert^2\,\tilde{G}_{\mathrm{meas}}(\delta) \\
&\approx \eta_0^2\,\tilde{G}_{\mathrm{meas}}(\delta).
\end{split}
\end{equation}
These net gain spectra are shown throughout this work and are used for the quantitative analysis of the gain characteristics.

Moreover, the approximate reference spectrum enables straightforward calibration of the measurement chain, which is used to determine the added noise of the combined FPJPA system (see the calibration procedures in more detail in Appendix~\ref{app:calib}).

\section{Experimental results}

Here, we quantitatively demonstrate how the theoretical model described in the above section can be applied to characterize the performance of a fabricated high-performance JPA device together with the external environment in which it is embedded in.

\subsection{Fabricated device}
\begin{figure}[t]
\centering
  \includegraphics[width=86mm]{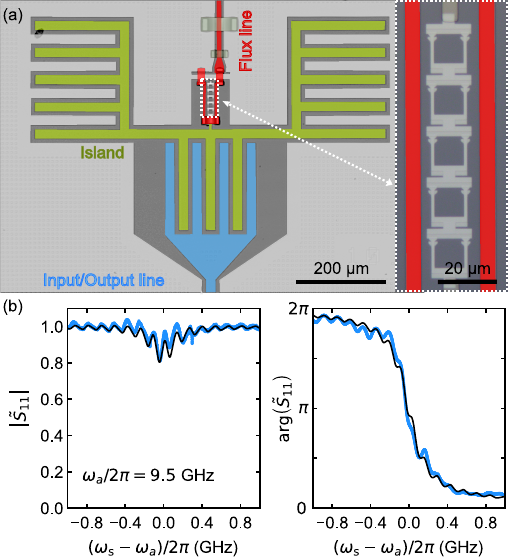} 
\caption{
Fabricated device and its reflection spectrum.
(a)~False-color microscope image of a fabricated JPA. 
The inset shows the SQUID array coupled to the flux line.
(b)~Measured reflection spectrum $\tilde{S}_{11}$ of the combined JPA-Fabry--P\'erot cavity system for $\omega_\mathrm{a}/2\pi= \SI{9.5}{\giga\hertz}$, normalized by the reference signal (blue dots). The solid black lines are fits based on our input--output model to the measured data.
} 
  \label{fig:f1_2nd} 
\end{figure}

We present a laser-microscope image of a fabricated JPA device in Fig.~\ref{fig:f1_2nd}(a). The JPA consists of an array of five SQUIDs (inset) fabricated in aluminum and aluminum oxide on a silicon wafer, with lumped-element circuit features patterned in a niobium film (blue and green). Aluminum airbridges connect the niobium sections to form the flux-line loop (red), designed to maximize the mutual inductance coupling to the SQUID loops (for a detailed description, see Appendix~\ref{app:fabrication}).

To characterize the JPA, we cool the device in a dilution refrigerator and measure its continuous-wave reflection signals using a microwave setup based on circulators under various operating conditions. The JPA is connected to the nearest circulator by normal-metal coaxial cables, which unintentionally form a lossy Fabry--Pérot cavity due to impedance mismatch at the circulator input port. A detailed schematic of the experimental setup is provided in Appendix~\ref{app:expsetup}.

%
\begin{figure*}[t]
\begin{center}
  \includegraphics[width=172mm]{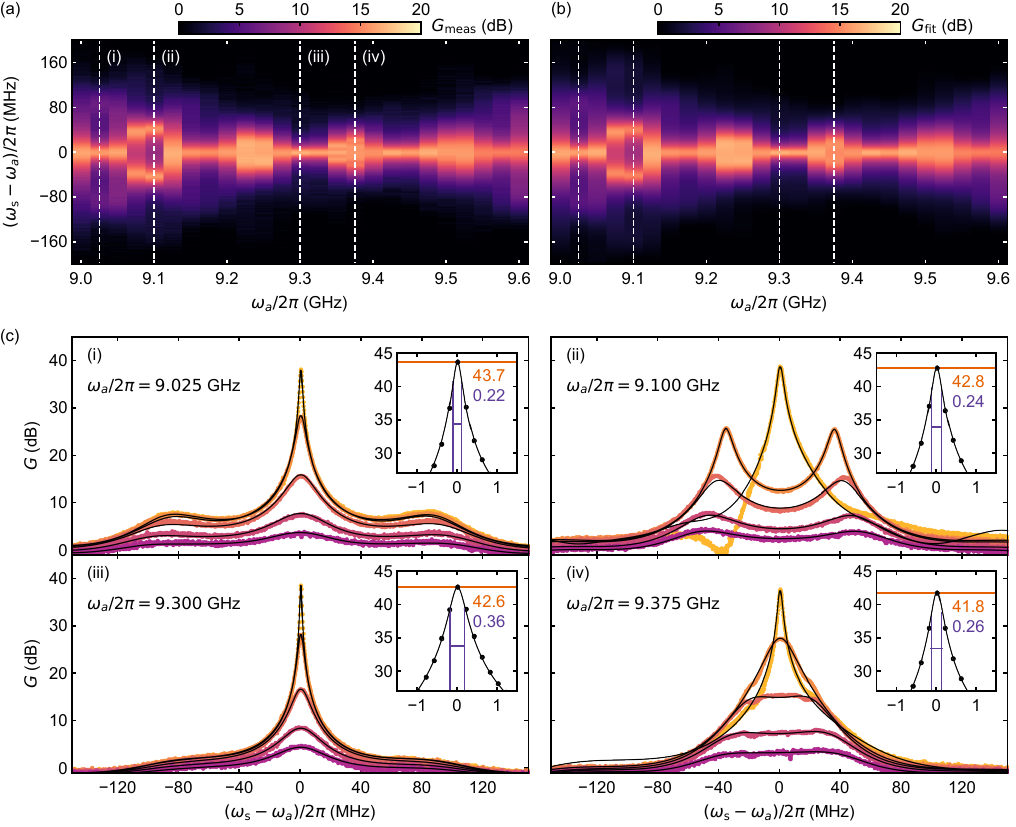} 
\caption{
Gain spectra influenced by Fabry--P\'erot interference.
(a)~Measured net gain spectra with maximum gain of approximately 20 \si{\decibel} as a function of signal detuning from half the pump frequency, or the JPA frequency, for different JPA frequencies between \SI{9.0}{\giga\hertz} and \SI{9.6}{\giga\hertz}. (b)~The fitting results to the data in panel (a). (c)~Measured net gain as a function of input signal frequency for four different JPA frequencies denoted by the dashed lines (i)-(iv) in panel (a). The dots correspond to measured data, while the solid lines are fits to the measured data. The insets show the gain spectra corresponding to the maximum observed gain for each JPA frequency together with the maximum net gain (in \si{\decibel}) and 3-\si{\decibel} bandwidth values (in \si{\mega\hertz}).
} 
  \label{fig:f2} 
\end{center}
\end{figure*}

\subsection{Reflection spectra}
We first characterize the parameters of the JPA, including its external environment, in the absence of flux pumping by measuring the reflection spectra of the JPA using a vector network analyzer. A coherent probe signal is applied to the input line, while a direct current is simultaneously applied through the flux line to tune the JPA resonance frequency. The measured reflection spectra are normalized by a reference signal obtained with the JPA far detuned, as described above in Eq.~\eqref{eq:jpagainbgsub}.

In Fig.~\ref{fig:f1_2nd}(b), we show an example of the measured \(\tilde{S}_{11}\) together with a fit of Eq.~\eqref{eq:jpagainbgsub} to the experimental data. 
Our analytical model of the reflection spectrum, incorporating Fabry--P\'erot interference, reproduces the experimental results well.
From this fit, we extract the JPA parameters
\(\omega_\mathrm{a}/2\pi=\SI{9.5}{\giga\hertz}\),
\(\kappa/2\pi=\SI{280}{\mega\hertz}\), and
\(\kappa_0/2\pi=\SI{22}{\mega\hertz}\),
as well as the Fabry--P\'erot cavity parameters
\(\eta=0.996\),
\(\eta_0=0.803\),
\(\Delta/2\pi=\SI{140}{\mega\hertz}\),
\(\varphi_0=-1.05\), and
\(\phi_\mathrm{r}=-0.048\), for a flux bias of approximately \(\phi_\mathrm{ex}^\mathrm{eff}=\pi/3\), corresponding to a dc bias current of about \(138~\mu\mathrm{A}\).
Moreover, the maximum JPA resonance frequency is found to be \(\omega_\mathrm{a}/2\pi=\SI{11.01}{\giga\hertz}\), obtained at \(\phi_\mathrm{ex}^\mathrm{eff}=0\).

\subsection{Gain spectra}
We pump the JPA through the flux line at different pump powers with a microwave source at frequency $\omega_\mathrm{p}/2\pi = 2\omega_\mathrm{a}/2\pi$, and measure the resulting gain spectra. The JPA resonance frequency \(\omega_\mathrm{a}\) is determined by fitting the reflection spectrum obtained without flux pumping at each flux bias. In addition, the gain spectra are measured with the probe power kept sufficiently weak to avoid saturation of the gain.

In Fig.~\ref{fig:f2}(a), we present the measured net gain spectra ($G_\mathrm{meas}(\delta)$) with a maximum gain of approximately \SI{20}{\decibel} over a range of JPA resonance frequencies between \SI{9.0}{\giga\hertz} and \SI{9.6}{\giga\hertz}. Owing to impedance mismatches in the cryogenic wiring, the gain spectra exhibit a strong dependence on the JPA resonance frequency and deviate markedly from a Lorentzian lineshape. In particular, the spectra can display two or three distinct peaks or, in some cases, a single flat-top profile.
In Fig.~\ref{fig:f2}(c), we additionally present the gain spectra measured at several pump powers for four selected JPA resonance frequencies exhibiting characteristically different interference patterns.
These measurements demonstrate that the combined FPJPA system achieves a maximum net gain of \SI{44}{\decibel}, corresponding to an on-chip gain of \SI{45}{\decibel}.

We fit our theoretical model, namely the modulus squared of Eq.~\eqref{eq:jpagainbgsub}, to the measured gain spectra in order to extract both the JPA internal parameters and the parameters characterizing the external environment. 
For each JPA operating frequency, we select the gain spectra that correspond to maximum gain points \SI{6}{\decibel}, \SI{10}{\decibel}, \SI{14}{\decibel}, \SI{18}{\decibel}, and \SI{22}{\decibel}, and fit them simultaneously. We assume that the Fabry--P\'erot cavity parameters are independent of pump power and thus only let the internal JPA parameters vary as a function of pump power in the fitting. We constrain the pump amplitudes to follow the relation $\Omega_\mathrm{p} = c_\mathrm{p}\sqrt{P_i}$ where $P_i$ is the output power value set at the microwave source and $c_\mathrm{p}$ is a common conversion factor only dependent on the pump frequency. Thus, the fitting parameters consist of a set of internal JPA parameters for each pump power $\{\kappa^{P_{i}}, \kappa_0^{P_{i}}\}, \forall i \in \{0, 1, 2, 3, 4\}$ and pump-power-independent Fabry--P\'erot-cavity parameters, $\{\eta, \eta_0, \Delta, \varphi_0, \phi_\mathrm{r}\}$, and pump-power conversion factor $c_\mathrm{p}$. 

The fitting results are shown in Figs.~\ref{fig:f2}(b) and (c). Remarkably, our theoretical model incorporating a Fabry--Pérot cavity reproduces the experimentally observed, strongly environment-dependent gain spectra with good agreement, including the periodic dependence on the JPA frequency (see the following section for a more detailed discussion).
Furthermore, we observe that the effective bandwidth of the JPA decreases with increasing pump power, accompanied by a corresponding change in the influence of Fabry--Pérot interference.

\subsection{Dependence on JPA operating frequency} 
%
\begin{figure}[t]
\begin{center}
  \includegraphics[width=\linewidth]{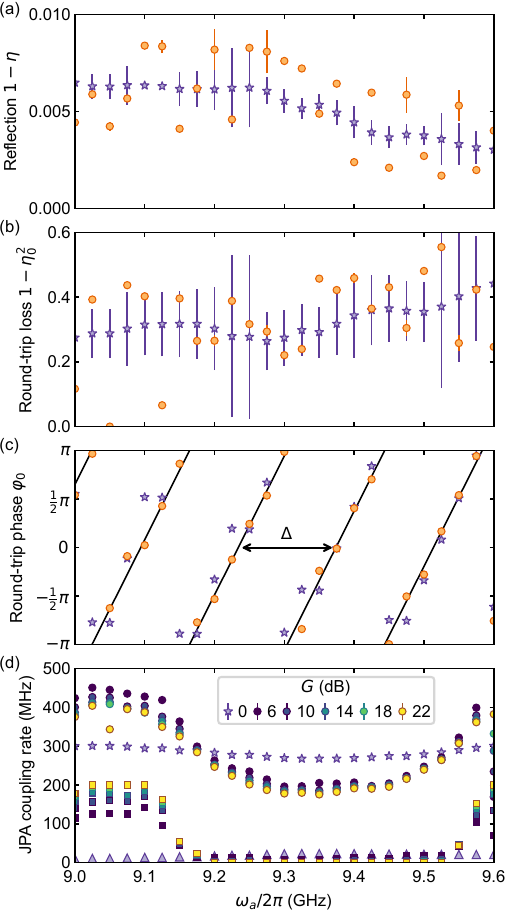} 
\caption{
Dependence of fitted model parameters on JPA frequency
(a)~Fabry--P\'erot-cavity mirror-reflection coefficient $1 - \eext$,
(b)~Round-trip propagation loss coefficient $1 - \eint^2$, and
(c)~Round-trip phase accumulation as a function of JPA frequency.
The orange dots show fit results of the gain spectra at five pump powers, while the purple stars are obtained from fits to $\tilde{S}_{11}$ without flux pumping. The solid line in panel (c) is a linear fit to the phase data.
(d)~External coupling rate (dots and stars) and intrinsic loss rate (squares and triangles) of the JPA as a function of the JPA frequency for different maximum gains.
} 
  \label{fig:f3} 
\end{center}
\end{figure}

Here, we discuss the dependence of the model parameters on the JPA operating frequency. Using the gain-spectrum fitting procedure described in the previous section, we extract the internal JPA parameters and the Fabry--Pérot cavity parameters and present them as functions of JPA operating frequency.

As shown in Fig.~\ref{fig:f3}(a), we extract a very low reflection coefficient for the Fabry--Pérot cavity mirror, reaching values as small as \SI{-25}{\decibel}. The commercial microwave circulator used in this work, manufactured by Quinstar, was independently characterized at room temperature and found to exhibit a reflection coefficient of approximately \SI{-22}{\decibel} over the same frequency range. This close agreement suggests that the circulator itself can act as the effective mirror of the Fabry--Pérot cavity, thereby significantly modifying the measured gain spectra. Other possible sources for the impedance mismatch are SMA connectors and a microwave switch used in the measurement setup, but these typically have smaller reflection coefficients than the circulator.

The round-trip propagation loss shown in Fig.~\ref{fig:f3}(b) is relatively high due to the Fabry--P\'erot cavity being formed mostly by a lossy coaxial cable containing a silver-plated copper center conductor. As an independent estimate of the round-trip propagation loss in the Fabry--P\'erot cavity, we also measured the propagation loss of coaxial cables equivalent to those forming the cavity. From this measurement, we infer a round-trip propagation loss of \(0.24\), assuming an effective cable length of approximately \(70~\mathrm{cm}\), which is very close to the average fit result. Both of these results are very similar to measured values of similar cables in literature~\cite{bib:wallraffcable}.

In Fig.~\ref{fig:f3}(c), we show the phase accumulated during the round trip inside the Fabry--P\'erot cavity at half the pump frequency, or the JPA frequency in this experimental setting. The solid line is a linear fit to the phase values as it is related to the free spectral range \(\Delta\) of the Fabry--P\'erot cavity, i.e., $\varphi_0 = \omega_\mathrm{a} \cdot (2\pi/\Delta) + \phi_\mathrm{R}$. From the linear fit, we obtain a free spectral range of \(\Delta/2\pi =\)~\SI{140}{\mega\hertz}. To account for this value, we hypothesize that the Fabry--P\'erot cavity is formed by the coaxial cables connecting the JPA to the nearest circulator. The length of approximately \SI{70}{\centi\meter} can consistently reproduce the above free spectral range if we consider the speed of microwave propagation in the cable to have a ratio of $2/3$ compared to the speed of light in vacuum. We note that this periodicity in phase also matches with the periodic qualitative features measured for the gain spectra in Fig.~\ref{fig:f2}.

In Fig.~\ref{fig:f3}(d), we show the external-coupling and intrinsic-loss rates of the JPA as a function of the JPA frequency, extracted from fits performed at different pump powers. The fitted decay rates exhibit a pronounced dependence on the JPA frequency that is not captured by our model based on a single Fabry--Pérot cavity.
By contrast, this frequency dependence is much less apparent in data acquired without flux pumping, suggesting that the strong frequency dependence observed under pumping may originate from a parametric process involving parasitic modes, for example, modes associated with the SQUID array or arising from impedance mismatches introduced by wire bonding or an SMP connector on the printed circuit board.
Importantly, although the origin of this frequency dependence remains unclear, our theoretical model successfully reproduces the complex gain spectra at each JPA frequency (see Fig.~\ref{fig:f2}).

\subsection{Gain-bandwidth products} 
%
\begin{figure}[t]
\begin{center}
  \includegraphics[width=86mm]{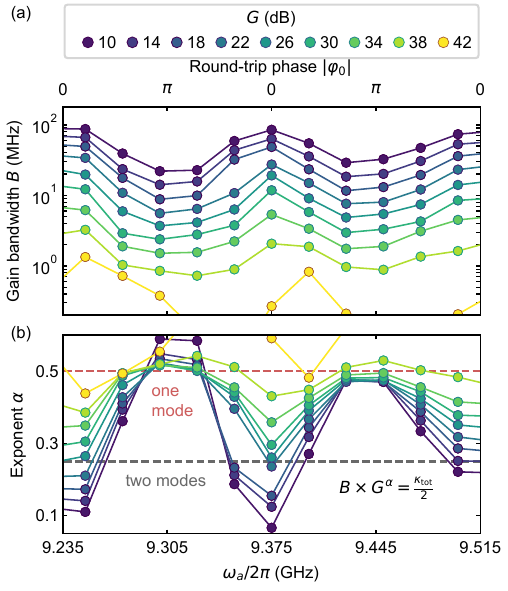} 
\caption{Gain-bandwidth products.
(a)~3-\si{\decibel} bandwidth of the FPJPA system for different maximum net gains (colored dots) in a range of JPA frequencies around \SI{9.375}{\giga\hertz} corresponding to twice the free spectral range of the Fabry--P\'erot cavity, $2\Delta = 2\times\SI{140}{\mega\hertz}$. 
(b)~Gain-bandwidth exponent $\alpha$ as a function of JPA frequency. The two dashed lines refer to values obtained when considering a single (red) or two (gray) effective linear modes in the amplification.  
The horizontal axis is additionally labeled by the corresponding round-trip phases.
} 
  \label{fig:f4} 
\end{center}
\end{figure}

As shown in Fig.~\ref{fig:f4}(a), we analyze the measured gain spectra of the combined FPJPA system in terms of the 3-\si{\decibel} gain bandwidth \(B\) as a function of the JPA operating frequency for different gains. We only consider the JPA frequencies in a range set by twice the free spectral range $[\omega_\mathrm{c} - \Delta, \omega_\mathrm{c} + \Delta]$ with $\omega_\mathrm{c}/2\pi = \SI{9.375}{\giga\hertz}$ where the gain spectra only have a single peak. 
As a function of the JPA frequency, the effective gain bandwidth varies periodically due to constructive or destructive interference between the signal amplified by the JPA and the signal reflected back by the Fabry--Pérot cavity.
At 20-\si{\decibel} net gain, we observe the gain bandwidths ranging between \SI{10}{\mega\hertz} and \SI{50}{\mega\hertz} depending on the JPA frequency. 
For the same frequency range, the bandwidth at 42-\si{\decibel} net gain is varying between \SI{0.1}{\mega\hertz} and \SI{1.5}{\mega\hertz}. 
At high gain, the bandwidth becomes unstable, most likely because the device performance becomes increasingly sensitive to the external environment.

%
\begin{figure}[t]
\begin{center}
  \includegraphics[width=86mm]{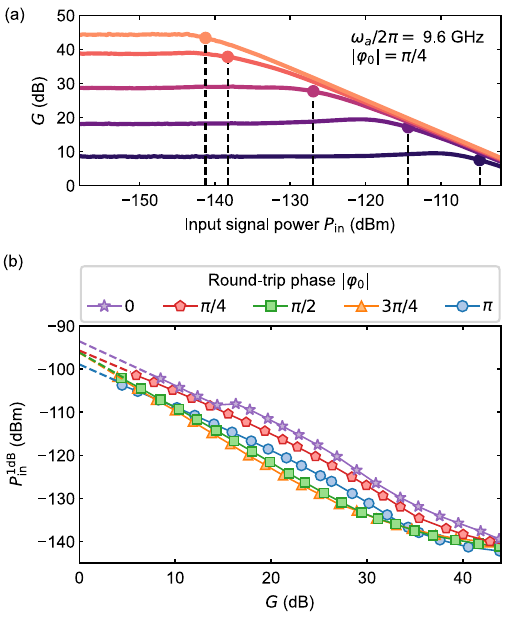} 
\caption{Dynamic range of amplification.
(a)~Net gain of the FPJPA system as a function of the input signal power for five different gain levels. The dots with dashed vertical lines indicate the 1-\si{\decibel} compression points of the input signal powers. (b)~1-\si{\decibel} compression point of the input signal power as a function of the gain for different round-trip phases inside the Fabry--P\'erot cavity. The round-trip phases correspond to the following JPA frequencies (from left to right): \SI{9.3}{\giga\hertz}, \SI{9.6}{\giga\hertz}, \SI{9.55}{\giga\hertz}, \SI{9.25}{\giga\hertz}, \SI{9.1}{\giga\hertz}.
The dashed lines indicate linear fits in the low-gain regime.
} 
  \label{fig:f5}
\end{center}
\end{figure}
%

To quantitatively evaluate the gain-bandwidth product of the FPJPA for different JPA frequencies and gains, we introduce the exponent \(\alpha\), defined through
\begin{equation}
\label{eq:BGalpha}
B \times G^\alpha = \frac{\kappa_\mathrm{tot}}{2},
\end{equation}
where \(B\) denotes the 3-dB gain bandwidth, \(G\) is the net gain evaluated at the JPA frequency, and \(\kappa_\mathrm{tot} = \kappa + \kappa_0 \) is the total decay rate of the JPA in the absence of flux pumping.
Typically, the value of 0.5 corresponds to a conventional single-pole JPA~\cite{bib:clerk2010introduction}, while the value of 0.25 corresponds to a two-pole JPA, commonly known as an impedance-matched parametric amplifier~\cite{bib:impa2}. 

We observe that, depending on the interference within the Fabry--Pérot cavity, the effective number of modes contributing to the amplification process switches between one and two. As a result, the device exhibits a Lorentzian gain spectrum at \(\varphi_0 = \pi\), where the Fabry--Pérot cavity mode is suppressed, and a flat-top gain spectrum at \(\varphi_0 = 0\), where the cavity mode becomes resonant. For low gain values, the observed exponent values appear to be lower than the value of 0.25, which might be due to effects not captured by our simplified gain-bandwidth model or due to more than two modes contributing to the amplification.
For gains exceeding \SI{34}{\decibel}, the results appear to converge toward a single-mode amplification regime over most of the frequency range. This behavior can be attributed to the narrowing of the gain bandwidth at high gain, which renders the external environment approximately frequency independent over the relevant bandwidth.
See more details in Appendix~\ref{app:gainbw}.

\subsection{1-dB compression points}
In order to study the effect of the nonlinearity engineering based on the SQUID array on the amplification process, we measure the gain at a frequency of $\omega_\mathrm{s}/2\pi = \omega_\mathrm{a}/2\pi + \SI{20}{\kilo\hertz}$ as a function of the input signal power, with the pump frequency set to twice the JPA resonance frequency, \(\omega_\mathrm{p} = 2\omega_\mathrm{a}\). The results of these measurements for the different gains are shown in Fig.~\ref{fig:f5}(a) for $\omega_\mathrm{a}/2\pi = \SI{9.6}{\giga\hertz}$. We observe that the gain starts to decrease due to saturation of the JPA past certain input powers. We define the input signal power where the gain has decreased by \SI{1}{\decibel} as the 1-\si{\decibel} compression point, as indicated by the dots and dashed lines in the figure.

In Fig.~\ref{fig:f5}(b), we show the measured 1-dB compression points as a function of the net gain for five different JPA frequencies, corresponding to five different round-trip phases. 
We observe that the round-trip phase has a pronounced effect on the 1-dB compression point.
For example, at a gain of \SI{20}{\decibel}, the saturation power varies from \SI{-121}{\dBm} to \SI{-111}{\dBm}. We attribute this variation to modifications of the effective JPA bandwidth induced by the interaction with the Fabry--Pérot cavity (see Appendix~\ref{app:gainsat}), which depends strongly on the round-trip phase, as discussed in the previous section.

In addition, while the 1-dB compression point scales approximately linearly with the FPJPA gain in the low-gain regime, we observe that at very high gain, the compression point decreases much more slowly with increasing gain than expected from this linear trend. As a consequence, for net gains exceeding \SI{42}{\decibel}, the compression point converges toward a value near \SI{-140}{\dBm}, largely independent of the round-trip phase.
At present, the underlying mechanism responsible for this behavior remains unclear, likely reflecting the effect of more complex dynamics at large gain, potentially involving higher-order nonlinear effects beyond the scope of our model.


\subsection{Added noise}
To characterize the amplifier-added noise of the combined FPJPA system with a pump at \(\omega_\mathrm{p} = 2\omega_\mathrm{a}\), we measure the signal-to-noise ratio (SNR) of a microwave signal applied from a microwave source at a frequency of $\omega_\mathrm{s}/2\pi = \omega_\mathrm{a}/2\pi + \SI{20}{\kilo\hertz}$, both with and without flux pumping, using a spectrum analyzer. All measurements are performed with a frequency span of \SI{40}{\kilo\hertz} and an intermediate-frequency (IF) bandwidth of \SI{1}{\kilo\hertz}.
The measured power spectra are fitted with a model consisting of a Gaussian peak, corresponding to the IF filter frequency response, superimposed on a constant offset. From this fit, the signal power is extracted from the Gaussian peak height, while the noise power is obtained from the offset level.

We develop a theoretical model of the measurement chain, including the JPA with and without flux pumping, and compare it with the measured SNR improvement. As described in Appendix~\ref{app:qubitcalibration}, the coherent response of a well-calibrated qubit coupled to the same measurement line as the JPA is used to calibrate the input powers at both the FPJPA and the subsequent output line. The measured SNR improvements, together with the calibrated power levels, are then used to extract the added noise in units of noise photons following the procedure outlined in Appendix~\ref{app:jpaaddednoise}. The cable loss between the JPA and the circulator is incorporated into the added noise of the FPJPA system, providing a metric relevant for the practical operation of reflection-type amplifiers.

%
\begin{figure}[t]
\begin{center}
  \includegraphics[width=86mm]{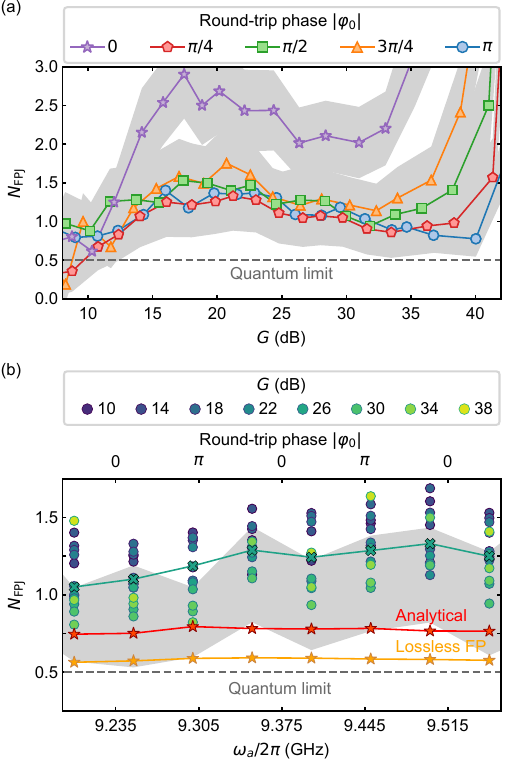} 
\caption{Nearly quantum-limited added noise of amplification.
(a)~Added noise of the FPJPA system in photons as a function of the gain for five different round-trip phases inside the Fabry--P\'erot cavity. The round-trip phases correspond to the following JPA frequencies (from left to right): \SI{9.3}{\giga\hertz}, \SI{9.6}{\giga\hertz}, \SI{9.55}{\giga\hertz}, \SI{9.25}{\giga\hertz}, \SI{9.1}{\giga\hertz}. The dashed line indicates the vacuum noise level, and the gray areas refer to the one-standard-deviation uncertainties of the measured noise values. 
(b)~Added noise as a function of the gain and frequency in roughly two periods of the Fabry--P\'erot-cavity free spectral range centered around \SI{9.375}{\giga\hertz}. The cyan solid line and crosses correspond to the average of the measured noise value over different gains. The shaded region corresponds to the one-standard-deviation uncertainty of the lowest measured noise value at each JPA frequency. The red (orange) solid lines and stars correspond to an analytical calculation of the added noise with (without) experimentally obtained Fabry--P\'erot-cavity propagation loss.
} 
  \label{fig:f6} 
\end{center}
\end{figure}

In Fig.~\ref{fig:f6}(a), we show the measured noise added by the FPJPA as a function of the measured gain for five JPA frequencies corresponding to different round-trip phases.
The results indicate a slight decrease in the added noise at gains exceeding \SI{20}{\decibel}; however, this trend remains within the estimated confidence intervals.
The lowest value we measure is 0.8 added noise photons, which is slightly higher than the quantum limit of 0.5 due to considerable propagation loss in the Fabry-Pérot cavity.
When the Fabry--P\'erot cavity and JPA are resonant (\(\varphi_0 = 0\)), it appears that the noise increases rapidly to a level that is multiple times higher than the noise level of the other operating points. 
This behavior may be associated with nontrivial effects not captured by our theoretical model, such as a parametric process with spurious modes of the JPA and the external environments.
We further note that the rapid increase in the JPA-added noise at high gain most likely arises from enhanced sensitivity to pump-field noise~\cite{bib:eichler2014controlling}.

In Fig.~\ref{fig:f6}(b), we show the measured added noise, expressed in the unit of photons, as a function of the JPA frequency (equivalently, the round-trip phase) for different gain values. In contrast to the bandwidth and saturation-power measurements discussed above, the added noise does not exhibit a pronounced dependence on the periodicity of the Fabry--Pérot cavity. This behavior may arise because the added noise is predominantly limited by propagation loss $\eint$.
In the same panel, we also present analytically calculated added-noise values based on the Fabry--Pérot cavity parameters extracted from the analysis shown in Fig.~\ref{fig:f3}, together with the noise model detailed in Appendix~\ref{app:noise_eff_model}. For the gain values that correspond to the lowest measured added noise, the measured noise values are mostly within one standard deviation of the analytical results.
In addition, we show a second set of analytical results obtained by setting $\eta_0=0$, corresponding to a lossless Fabry--Pérot cavity.
These comparisons indicate that propagation loss within the Fabry--Pérot-cavity contributes significantly to the added noise, but does not fully account for the experimentally observed noise level at all gain values. Moreover, they confirm that a JPA combined with a lossless Fabry--Pérot cavity can achieve quantum-limited parametric amplification.

\section{Discussion}

In this work, we realized a flux-driven lumped-element Josephson parametric amplifier based on a SQUID array and demonstrated that high gain and large bandwidth for resonator-based amplifiers can be achieved through careful device modeling and parameter optimization. Our device exhibits near-quantum-limited, phase-preserving amplification with a gain of 20~dB and a 3-dB bandwidth of approximately 50~MHz. We also observe a maximum net gain of \SI{44}{\decibel} with a bandwidth of \SI{220}{\kilo\hertz}, highlighting the large dynamic range enabled by nonlinear engineering using SQUID arrays. The high performance of the single-pole lumped-element Josephson parametric amplifiers developed in this work establishes a solid foundation for achieving even larger gain--bandwidth products and dynamic range through integration with externally engineered environments, including impedance-transformer structures~\cite{bib:impa1,bib:impa2} (see Appendix~\ref{app:table} for a performance summary of superconducting parametric amplifiers).

In addition, we observe a pronounced dependence of the gain profile on the JPA operating frequency, with characteristic shapes ranging from standard Lorentzian curves to spectra containing multiple peaks. Rather than relying on a full circuit-theory treatment~\cite{bib:impa1}, we explain these features using a quantum-optical model in which a weakly reflecting element in the input/output waveguide forms a Fabry--P\'erot cavity coupled to the JPA. Despite its simplicity, this minimal model reproduces the experimentally observed gain spectra well and enables reliable fitting over a wide range of JPA operating frequencies. From these fits, we extract a Fabry--P\'erot mirror reflectivity below \(0.01\) and a round-trip propagation loss of approximately \(0.25\), indicating that the cavity is most likely formed by the normal-metal coaxial cables between the JPA and the nearest microwave circulator, with imperfections in the circulator acting as the reflecting mirror.



\begin{figure}[t]
\begin{center}
  \includegraphics[width=79.769mm]{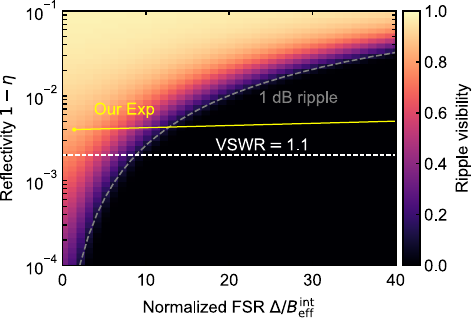} 
\caption{
Fabry--P\'erot interference effects on the gain spectra of parametric amplifiers.
Visibility of gain ripples as a function of the Fabry--P\'erot cavity reflectivity, \(1-\eext\), and the free spectral range normalized by the effective gain bandwidth, \(\Delta / B_\mathrm{eff}^\mathrm{int}\). The effective bandwidth is defined as the full width at half maximum of the Lorentzian gain spectrum of a resonator-based parametric amplifier in the absence of Fabry--P\'erot interference (\(\eext = 1\)). The data are calculated numerically using \(\kappa_0/2\pi = \SI{10}{\mega\hertz}\), \(\eint = 0.9\), and \(\varphi_0 = 0\). The yellow dot and line indicate the parameter range observed in our experiments, while the white dashed line indicates the reflectivity typical of commercially available cryogenic circulators (VSWR \(=1.1\), corresponding to a return loss of \SI{-26}{\decibel}). The gray dashed contour corresponds to gain ripples of \SI{1}{\decibel}.
} 
  \label{fig:fdisc} 
\end{center}
\end{figure}


These observations indicate that even weak reflections in the input/output waveguide can significantly affect the gain profile, bandwidth, and saturation properties of high-performance parametric amplifiers. To further quantify the parameter range over which these interference effects become important, we analytically investigate the gain profiles under varying environmental conditions.

For a quantitative analysis, we focus on the resonant case between the parametric amplifier and the Fabry--P\'erot cavity modes, \(\varphi_0 = 0\), and define the ripple visibility \(V_\mathrm{ripple}\) of the Fabry--P\'erot interference-induced gain ripples in terms of the maximum gain, \(\max\!\left\{\lvert S_{11}(\omega_\mathrm{s}) \rvert^2\right\}\), and the ripple minimum at the center frequency of the gain spectrum, \(\lvert S_{11}(\omega_\mathrm{a}) \rvert^2\), as
\begin{equation}
  V_\mathrm{ripple} = \frac{\max\!\left\{\lvert S_{11}(\omega_\mathrm{s}) \rvert^2\right\} - \lvert S_{11}(\omega_\mathrm{a}) \rvert^2}{\max\!\left\{\lvert S_{11}(\omega_\mathrm{s}) \rvert^2\right\} + \lvert S_{11}(\omega_\mathrm{a}) \rvert^2}.
\end{equation}
As an example, the ripple visibility for the 4 panels in Fig.~\ref{fig:f1_1st_b}(b) are 0, 0.01, 0.73, and 0.85 from left to right. 

Figure~\ref{fig:fdisc} plots this metric as a function of the Fabry--P\'erot cavity reflectivity and the free spectral range normalized by the effective gain bandwidth. A larger Fabry--P\'erot cavity reflectivity and a smaller free spectral range enhance the gain ripples, thereby strengthening the effect of the Fabry--P\'erot interference. This suggests that a larger \(\Delta\), or equivalently a shorter distance between the amplifier and the nearest circulator, is preferable for the reliable operation of superconducting parametric amplifiers. In addition, a narrower effective gain bandwidth at higher gain suppresses the gain ripples, consistent with the experimental observations in Fig.~\ref{fig:f4}(b).

We find that a free spectral range approximately ten times larger than the effective gain bandwidth is required to sufficiently suppress gain ripples, even when using high-quality commercially available circulators with a finite impedance mismatch characterized by a VSWR of \(1.1\), corresponding to a return loss of \SI{-26}{\decibel}. These results highlight the strong impact of the external microwave environment on the operation of high-performance amplifiers.

For a broader effective gain bandwidth of \SI{500}{\mega\hertz}, this condition corresponds to an FSR of \SI{5}{\giga\hertz}, or equivalently a cable length shorter than roughly \SI{2.5}{\centi\meter}. It is experimentally challenging to implement such a short distance between the amplifier and circulator, which is why it is practically difficult to have flat high gain and high bandwidth with current cryogenic microwave components. This highlights the need for development of cryogenic microwave components with lower return loss, ideally implemented on the same chip as the amplifier.

Our theoretical framework provides a practical method for separating the intrinsic dynamics of the amplifier from environmental effects, enabling independent characterization and optimization of both. This separation makes it possible to identify unwanted reflections in cryogenic wiring and optimize the external environment for high-gain operation.

At the same time, the controlled incorporation of environmental modes opens new possibilities for intentionally shaping the gain spectrum through engineered interference. Together, these capabilities establish a versatile approach for designing and operating parametric amplifiers across a wide range of practical applications. Looking ahead, the high gain and sufficiently large bandwidth achieved in this work may facilitate single-shot readout of superconducting qubits without the need for cryogenic semiconductor amplifiers.

\section{Acknowledgements}
We acknowledge fruitful discussions with José Aumentado.
This work was supported in part by JSPS KAKENHI (No.~JP22H04937), and MEXT Quantum Leap Flagship Program (MEXT Q-LEAP No.~JPMXS0118068682). J.I. was supported by the Oskar Huttunen Foundation.

\section{Author contributions}

S.K. developed the theoretical model, designed the devices, and established the fabrication process as well as the basic measurement setup and protocols.
S.K., Y.S., and Y.T. fabricated the devices. 
J.I. and S.K. performed the experiments and analyzed the data with the help of A.F.v.L. 
K.Y. provided the qubit--qubit device used for power calibration. 
C.W.S.C and K.K. helped with the noise characterization and the theoretical modeling, respectively. 
T.M. and K.M. optimized and maintained the experimental environment. 
J.I., S.K., and A.F.v.L. wrote the manuscript with feedback from the other authors. 
Y.N. and S.K. supervised the project.

\appendix

\section{Comparison of Josephson parametric amplifier performance}\label{app:table}
As a reference, Table~\ref{tab:jpa_survey} summarizes various parametric amplifiers along with their characteristic operating mode, type of environmental and non-linear engineering, number of Josephson junctions, bandwidths, and saturation powers. Although not comprehensive, this comparison helps clarify the relationship between saturation power and junction number, illustrates the impact of impedance-matched multi-pole architectures on achievable bandwidth, and highlights how our simple lumped-element devices can be further enhanced through integration with externally engineered environments.

\begin{table*}[!t]
\centering
\caption{Performance comparison between different types of superconducting parametric amplifiers found in the literature. EE refers to the environmental or impedance engineering present in the study while NLE refers to the type of the non-linear elements used to engineer the non-linearity. $N_\mathrm{JJ}$, $B$, and $P_\mathrm{in}^{\SI{1}{\decibel}}$ refer to the total number of junctions in the device, 3-\si{\decibel} bandwidth, and input signal power at which 1 \si{\decibel} of compression can be observed, respectively. $B$, and $P_\mathrm{in}^{\SI{1}{\decibel}}$ refer to values measured near \SI{20}{\decibel} gain in each study. We have also divided the references into different groups based on their characteristic structure: single-pole resonant mode JPAs, impedance-matched JPAs, kinetic-inductance-based parametric amplifiers, Josephson array-mode parametric amplifiers, and Josephson travelling wave parametric amplifiers. The list of references is not comprehensive but aims to show a representative set of studies.}
\label{tab:jpa_survey}
\begin{tabular}{l c c c c c r}
\hline\hline
\textbf{Author (Year)} & \textbf{Mode} & \textbf{EE} & \textbf{NLE} & \textbf{$N_\mathrm{JJ}$} & \textbf{$B$} & \textbf{$P_\mathrm{in}^{\SI{1}{\decibel}}$} \\
\hline
Yamamoto \textit{et al.}~\cite{bib:fluxJPA1} (2008) & Resonant & - & dc-SQUID & 2 & 1 MHz & \SI{-135}{\dBm} \\
Mutus \textit{et al.}~\cite{bib:ljpa1} (2013) & Resonant & - & dc-SQUID & 2 & 50 MHz & \SI{-120}{\dBm} \\
Zhou \textit{et al.}~\cite{bib:zhou2014high} (2014)  & Resonant & - & \textcolor{red}{dc-SQUID array} & 16 & 5 MHz & \SI{-123}{\dBm} \\
Frattini \textit{et al.}~\cite{bib:frattini2018optimizing} (2018) & Resonant & - & \textcolor{RawSienna}{SNAIL array} & 80 & 40 MHz & \SI{-102}{\dBm} \\
Sivak \textit{et al.}~\cite{bib:sivak2019kerr} (2019) & Resonant & - & \textcolor{RawSienna}{SNAIL array} & 80 & 10 MHz & \SI{-102}{\dBm} \\
Planat \textit{et al.}~\cite{bib:sqarray2} (2019) & Resonant & - & \textcolor{red}{dc-SQUID array} & 160 & 45 MHz & \SI{-117}{\dBm} \\
Esposito \textit{et al.}~\cite{bib:esposito2019development} (2019) & Resonant & - & \textcolor{red}{dc-SQUID array} & 8 & 15 MHz & \SI{-115}{\dBm} \\
Mahboob \textit{et al.}~\cite{bib:mahboob2022} (2022) & Resonant & - & dc-SQUID & 2 & 0.4 MHz & \SI{-115}{\dBm} \\
Kaufman \textit{et al.}~\cite{bib:kaufman2025simple} (2025) & Resonant & - & \textcolor{ForestGreen}{rf-SQUID array} & 25 & 28 MHz & \SI{-91}{\dBm} \\
\textbf{Our work} & Resonant & - & \textcolor{red}{dc-SQUID array} & 10 & 50 MHz & \SI{-111}{\dBm} \\
\\
Mutus \textit{et al.}~\cite{bib:impa1} (2014) & Resonant & \textcolor{blue}{Taper} & dc-SQUID & 2 & 700 MHz & \SI{-100}{\dBm} \\
Roy \textit{et al.}~\cite{bib:impa2} (2015) & Resonant & \textcolor{blue}{Transformer} & dc-SQUID & 2 & 640 MHz & \SI{-110}{\dBm} \\
Duan \textit{et al.}~\cite{bib:duan2021broadband} (2021) & Resonant & \textcolor{blue}{Transformer} & dc-SQUID & 2 & 500 MHz & \SI{-110}{\dBm} \\
Grebel \textit{et al.}~\cite{bib:grebel2021flux} (2021) & Resonant & \textcolor{blue}{Transformer} & dc-SQUID & 2 & 300 MHz & \SI{-116}{\dBm} \\
Ranzani \textit{et al.}~\cite{bib:ranzani2022wideband} (2022) & Resonant & \textcolor{blue}{Transformer} & dc-SQUID & 2 & 300 MHz & \SI{-126}{\dBm} \\
Ezenkova \textit{et al.}~\cite{bib:ezenkova2022broadband} (2022) & Resonant & \textcolor{blue}{Transformer} & \textcolor{RawSienna}{SNAIL array} & 268 & 300 MHz & \SI{-100}{\dBm} \\
White \textit{et al.}~\cite{bib:googlero} (2023) & Resonant & \textcolor{blue}{Taper} & \textcolor{ForestGreen}{rf-SQUID array} & 40 & 300 MHz & \SI{-95}{\dBm} \\
Kaufman \textit{et al.}~\cite{bib:googleimpa2} (2023) & Resonant & \textcolor{blue}{Transformer} & \textcolor{ForestGreen}{rf-SQUID array} & 40 & 500 MHz & \SI{-90}{\dBm} \\
Qing \textit{et al.}~\cite{bib:impa3} (2024) & Resonant & \textcolor{blue}{Taper} & dc-SQUID & 2 & 200 MHz & \SI{-110}{\dBm} \\
\\
Parker \textit{et al.}~\cite{bib:parker2022degenerate} (2022) & Resonant & - & \textcolor{magenta}{High-kinetic-inductance film} & - & 5 MHz & \SI{-70}{\dBm} \\
Frasca \textit{et al.}~\cite{bib:frasca2024three} (2024) & Resonant & - & \textcolor{magenta}{High-kinetic-inductance film} & - & 2 MHz & \SI{-86}{\dBm} \\
Mohamed \textit{et al.}~\cite{bib:mohamed2024selective} (2024) & Resonant & - & \textcolor{magenta}{High-kinetic-inductance film} & - & 1 MHz & \SI{-73}{\dBm} \\
Zapata \textit{et al.}~\cite{bib:zapata2024granular} (2024) & Resonant & - & \textcolor{magenta}{High-kinetic-inductance film} & - & 3 MHz & \SI{-110}{\dBm} \\
Hung \textit{et al.}~\cite{bib:hung2025broadband} (2025) & Resonant & \tblue{Transformer} & \textcolor{magenta}{High-kinetic-inductance film} & - & 450 MHz & \SI{-68}{\dBm} \\
\\
Sivak \textit{et al.}~\cite{bib:sivak2020josephson} (2020) & \textcolor{orange}{Josephson array mode} & - & \textcolor{RawSienna}{SNAIL array} & 4000 & 11 MHz & \SI{-93}{\dBm} \\
Winkel \textit{et al.}~\cite{bib:winkel2020nondegenerate} (2020) & \textcolor{orange}{Josephson array mode} & - & \textcolor{red}{dc-SQUID array} & 3600 & 14 MHz & \SI{-118}{\dBm} \\
\\
Macklin \textit{et al.}~\cite{bib:macklin2015near} (2015) & \textcolor{RoyalPurple}{Travelling wave} & - & Junction array & 2000 & 3000 MHz & \SI{-99}{\dBm} \\
Planat \textit{et al.}~\cite{bib:planat2020photonic} (2020) & \textcolor{RoyalPurple}{Travelling wave} & - & \textcolor{red}{dc-SQUID array} & 5320 & 3000 MHz & \SI{-100}{\dBm} \\
Ranadive \textit{et al.}~\cite{bib:ranadive2022kerr} (2022) & \textcolor{RoyalPurple}{Travelling wave} & - & \textcolor{RawSienna}{SNAIL array} & 2800 & 4000 MHz & \SI{-98}{\dBm} \\
Wang \textit{et al.}~\cite{bib:wang2025high} (2025) & \textcolor{RoyalPurple}{Travelling wave} & - & Junction array & 9024 & 3000 MHz & \SI{-107}{\dBm} \\
Chang \textit{et al.}~\cite{bib:sandbotwpa} (2025) & \textcolor{RoyalPurple}{Travelling wave} & - & Junction array & 2400 & 5000 MHz & \SI{-99}{\dBm} \\
\hline \hline
\end{tabular}
\end{table*}


\section{Experimental setup and device fabrication}

\subsection{Experimental setup}\label{app:expsetup}

%
\begin{figure}[t]
\begin{center}
  \includegraphics[width=85.1mm]{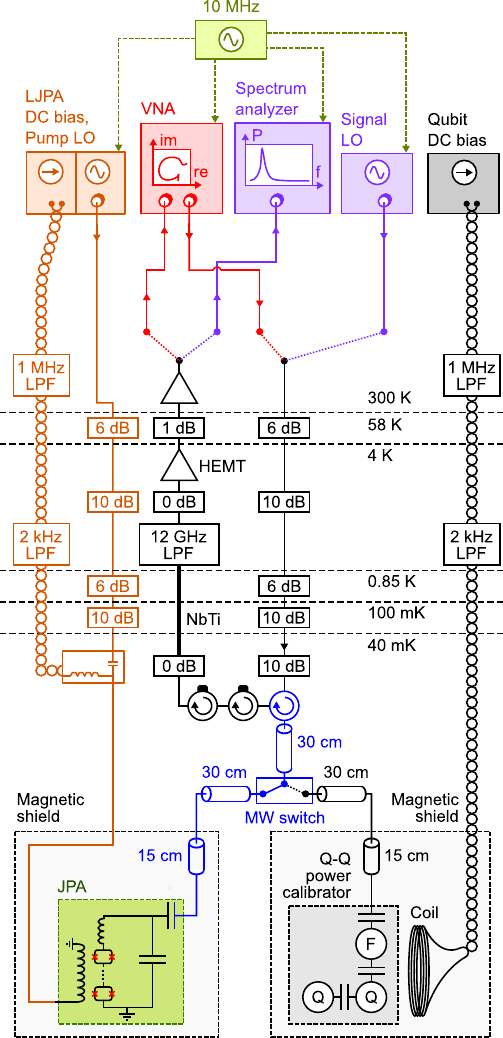} 
\caption{
Detailed schematic of the experimental setup used in the experiment.
} 
  \label{fig:fs1}
\end{center}
\end{figure}

Figure~\ref{fig:fs1} illustrates the experimental setup, including cryogenic wiring in the dilution refrigerator and room-temperature electronics. Coherent measurements are performed using a vector network analyzer (VNA) (red), while power spectral measurements use a signal microwave source and a spectrum analyzer (purple); the two configurations are selected by reconfiguring room-temperature cabling. Pump fields from a microwave source (brown) are DC-biased via a bias-T at the base temperature and applied to the JPA through the flux line. 
Signal lines connect the JPA to a microwave switch and circulators using silver-plated copper coaxial cables of varying lengths in three sections (blue).
Moreover, a qubit-based power calibrator (gray) can be switched to the output line.
All microwave instruments are phase-locked to a Rb frequency standard (light green).

\subsection{Fabrication process}\label{app:fabrication}
The fabrication of the JPA devices is summarized as follows.
First, a \SI{200}{\nano\meter}-thick Nb film sputtered on a high-resistivity Si wafer is patterned by photolithography and $\mathrm{CF}_4$ plasma etching to define the lumped-element resonator, signal line, and flux-line structures.

Next, the SQUID array is fabricated using electron-beam lithography based on Dolan-bridge structures formed with a bilayer resist stack consisting of a \SI{600}{\nano\meter} copolymer EL-11 bottom layer and a \SI{300}{\nano\meter} ZEP520A-7 top layer. The Josephson junctions are realized by double-angle Al evaporation with thicknesses of \SI{30}{\nano\meter} and \SI{50}{\nano\meter}, combined with an intermediate in situ oxidation step, followed by lift-off. Electrical contact between the Nb and Al layers is ensured by in situ Ar ion milling prior to Al evaporation.

Finally, airbridges are fabricated in-situ on the patterned Nb film based on the process detailed in Ref.~\cite{bib:chen2014fabrication}. The support-pillar mask is defined on a \SI{4}{\micro\meter}-thick layer of photoresist (SPR220-3.0) by photolithography and reflow baked, followed by evaporation of a \SI{300}{\nano\meter}-thick Al layer. The Al layer is then patterned to define the bridge structures.

After these processes, the wafer is coated with a protective resist, diced into chips, and subsequently released and cleaned using solvent rinses and $\mathrm{O}_2$ plasma.

\section{Calibration procedures} \label{app:calib}
Accurate calibration of the signal power and environmental noise levels at the input of the circulator used for the reflection-type amplifier is crucial for determining the added noise of the combined Fabry--Pérot cavity–JPA (FPJPA) system by comparing the signal-to-noise ratio with the JPA gain turned on and off.
We perform the following set of calibrations for the combined FPJPA system, as shown in the schematic in Fig.~\ref{fig:fsnoise}. 
We use a microwave switch to create identical wiring configurations for the two different devices, one leading to the JPA and the other to a calibration system consisting of two coupled qubits (Q-Q) and a Purcell filter (F). 
We use one of the qubits, a frequency-tunable transmon used as a readout "resonator", to calibrate the attenuation inside the dilution refrigerator and thereby determine the signal power level at the input of the circulator. The second qubit, a fixed-frequency transmon functioning as a data qubit, is used to calibrate the environmental noise photons.
Here, we write in detail how the above calibration steps are performed.

\subsection{Calibration based on qubits}\label{app:qubitcalibration}

We show a simplified schematic of the calibration setup for the signal power level in Fig.~\ref{fig:fsnoise}(a). The system parameters of the qubit--qubit power calibrator system are shown in Table~\ref{tbl:qqparams}.

\begin{table}[tp]
\bec
\caption{Measured qubit--qubit system parameters.}
  \begin{tabular}{lcl} \hline \hline
\textbf{Readout qubit} \\
\hline
$\ket{\textnormal{g}}$--$\ket{\textnormal{e}}$ transition frequency & $\omega_{\textnormal{q}}/2\pi$ & \SI{9.798}{\giga\hertz} \\
Maximum radiative coupling rate & $\Gamma_\mathrm{r}/2\pi$ & \SI{5.8}{\mega\hertz} \\
Non-radiative coupling rate & $\Gamma_\mathrm{nr}/2\pi$ & $\lesssim$ \SI{0.1}{\mega\hertz} \\
Pure dephasing rate & $\Gamma_\mathrm{p}/2\pi$ & $\lesssim$ \SI{0.1}{\mega\hertz} \\ \\
\textbf{Data qubit} \\
\hline
$\ket{\textnormal{g}}$--$\ket{\textnormal{e}}$ transition frequency & $\omega_{\textnormal{ge}}/2\pi$ & \SI{7.958}{\giga\hertz} \\ 
Anharmonicity & $\alpha/2\pi$ & \SI{-340}{\mega\hertz} \\
$\ket{\textnormal{g}}$--$\ket{\textnormal{e}}$ energy relaxation time & $T_1$ & \SI{14}{\micro\second} \\ 
$\ket{\textnormal{g}}$--$\ket{\textnormal{e}}$ coherence time & $T_2^{*}$ & \SI{3}{\micro\second} \\ \\
\textbf{Purcell filter} \\
\hline
Filter frequency & $\omega_{\textnormal{f}}/2\pi$ & \SI{9.995}{\giga\hertz} \\ 
External coupling rate & $\gamma_\mathrm{f}/2\pi$ & \SI{100}{\mega\hertz} \\ 
\\ \\
Qubit--qubit coupling rate & $g/2\pi$ & \SI{282.6}{\mega\hertz}\\
Qubit--qubit dispersive shift & $\chi/2\pi$ & \SI{6.5}{\mega\hertz} \\ \\
\hline \hline
  \end{tabular}\
\label{tbl:qqparams}
\enc  
\end{table}

By measuring the reflection coefficient of the readout qubit as a function of probe frequency and power, we can estimate the input power and thus attenuation at the input port to the readout qubit close to its resonance frequency. Since the readout qubit has a SQUID, we can tune its frequency with a flux bias and repeat the above measurement to obtain the attenuation at different frequencies.

We can analytically calculate the reflection coefficient for the readout-qubit system by considering the input--output relations of a waveguide terminated by a qubit, resulting in the final expression~\cite{bib:qubitspectrum}:
\begin{equation}
\label{eq:qubits11}
S_{11}^\mathrm{q}(\omega) = 1 - \frac{i\Gamma_\mathrm{r}\Gamma_1\left[\left(\omega - \omega_\mathrm{q}\right) - i\Gamma_2\right]}{\Omega^2\Gamma_2 + \Gamma_1\left[\left(\omega - \omega_\mathrm{q}\right)^2+\Gamma_2^2\right]},
\end{equation}
where $\omega_\mathrm{q}$ is the resonance frequency, $\Omega$ is the probe-induced driving rate, $\Gamma_1 = \Gamma_\mathrm{r} + \Gamma_\mathrm{nr}$ is a sum of radiative and non-radiative decay rates, and $\Gamma_2 = \Gamma_1/2 + \Gamma_\mathrm{p}$ is the total dephasing rate, with a pure dephasing rate of $\Gamma_\mathrm{p}$.

\begin{figure}[t]
\begin{center}
  \includegraphics[width=84.949mm]{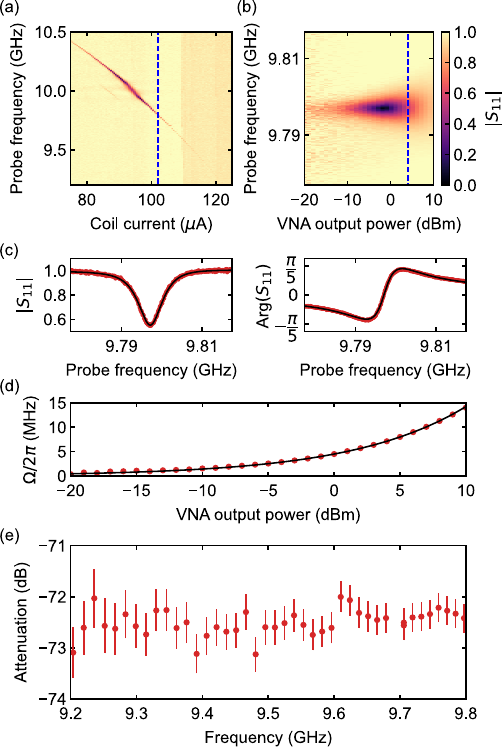} 
\caption{
Measured qubit $|S_{11}|$ as a function of the VNA probe frequency, (a)~readout-qubit bias-coil current, and (b)~VNA probe output power. (c)~An example of a measured qubit reflection spectrum magnitude and phase as a function of probe frequency (dots) with a fit shown as a solid line. The coil current and VNA output power used to measure the spectrum correspond to the blue dashed lines in panels (a) and (b). (d)~Input-probe-induced qubit driving rate obtained from fits to the qubit reflection spectrums measured as a function of the VNA probe output power. (e)~Calibrated values for the attenuation inside the dilution refrigerator obtained from the qubit reflection spectrum fits as a function of frequency.
} 
  \label{fig:fs4} 
\end{center}
\end{figure}

In Figs.~\ref{fig:fs4}(a) and (b), we show measured $|S_{11}|$ as a function of the VNA probe frequency for different readout qubit frequencies (tuned via external flux applied through a coil) as well as the dependence on the VNA probe power at a certain readout-qubit frequency. By fitting Eq.~\eqref{eq:qubits11} to the qubit spectra at a fixed readout-qubit frequency as shown in Fig.~\ref{fig:fs4}(c), we obtain the probe-induced driving rate as a function of the VNA output power, as shown in Fig.~\ref{fig:fs4}(d). Here, we note that the nonradiative decay rate extracted from the fitting is generally much smaller than the radiative decay rate, allowing us to neglect it. 
With this consideration, we can convert the qubit drive rate to the probe input power at the qubit frequency as $P_\mathrm{probe}(\omega_\mathrm{q}) = \hbar \omega_\mathrm{q} \Omega^2 / (4\Gamma_\mathrm{r})$.
By repeating the above process at different qubit frequencies, we obtain the attenuation inside the dilution refrigerator as a function of frequency, as shown in Fig.~\ref{fig:fs4}(d). This enables calibration of the signal power at the input of the qubit-based calibrator by averaging the attenuation over the frequency range from \SI{9.0}{\giga\hertz} to \SI{9.6}{\giga\hertz}, yielding a value of $72.5 \pm \SI{0.3}{\decibel}$. Using the calibrated attenuation, we determine the signal power at the JPA input, denoted by \(P^\mathrm{S}_\mathrm{calib}\) in Fig.~\ref{fig:fsnoise}.

%

\subsection{Calculation for Fabry--P\'erot-cavity--JPA system}\label{app:jpaaddednoise}

\begin{figure}[t]
\begin{center}
  \includegraphics[width=85.958mm]{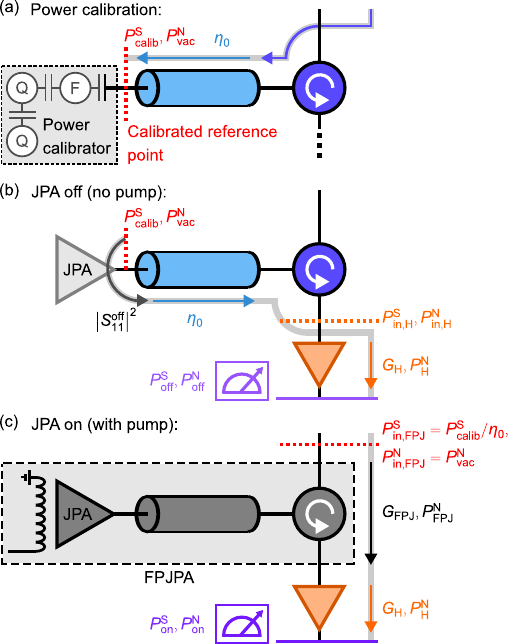} 
\caption{
Schematics of the noise calibration process and related components. (a)~A reference qubit--qubit system used to determine the input power level at the JPA input port, (b)~FPJPA system without pump, and (c)~FPJPA system with the JPA activated by flux-pumping.
} 
  \label{fig:fsnoise} 
\end{center}
\end{figure}

In contrast to characterizing the on-chip performance of quantum-limited amplifiers, we explicitly account for the effects of the nearest circulator required for reflection-type amplifiers and the cable loss between the circulator and the JPA device. 
As described in Fig.~\ref{fig:fsnoise}c, this modeling yields more practical figures of merit for signal amplification, namely the effective gain and added noise power of the combined FPJPA system, denoted by \(G_\mathrm{FPJ}\) and \(P^N_\mathrm{FPJ}\), respectively. Here, the signal input to the FPJPA is defined as the input at the circulator.

After the FPJPA, the signal is amplified by a HEMT amplifier and subsequently routed to the room-temperature measurement apparatus. The measurement chain following the FPJPA (i.e., after the circulator), including the losses of cables and isolators preceding the HEMT amplifier (see Fig.~\ref{fig:fs1}), is effectively characterized by an overall gain and an effective added noise, denoted by \(G_\mathrm{H}\) and \(P^N_\mathrm{H}\), respectively.

To characterize these quantities, well-calibrated signal and noise power levels are required at the input of each reference point, shown in Figs.~\ref{fig:fsnoise}(b) and (c), respectively. 
The input noise powers for the subsequent measurement chain (\(P^\mathrm{N}_\mathrm{in, H}\)) and the FPJPA (\(P^\mathrm{N}_\mathrm{in,FPJ}\)) can be safely assumed to be at the vacuum level ($P^\mathrm{N}_\mathrm{vac}$), respectively.
However, the corresponding input signal power levels ($P^\mathrm{S}_\mathrm{in,H}$ and $P^\mathrm{S}_\mathrm{in,FPJ}$) cannot be directly calibrated in our setup. 
Nevertheless, by using the calibrated power at the input of the qubit-based calibrator (\(P^\mathrm{S}_\mathrm{calib}\), determined in the previous section), we can infer the input signal power levels under the assumption \(\eext \approx 1\). This assumption is supported by the fitting results shown in Fig.~\ref{fig:f3}(a) and its effect should be within the fitting error of $\eta_0$.

Under the condition of \(\eext \approx 1\), using the calibrated input signal power (\(P^S_\mathrm{calib}\)) and accounting for the one-way propagation loss between the circulator and the JPA, denoted by \(\eint\), we approximately infer the signal power at the input of the subsequent measurement chain (after the circulator) and at the input of the FPJPA (before the circulator) as \(P^\mathrm{S}_\mathrm{in,H} = |S_{11}^\mathrm{off}|^2 \eint P^\mathrm{S}_\mathrm{calib}\) and \(P^\mathrm{S}_\mathrm{in,FPJ} = P^\mathrm{S}_\mathrm{calib}/\eint\), respectively, where \(|S_{11}^\mathrm{off}|^2=(\kappa-\kappa_0)^2/(\kappa+\kappa_0)^2\) is the power reflectance of the JPA at the resonant signal frequency when it is not pumped. In the following, we describe the calibration procedures based on this model, using the calibrated signal power levels \(P^\mathrm{S}_\mathrm{in,H}\) and \(P^\mathrm{S}_\mathrm{in,FPJ}\), together with the input noise powers, which can be assumed to be at the vacuum level in both cases (\(P^\mathrm{N}_\mathrm{in,H} =P^\mathrm{N}_\mathrm{in,FPJ}= P^{N}_\mathrm{vac}\)).

%

In our experiment, we use a spectrum analyzer combined with a microwave signal source to measure the power spectral density with flux-pumping on and off, over a frequency span containing both the coherent signal and the noise floor.
As detailed in the main text, each measured power spectrum provides the signal and noise powers for each experimental setting.
When the pump is off, the measured signal and noise powers are explicitly described as
\begin{equation}
\label{eq:soff}
\begin{split}
P_\mathrm{off}^\mathrm{S} = G_\mathrm{H}P^{S}_\mathrm{in,H} 
=G_\mathrm{H}|S_{11}^\mathrm{off}|^2\eint P^{S}_\mathrm{calib},
\end{split}
\end{equation}
and
\begin{equation}
\label{eq:noff}
\begin{split}
P_\mathrm{off}^\mathrm{N} = G_\mathrm{H}\left(P_\mathrm{in,H}^\mathrm{N} + P_\mathrm{H}^\mathrm{N}\right) 
= G_\mathrm{H}\left(P_\mathrm{vac}^\mathrm{N} + P_\mathrm{H}^\mathrm{N}\right),
\end{split}
\end{equation}
respectively. When the pump is on, in contrast, the measured signal and noise powers are described as
\begin{equation}
\label{eq:son}
\begin{split}
P_\mathrm{on}^\mathrm{S} = G_\mathrm{H}G_\mathrm{FPJ}P^{S}_\mathrm{in,FPJ} 
= G_\mathrm{H}G_\mathrm{FPJ}\frac{ P^{S}_\mathrm{calib}}{\eint}
\end{split}
\end{equation}
and
\begin{equation}
\label{eq:non}
\begin{split}
P_\mathrm{on}^\mathrm{N} &= G_\mathrm{H}\left[G_\mathrm{FPJ}\left(P_\mathrm{in,FPJ}^\mathrm{N} + P_\mathrm{FPJ}^\mathrm{N}\right) + P_\mathrm{H}^\mathrm{N}\right] \\
&=G_\mathrm{H}\left[G_\mathrm{FPJ}\left(P_\mathrm{vac}^\mathrm{N} + P_\mathrm{FPJ}^\mathrm{N}\right) + P_\mathrm{H}^\mathrm{N}\right],
\end{split}
\end{equation}
respectively.

\begin{figure}[t]
\begin{center}
  \includegraphics[width=86mm]{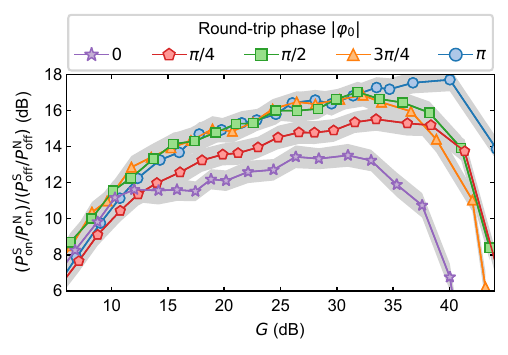} 
\caption{
Measured signal-to-noise ratio improvement as a function of gain and round-trip phase inside the Fabry--P\'erot cavity.
} 
  \label{fig:fs5} 
\end{center}
\end{figure}

By combining these expressions, the added noise of the FPJPA can be determined in terms of directly measurable and calibrated quantities as
\begin{equation}
\label{eq:njpa}
\begin{split}
P_\mathrm{FPJ}^\mathrm{N} = &\frac{P^\mathrm{S}_\mathrm{calib}}{\eint}\left(\frac{P_\mathrm{on}^\mathrm{N}}{P_\mathrm{on}^\mathrm{S}} - \frac{P_\mathrm{off}^\mathrm{N}}{P_\mathrm{on}^\mathrm{S}}\right) \\ &- P_\mathrm{vac}^\mathrm{N}\left(1 - \frac{1}{\eint^2|S_{11}^\mathrm{off}|^2}\frac{P_\mathrm{off}^\mathrm{S}}{P_\mathrm{on}^\mathrm{S}}\right).
\end{split}
\end{equation}
%
As a reference, we show the measured improvement in signal-to-noise ratio as a function of gain, for different round-trip phases inside the Fabry--P\'erot cavity in Fig.~\ref{fig:fs5}.

The conversion of the obtained noise power in Eq.~\eqref{eq:njpa} to the added noise photons, as shown in Fig.~\ref{fig:f6}, can be performed as 
\begin{equation}
\label{eq:jpa_noise_photons}
N_\mathrm{FPJ} = \frac{P_\mathrm{N}^\mathrm{FPJ}}{\hbar\omega_\mathrm{s}B_\mathrm{IF}},
\end{equation}
where $\omega_\mathrm{s}$ is the signal frequency and $B_\mathrm{IF}$ is the IF bandwidth of the spectrum analyzer. 
In this process, we can convert the vacuum noise power \(P_\mathrm{vac}^\mathrm{N}\) into noise quanta using Eq.~\eqref{eq:jpa_noise_photons}, yielding \(N_\mathrm{vac}^\mathrm{N}=0.5\). We independently verify this assumption using the data qubit in the qubit--qubit calibration system, where the thermal photon number of the input noise is extracted from the thermal-photon-induced dephasing rate of the data qubit. This analysis shows that the residual thermal input noise level is within \(10\%\) of the vacuum noise. Moreover, we use the fitting parameter values of the FPJPA gain spectra shown in Fig.~\ref{fig:f3} to determine $\eint$ and $|S_{11}^\mathrm{off}|^2$ for each JPA operating frequency in the above equations.

\section{Gain-bandwidth products}\label{app:gainbw}
The dependence of the gain on the effective number of resonator modes involved in the amplification process can be examined in more detail via the gain--bandwidth product, or more precisely, through the ratio between the logarithm of the bare JPA bandwidth normalized by the gain bandwidth and the logarithm of the gain, which is given by
\begin{equation}
\alpha = {\log \left( \kappa_\mathrm{tot}/(2B) \right)}/{\log \left( G \right)}.
\end{equation}
The ratio, or equivalently the exponent defined in Eq.~\eqref{eq:BGalpha}, typically takes the value \(\alpha=0.5\) for a single-pole JPA and \(\alpha=0.25\) for a two-pole JPA (i.e., IMPA), independent of the gain.

%
\begin{figure}[t]
\begin{center}
  \includegraphics[width=86mm]{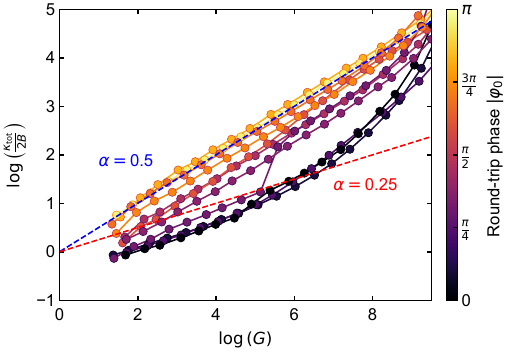} 
\caption{
Bandwidth-gain data used to determine the exponent $\alpha = {\log \left( \kappa_\mathrm{tot}/(2B) \right)}/{\log \left( G \right)}$ shown in logarithmic scale.
The red and blue dashed lines correspond to \(\alpha = 0.25\) and \(\alpha = 0.5\), respectively.
} 
  \label{fig:fsup4} 
\end{center}
\end{figure}

In Fig.~\ref{fig:fsup4}, we plot the logarithm of the gain bandwidth divided by the JPA bandwidth as a function of the logarithm of the gain to study the gain dependence of the gain-bandwidth product. 
For data acquired near a round-trip phase of \(\varphi_0=\pi\), the logarithmic relation exhibits a clear linear trend over a wide gain range with \(\alpha=0.5\), consistent with single-mode amplification. In contrast, around \(\varphi_0=0\), the relationship between the two logarithmic quantities does not show a clear linear dependence, and some datasets exhibit a crossover between \(\alpha=0.25\) and \(\alpha=0.5\) as the gain is increased. This behavior can be understood as a consequence of gain-induced narrowing of the gain bandwidth, which progressively suppresses the influence of the Fabry--P\'erot cavity and drives the system toward effective single-mode operation even near \(\varphi_0=0\).


\section{Design theory}\label{app:designtheory}
\subsection{Relation between charge and flux}
Two-port circuit elements, such as capacitors, inductors, and Josephson junctions, can be characterized by the relation between electric charge and generalized magnetic flux.
Here, the charge ($Q$) and flux ($\Phi$) of a two-port element are defined as
\begin{equation}
\begin{split} 
Q = \int \di t\: I, 
\quad \mathrm{and} \quad
\Phi = \int \di t\: V,
\end{split}
\end{equation}
where $I$ is the current through the element and $V$ is the differential voltage between the two ports, respectively. 
For instance, the charge-flux relations of a capacitor, an inductor, and a Josephson junction are described using their corresponding characteristic parameters, i.e., the capacitance $C$, inductance $L$, and Josephson inductance $L_\mathrm{J}$, as
\begin{equation}
\label{eq:QPhiL}
\begin{split} 
Q = C\dot{\Phi}, \quad
\dot{Q} = \frac{\Phi}{L}, 
\quad \mathrm{and} \quad
\dot{Q} &= \frac{\Phi_0}{L_\mathrm{J}} \sin\left(\frac{\Phi}{\Phi_0}\right),
\end{split}
\end{equation}
where $\dot{X}$ is the time derivative of variable $X$.
The Josephson junction is characterized by Josephson inductance $L_\mathrm{J}$, which can be associated with the critical current $I_\mathrm{c}$ as $ L_\mathrm{J} = \Phi_0/I_\mathrm{c}$, where $\Phi_0 = \hbar/(2e)$ is the reduced quantum of magnetic flux.
Moreover, the junction's generalized magnetic flux $\Phi$ is associated with the differential superconducting phases $\phi$ between the two isolated islands, as $\Phi = \Phi_0\phi$.
Compared to the inductor, the Josephson junction can be regarded as an inductive element with a strong nonlinearity.
Note that, in the following discussion, the Josephson capacitances are neglected within the Born--Oppenheimer approximation~\cite{kafri2017tunable}.

By considering the flux degree of freedom as the counterpart of position in canonical mechanics, a capacitor gives kinetic energy, 
\begin{equation}
K = \frac{C \dot{\Phi}^2}{2}.
\end{equation}
On the other hand, the inductive elements, such as inductors and Josephson junctions, give the potential energies,
\begin{equation}
\label{eq:U_ind}
\begin{split} 
U = \frac{\Phi^2}{2L} 
\quad \mathrm{and} \quad
U = \frac{{\Phi_0}^2}{2L_\mathrm{J}} (-2)\!\cos\!\left(\frac{\Phi}{\Phi_0}\right),
\end{split}
\end{equation}
respectively.

When considering the quantization of circuits, it may be useful to introduce dimensionless conjugate variables for charge and flux, defined as the number of Cooper pairs ($n=Q/2e$) and the number of reduced flux quanta ($\phi = \Phi/\Phi_0$). 
Since the conjugate variable of $\Phi$ corresponds to charge, obtained through the Legendre transformation as $Q = C\dot{\Phi}$ for a parallel circuit of a capacitor ($C$) and an inductive element, the kinetic (capacitive) energy is rewritten as 
\begin{equation}
K = E_C \, n^2,
\end{equation}
where $E_C = (2e)^2/(2C)$ is the capacitive energy of a single cooper pair.
In contrast, the potential energies for the inductor and the Josephson junction are rewritten as
\begin{equation}
\label{eq:U_ind_dimless}
\begin{split} 
U = E_L \phi^2
\quad \mathrm{and} \quad
U =  E_{L_\mathrm{J}}\, (-2)\!\cos\!\left(\phi\right),
\end{split}
\end{equation}
respectively, where $E_L = {\Phi_0}^2/(2L)$ and $E_{L_\mathrm{J}} = {\Phi_0}^2/(2L_\mathrm{J})$ are the inductive energies of a single reduced flux quantum. 

\subsection{Superconducting Quantum Interference Device (SQUID) with a loop inductance}\label{app:squidtheory}
In the following, we will derive the charge-flux relation of a superconducting quantum interference device (SQUID), i.e. a parallel circuit of two identical Josephson junctions, while taking into account a loop inductance. 

\subsubsection{Lagrange inversion theorem}
Deriving the charge-flux relation of a SQUID with a loop inductance is not trivial due to the non-linearity of the Josephson junctions. 
Here, we find the analytical solution by a perturbative approach considering only up to the leading terms of nonlinearity, which may be sufficient to understand the saturation effect of parametric amplifiers.

When the charge-flux relation of a two-port inductive element is analytically found in the form of $\dot{Q}$ as an explicit function of flux $\Phi$, it can be expanded in the Maclaurin series around $\Phi=0$ as
\begin{equation}
\label{eq:QMaclaurin}
\dot{Q} = \dot{Q}'\!(0)\Phi + \dot{Q}''\!(0)\frac{\Phi^2}{2} + \dot{Q}'''\!(0)\frac{\Phi^3}{6} + \cdots,
\end{equation}
where $\dot{Q}^{(k)}\!(0)$ denotes the $k$-th $\Phi$ derivative of $\dot{Q}$ evaluated at $\Phi = 0$.
We note that $\dot{Q}(0)=0$ is assumed, which is satisfied for conventional inductive elements and all the circuits under our consideration throughout the paper.

It is useful to rewrite the charge-flux relation in the form of $\Phi$ as an explicit function of $\dot{Q}$, especially when a series circuit of inductive elements is considered.
We can analytically find the inverse function in the form of the Maclaurin series by using the Lagrange inversion theorem.
When $\dot{Q}'\!(0)\neq 0$ is satisfied, which is the case for all circuits considered throughout our paper, $\Phi$ can be written as
\begin{equation}
\label{eq:PhiMaclaurin}
\Phi = \Phi'\!(0)\dot{Q} + \Phi''\!(0)\frac{\dot{Q}^2}{2} + \Phi'''\!(0)\frac{\dot{Q}^3}{6} + \cdots,
\end{equation}
where $\Phi^{(k)}(0)$ denotes the $k$-th $\dot{Q}$ derivative of $\Phi$ evaluated at $\dot{Q} = 0$.
The inversion theorem relates the derivatives of the inverse function to those of the original function as
\begin{equation}
\label{eq:LIT}
\begin{split}
\Phi'\!(0) &= {\dot{Q}'\!(0)}^{-1},\\
\Phi''\!(0) & = - \dot{Q}''\!(0){\dot{Q}'\!(0)}^{-3},\\
\Phi'''\!(0) & =  - \dot{Q}'''\!(0){\dot{Q}'\!(0)}^{-4}  + 3 {\dot{Q}''\!(0)}^2{\dot{Q}'\!(0)}^{-5},\\
&\,\,\,\vdots
\end{split}
\end{equation}

When a parallel circuit of inductive elements is considered or the potential energy for an inductive element is calculated, it is useful to rewrite the charge-flux relation from an explicit function of $\dot{Q}$ to that of $\Phi$.
Similarly to Eq.~(\ref{eq:LIT}), through the Lagrange inversion theorem, $\dot{Q}^{(n)}\!(0)$ can be described using $\Phi^{(n)}\!(0)$, which makes it possible to analytically describe $\dot{Q}$ in an explicit functional form of $\Phi$ in the form of the Maclaurin series. 

In summary, when the nonlinearity of an inductive element is restricted to leading-order terms, the Lagrange inversion theorem enables an analytical inversion of the charge--flux relation, allowing \(\dot{Q}\) (\(\Phi\)) to be expressed explicitly as a function of \(\Phi\) (\(\dot{Q}\)).

\subsubsection{Series circuit of an inductor and a Josephson junction}

%
\begin{figure}[t]
\begin{center}
  \includegraphics[width=85.092mm]{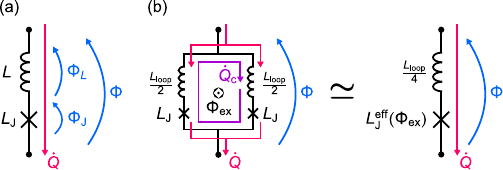} 
\caption{
Charge--flux relations for inductive elements.
(a)~Series circuit consisting of a linear inductance and a Josephson junction.
(b)~SQUID including a finite loop inductance and its corresponding effective circuit model.
} 
  \label{fig:app_circuit2} 
\end{center}
\end{figure}

Before deriving the charge-flux relation of a SQUID with a loop inductance, as shown in Fig.~\ref{fig:app_circuit2}(a), we discuss a series circuit of a linear inductor and a Josephson junction, characterized by $L$ and $L_\mathrm{J}$, respectively.
By taking into account Kirchhoff's current law, we introduce a shared current flowing through both elements ($\dot{Q}$), while we define individual fluxes for the inductor and the junction ($\Phi_L$ and $\Phi_\mathrm{J}$), respectively.
The charge-flux relations of the individual elements are given by
\begin{equation}
\label{eq:LLJ_series}
\begin{split} 
\dot{Q} = \frac{\Phi_L}{L} \quad \mathrm{and} \quad
\dot{Q} = \frac{\Phi_0}{L_\mathrm{J}} \sin\left(\frac{\Phi_\mathrm{J}}{\Phi_0}\right).
\end{split}
\end{equation}
In the series configuration, we have
\begin{equation}
\label{eq:Laws_series}
\Phi = \Phi_L + \Phi_\mathrm{J},
\end{equation}
where $\Phi$ is the total flux of the series circuit.
This corresponds to effectively considering the two elements in series as one two-port inductive element characterized by $\dot{Q}$ and $\Phi$.
Note that, for $\Phi_\mathrm{J}$ to remain a single-value variable with a given $\Phi$ and thereby to avoid bistability, we assume $L_\mathrm{J} \geq L$ in the following discussions~\cite{rymarz2023consistent}. 

The next step is to explicitly derive the charge-flux relation for the effective circuit element, i.e., the relation between $\dot{Q}$ and $\Phi$.
Eqs.~(\ref{eq:LLJ_series}) and (\ref{eq:Laws_series}) are used to derive
\begin{equation}
\label{eq:PhiQLLJ_series}
\Phi = L \dot{Q}  + \Phi_0 \arcsin\left(\frac{L_\mathrm{J}\dot{Q}}{\Phi_0}\right).  
\end{equation}
Assuming \(\Phi\) and \(\dot{Q}\) are small, \(\Phi\) can be expanded as a Maclaurin series around \(\dot{Q} = 0\), leading to
\begin{equation}
\label{eq:PhiMacLLJ_series}
\Phi = \left(L+L_\mathrm{J}\right) \dot{Q} + \frac{{L_\mathrm{J}}^3}{{\Phi_0}^2} \frac{\dot{Q}^3}{6} + \cdots.
\end{equation}
This yields
\begin{equation}
\label{eq:DerivLLJ_series}
\begin{split}
\Phi'\!(0) &= L+L_\mathrm{J},\\
\Phi''\!(0) & = 0,\\
\Phi'''\!(0) & = \frac{{L_\mathrm{J}}^3}{{\Phi_0}^2}, \\
&\,\,\,\vdots
\end{split}
\end{equation}
As explained above, the Lagrange inversion theorem can be used to derive the derivatives of the inverse function for Eq.~(\ref{eq:PhiMacLLJ_series}) evaluated at $\Phi=0$, leading to
\begin{equation}
\label{eq:QMacLLJ_series}
\dot{Q} = \frac{\Phi}{L+L_\mathrm{J}} - \frac{{L_\mathrm{J}}^3}{{\Phi_0}^2(L+L_\mathrm{J})^4}\frac{\Phi^3}{6} + \cdots.
\end{equation}

When only up to third-order terms are taken into account, the expression can be rewritten as
\begin{equation}
\label{eq:QPhiLLJ_series}
\begin{split}
\dot{Q} = \frac{\Phi_0}{L+L_\mathrm{J}}\bigg[\!\bigg(\!\frac{L}{L+L_\mathrm{J}}\frac{\Phi}{\Phi_0}\bigg) + \sin\!\left(\!\frac{L_\mathrm{J}}{L+L_\mathrm{J}}\frac{\Phi}{\Phi_0}\!\right)\!\bigg].
\end{split}
\end{equation}

In summary, Eqs.~(\ref{eq:PhiQLLJ_series}) and (\ref{eq:QPhiLLJ_series}) are the charge-flux relation of a series circuit of a linear inductor and a Josephson junction in explicit functional forms of $\dot{Q}$ and $\Phi$, respectively.

\subsubsection{SQUID with a loop inductance}
As shown in Fig.~\ref{fig:app_circuit2}(b), a SQUID can be considered as a superconducting loop with loop inductance $L_\mathrm{loop}$ containing identical Josephson junctions in each branch, each characterized by an inductance $L_\mathrm{J}$. 
As described in the following, applying an external magnetic flux through the SQUID loop can tune the effective inductance of the SQUID.

By taking into account Kirchhoff's current law, one can define two independent current degrees of freedom: a current flowing across the SQUID ($\dot{Q}$) and a current circulating in the loop ($\dot{Q}_\mathrm{c}$).
Moreover, current $\dot{Q}$ can be defined to be symmetrically split between the two identical branches of the SQUID loop, giving the branch currents $\dot{Q}/2 + \dot{Q}_\mathrm{c}$ and $\dot{Q}/2 - \dot{Q}_\mathrm{c}$, respectively.
See the schematic in Fig.~\ref{fig:app_circuit2}(b) for the definitions of the currents.
For convenience, we consider the loop inductance as a lumped-element inductor, symmetrically divided into two parts ($L_\mathrm{h} = L_\mathrm{loop}/2$) and positioned at the two identical branches.
Thus, each branch contains a series circuit of half the loop inductance ($L_\mathrm{h}$) and the Josephson inductance ($L_\mathrm{J}$).
The flux across the SQUID is denoted by $\Phi$, while the fluxes across the two branches are denoted as $\Phi_a$ and $\Phi_b$, respectively. 
Kirchhoff's voltage law and flux quantization impose constraints on these fluxes below.
\begin{equation}
\label{eq:PhiPhiex_SQUID}
\begin{split}
\Phi &= \frac{\Phi_a + \Phi_b}{2} + n\pi  \Phi_0 \\
\Phi_\mathrm{ex} &= \Phi_a - \Phi_b + 2m\pi \Phi_0,
\end{split}
\end{equation}
where \(\Phi_\mathrm{ex}\) is the external magnetic flux threading the SQUID loop, and \(n\) and \(m\) are integers.
It is assumed that \(L_\mathrm{J} \geq L_\mathrm{h}\) in order to avoid bistability~\cite{rymarz2023consistent}.

Since each branch consists of a series circuit of the linear inductor ($L_\mathrm{h}$) and the Josephson junction ($L_\mathrm{J}$), the fluxes for the individual branches are associated with the corresponding currents using Eq.~(\ref{eq:PhiQLLJ_series}) as 
\begin{equation}
\label{eq:PhiaPhib_SQUID}
\begin{split}
\frac{\dot{Q}}{2}+\dot{Q}_\mathrm{c} & = \frac{\Phi_0}{L_\mathrm{J}}\!\sin\left(\frac{\Phi_a}{\Phi_0} - \frac{L_\mathrm{h}}{\Phi_0}\!\left(\frac{\dot{Q}}{2}+\dot{Q}_\mathrm{c}\right)\right), \\
\frac{\dot{Q}}{2}-\dot{Q}_\mathrm{c} & = \frac{\Phi_0}{L_\mathrm{J}}\!\sin\left(\frac{\Phi_b}{\Phi_0} - \frac{L_\mathrm{h}}{\Phi_0}\!\left(\frac{\dot{Q}}{2}-\dot{Q}_\mathrm{c}\right)\right).
\end{split}
\end{equation}
By using Eqs. (\ref{eq:PhiPhiex_SQUID}) and (\ref{eq:PhiaPhib_SQUID}), we have
\begin{widetext}
\begin{equation}
\label{eq:QPhi_SQUID}
\begin{split}
\dot{Q} = \frac{2\Phi_0}{L_\mathrm{J}}&\cos\!\left(\frac{\Phi_\mathrm{ex}}{2\Phi_0} -m\pi - \frac{L_\mathrm{h}\dot{Q}_\mathrm{c}}{\Phi_0}\right)\sin\!\left(\frac{\Phi}{\Phi_0} -n\pi -  \frac{L_\mathrm{h}\dot{Q}}{2\Phi_0}\right),\\
2\dot{Q}_\mathrm{c} = \frac{2\Phi_0}{L_\mathrm{J}}&\sin\!\left(\frac{\Phi_\mathrm{ex}}{2\Phi_0} -m\pi - \frac{L_\mathrm{h}\dot{Q}_\mathrm{c}}{\Phi_0}\right)\cos\!\left(\frac{\Phi}{\Phi_0} -n\pi-  \frac{L_\mathrm{h}\dot{Q}}{2\Phi_0}\right).
\end{split}
\end{equation}
Here, the integers \(n\) and \(m\) are chosen such that the potential energy minimum of the SQUID is located at \(\Phi = 0\), which requires a positive linear relation between \(\dot{Q}\) and \(\Phi\) in the vicinity of \(\Phi = 0\) in the first equation.
This gives
\begin{equation}
\label{eq:QPhi_SQUID2}
\begin{split}
\dot{Q} &=  \frac{2\Phi_0}{L_\mathrm{J}}\left|\cos\!\left(\frac{\Phi_\mathrm{ex}}{2\Phi_0}  - \frac{L_\mathrm{h}\dot{Q}_\mathrm{c}}{\Phi_0}\right)\right|\sin\!\left(\frac{\Phi}{\Phi_0}  -  \frac{L_\mathrm{h}\dot{Q}}{2\Phi_0}\right),\\
2\dot{Q}_\mathrm{c} &= \pm \frac{2\Phi_0}{L_\mathrm{J}}\sin\!\left(\frac{\Phi_\mathrm{ex}}{2\Phi_0} - \frac{L_\mathrm{h}\dot{Q}_\mathrm{c}}{\Phi_0}\right)\cos\!\left(\frac{\Phi}{\Phi_0} -  \frac{L_\mathrm{h}\dot{Q}}{2\Phi_0}\right).
\end{split}
\end{equation}
\end{widetext}
Here, we use the sign notations \(\pm\) and \(\mp\), where the upper (lower) sign is taken when
$
\cos\!\left(\frac{\Phi_\mathrm{ex}}{2\Phi_0}
            - \frac{L_h \dot{Q}_\mathrm{c}}{\Phi_0}\right) > 0
$
$
\left(< 0\right),
$
respectively.

Given a controlled external flux $\Phi_\mathrm{ex}$, $\dot{Q}$ could be described in an explicit functional form of only $\Phi$ by solving the simultaneous equations of Eq.~(\ref{eq:QPhi_SQUID}).
While a fully analytical treatment is difficult due to the intrinsic nonlinearity, we derive approximate analytical expressions using Maclaurin expansions.

For convenience, Eqs. (\ref{eq:QPhi_SQUID}) are simplified as
\begin{equation}
\label{eq:QPhi_SQUID_sim}
\begin{split}
\theta = |\mathcal{C}_\mathrm{c}|\mathcal{S}
\quad \mathrm{and} \quad
\theta_\mathrm{c} = \pm \mathcal{S}_\mathrm{c}\mathcal{C},
\end{split}
\end{equation}
where 
$\mathcal{S} = \sin\!\left(\phi - \beta \theta\right)$,
$\mathcal{C} = \cos\!\left(\phi - \beta \theta\right)$,
$\mathcal{S}_\mathrm{c} = \sin\!\left(\phi_\mathrm{c} - \beta \theta_\mathrm{c}\right)$, and
$\mathcal{C}_\mathrm{c} = \cos\!\left(\phi_\mathrm{c} - \beta \theta_\mathrm{c}\right)$.
Moreover, we redefine the variables as $\theta = L_\mathrm{J}\dot{Q}/(2\Phi_0)$, $\theta_\mathrm{c} = 2L_\mathrm{J}\dot{Q}_\mathrm{c}/(2\Phi_0)$, $\phi=\Phi/\Phi_0$, $\phi_\mathrm{c}=\Phi_\mathrm{ex}/(2\Phi_0)$ while defining the inductance ratio as 
\begin{equation}
\beta = \frac{L_\mathrm{h}}{L_\mathrm{J}}.
\end{equation}

Here, Eqs.~(\ref{eq:QPhi_SQUID_sim}) can be expected to find the solutions for $\theta$ and $\theta_\mathrm{c}$ expanded around $\phi=0$ as
\begin{equation}
\label{eq:QPhi_SQUID_sim_Mac}
\begin{split}
\theta &= \theta\!(0) + \theta'\!(0) \phi + \theta''\!(0)\frac{\phi^2}{2} + \theta'''\!(0)\frac{\phi^3}{6} + \cdots, \\
\theta_\mathrm{c}\ &= \theta_\mathrm{c}\!(0) + \theta_\mathrm{c}'\!(0) \phi + \theta_\mathrm{c}''\!(0)\frac{\phi^2}{2} + \theta_\mathrm{c}'''\!(0) \frac{\phi^3}{6} + \cdots, \\
\quad
\end{split}
\end{equation}
where ${\theta}^{(k)}\!(0)$ and ${\theta}_\mathrm{c}^{(k)}\!(0)$ denote the $k$-th $\phi$ derivatives of $\theta$ and $\theta_\mathrm{c}$ evaluated at $\phi = 0$, respectively. 

To determine these coefficients, we differentiate both sides of Eq.~\eqref{eq:QPhi_SQUID_sim} with respect to \(\phi\).
The first-order derivatives are obtained as
\begin{equation}
\label{eq:QPhi_SQUID_sim_1}
\begin{split}
\theta' &= \pm \beta \theta_\mathrm{c}'\mathcal{S}_\mathrm{c}\mathcal{S} + (1-\beta \theta')|\mathcal{C}_\mathrm{c}|\mathcal{C}, \\
\theta_\mathrm{c}' &= - \beta \theta_\mathrm{c}'|\mathcal{C}_\mathrm{c}|\mathcal{C} \mp (1-\beta \theta')\mathcal{S}_\mathrm{c}\mathcal{S},
\end{split}
\end{equation}
where we use the chain rules, i.e.,
$\mathcal{S}' = (1-\beta \theta')\mathcal{C}$,
$\mathcal{C}' = -(1-\beta \theta')\mathcal{S}$,
$\mathcal{S}_\mathrm{c}' = -\beta \theta'_\mathrm{c}\mathcal{C}_\mathrm{c}$, and
$\mathcal{C}_\mathrm{c}' = + \beta \theta'_\mathrm{c}\mathcal{S}_\mathrm{c}$.
Note that we follow the same sign conventions as in Eq.~\eqref{eq:QPhi_SQUID2}, which yields
\(|\mathcal{C}_\mathrm{c}|' = \pm \beta\,\theta'_\mathrm{c}\mathcal{S}_\mathrm{c}\)
and
\(\pm \mathcal{C}_\mathrm{c} = |\mathcal{C}_\mathrm{c}|\).

\begin{widetext}
The second-order derivatives are obtained from Eq.~(\ref{eq:QPhi_SQUID_sim_1}) as
\begin{equation}
\label{eq:QPhi_SQUID_sim_2}
\begin{split}
\theta'' &= -(\beta \theta_\mathrm{c}')^2|\mathcal{C}_\mathrm{c}|\mathcal{S} 
\pm2\beta \theta_\mathrm{c}'(1-\beta \theta')\mathcal{S}_\mathrm{c}\mathcal{C} -(1-\beta \theta')^2|\mathcal{C}_\mathrm{c}|\mathcal{S}\pm\beta \theta_\mathrm{c}''\mathcal{S}_\mathrm{c}\mathcal{S} - \beta \theta''|\mathcal{C}_\mathrm{c}|\mathcal{C}, \\
\theta_\mathrm{c}'' &= \mp(\beta \theta_\mathrm{c}')^2\mathcal{S}_\mathrm{c}\mathcal{C} 
+2\beta \theta_\mathrm{c}'(1-\beta \theta')|\mathcal{C}_\mathrm{c}|\mathcal{S} \mp (1-\beta \theta')^2\mathcal{S}_\mathrm{c}\mathcal{C} - \beta \theta_\mathrm{c}''|\mathcal{C}_\mathrm{c}|\mathcal{C} 
\pm\beta \theta''\mathcal{S}_\mathrm{c}\mathcal{S},
\end{split}
\end{equation}
and the third-order derivatives are given by
\begin{equation}
\label{eq:QPhi_SQUID_sim_3}
\begin{split}
\theta''' &= \mp(\beta \theta_\mathrm{c}')^3\mathcal{S}_\mathrm{c}\mathcal{S} 
-(1-\beta \theta')^3|\mathcal{C}_\mathrm{c}|\mathcal{C} \pm \beta \theta'''_\mathrm{c}\mathcal{S}_\mathrm{c}\mathcal{S} 
-\beta \theta'''|\mathcal{C}_\mathrm{c}|\mathcal{C} - 3(\beta \theta_\mathrm{c}')^2(1-\beta \theta')|\mathcal{C}_\mathrm{c}|\mathcal{C} 
\\ 
&\quad\mp3\beta \theta_\mathrm{c}'(1-\beta \theta')^2\mathcal{S}_\mathrm{c}\mathcal{S}
-3\beta^2 \theta_\mathrm{c}''\theta_\mathrm{c}'|\mathcal{C}_\mathrm{c}|\mathcal{S} 
\pm3\beta \theta_\mathrm{c}''(1-\beta \theta')\mathcal{S}_\mathrm{c}\mathcal{C} 
\\ 
&\quad\mp3\beta^2 \theta_\mathrm{c}' \theta''\mathcal{S}_\mathrm{c}\mathcal{C} 
+3\beta \theta''(1-\beta \theta')|\mathcal{C}_\mathrm{c}|\mathcal{S},
\\
\theta'''_\mathrm{c} &= (\beta \theta_\mathrm{c}')^3|\mathcal{C}_\mathrm{c}|\mathcal{C} 
\pm(1-\beta \theta')^3\mathcal{S}_\mathrm{c}\mathcal{S} -\beta \theta'''_\mathrm{c}|\mathcal{C}_\mathrm{c}|\mathcal{C} 
\pm\beta \theta'''\mathcal{S}_\mathrm{c}\mathcal{S} \pm3(\beta \theta_\mathrm{c}')^2(1-\beta \theta')\mathcal{S}_\mathrm{c}\mathcal{S}
\\ 
&\quad+3\beta \theta_\mathrm{c}'(1-\beta \theta')^2|\mathcal{C}_\mathrm{c}|\mathcal{C}\mp3\beta^2 \theta_\mathrm{c}''\theta_\mathrm{c}'\mathcal{S}_\mathrm{c}\mathcal{C}
+3\beta \theta_\mathrm{c}''(1-\beta \theta')|\mathcal{C}_\mathrm{c}|\mathcal{S}
\\ 
&\quad-3\beta^2 \theta_\mathrm{c}' \theta''|\mathcal{C}_\mathrm{c}|\mathcal{S} 
\pm 3\beta \theta''(1-\beta \theta')\mathcal{S}_\mathrm{c}\mathcal{C}.
\end{split}
\end{equation}
\end{widetext}

Using these relations, we can evaluate the differential coefficients of Eqs.~(\ref{eq:QPhi_SQUID_sim_Mac}) from the lowest orders.
First, we can determine the 0th-order differential coefficients, namely $\theta\!(0)$ and $\theta_\mathrm{c}\!(0)$. 
Using Eqs.~(\ref{eq:QPhi_SQUID_sim}), we find the solution of $\theta$ at $\phi = 0$, i.e., 
\begin{equation}
\theta\!(0) =0.
\end{equation}
This leads to 
\begin{equation}
\begin{split}
\mathcal{S}\!(0)=0
\quad \mathrm{and} \quad
\mathcal{C}\!(0)=1,
\end{split}
\end{equation}
where $\mathcal{S}\!(0)$ and $\mathcal{C}\!(0)$ are $\mathcal{S}$ and $\mathcal{C}$ evaluated at $\phi=0$, respectively.
Moreover, $\theta_\mathrm{c}\!(0)$ is determined by solving
\begin{equation}\label{eq:zero}
\theta_\mathrm{c}\!(0) = \pm\mathcal{S}_\mathrm{c}\!(0),
\end{equation}
or explicitly
\begin{equation}
\theta_\mathrm{c}\!(0) = \pm \sin\!\left(\phi_\mathrm{c} - \beta \theta_\mathrm{c}\!(0)\right),
\end{equation}
where we use $\mathcal{C}\!(0)=1$.
This condition also determines $\mathcal{S}_\mathrm{c}\!(0)$ and $\mathcal{C}_\mathrm{c}\!(0)$ evaluated at $\phi=0$.

For the first-order coefficients evaluated at $\phi=0$, we substitute the 0th-order coefficients into Eqs.~(\ref{eq:QPhi_SQUID_sim_1}), leading to
\begin{equation}
\label{eq:thetathetac0}
\begin{split}
\theta'(0)=\frac{|\mathcal{C}_\mathrm{c}\!(0)|}{1+\beta|\mathcal{C}_\mathrm{c}\!(0)|}
\quad \mathrm{and} \quad
\theta_\mathrm{c}'(0)=0,
\end{split}
\end{equation}
where we assume $1+\beta|\mathcal{C}_\mathrm{c}\!(0)|\neq0$, which is always the case for our case ($\beta = L_\mathrm{h}/L_\mathrm{J} \leq 1$).
In order to evaluate the higher-order derivatives, it is convenient to have
\begin{equation}
1-\beta\theta'(0) = \frac{1}{1+\beta|\mathcal{C}_\mathrm{c}\!(0)|},
\end{equation}
where Eq.~\eqref{eq:thetathetac0}.
By substituting the 0th and first-order derivatives evaluated at $\phi=0$ into Eqs.~(\ref{eq:QPhi_SQUID_sim_2}), we can evaluate the second-order derivatives as
\begin{equation}
\begin{split}
\theta''(0)=0
\quad \mathrm{and} \quad
\theta_\mathrm{c}''(0) = \mp\frac{\mathcal{S}_\mathrm{c}\!(0)}{\left(1+\beta|\mathcal{C}_\mathrm{c}\!(0)|\right)^3}.
\end{split}
\end{equation}
Successively, the lower-order derivatives are substituted into Eqs.~(\ref{eq:QPhi_SQUID_sim_3}), leading to
\begin{equation}
\begin{split}
\theta'''(0)=-\frac{|\mathcal{C}_\mathrm{c}\!(0)|(1 + \epsilon)}{\left(1+\beta|\mathcal{C}_\mathrm{c}\!(0)|\right)^4}
\quad \mathrm{and} \quad
\theta_\mathrm{c}'''(0)=0,
\end{split}
\end{equation}
where the correction coefficient $\epsilon$ of the self-Kerr nonlinearity is defined as
\begin{equation}
\label{eq:epsilonKerr}
\epsilon = \frac{3\beta\mathcal{S}_\mathrm{c}\!(0)^2}{|\mathcal{C}_\mathrm{c}\!(0)|\left(1+\beta|\mathcal{C}_\mathrm{c}\!(0)|\right)}.
\end{equation}

Taking into account these derivative coefficients, the Maclaurin expansion of $\theta$ with respect to $\phi$ is obtained as
\begin{equation}
\label{eq:thetaphi_Mac}
    \theta = \frac{|\mathcal{C}_\mathrm{c}\!(0)|}{1+\beta|\mathcal{C}_\mathrm{c}\!(0)|}\phi -\frac{|\mathcal{C}_\mathrm{c}\!(0)|( 1 + \epsilon)}{\left(1+\beta|\mathcal{C}_\mathrm{c}\!(0)|\right)^4} \frac{\phi^3}{6} + \cdots.
\end{equation}

Going back to canonical conjugate variables $Q$ and $\Phi$, the charge-flux relation of the SQUID with Eq.~(\ref{eq:thetaphi_Mac}) is rewritten as 
\begin{widetext}
\begin{equation}
\label{eq:QPhi_SQUID_Mac}
    \dot{Q} = \frac{2}{ L_\mathrm{h} + L_\mathrm{J}\!/|\!\cos\!\phi^\mathrm{eff}_\mathrm{ex}|}\left[\Phi - \frac{1+\epsilon}{6}\left(\frac{L_\mathrm{J}\!/|\!\cos\!\phi^\mathrm{eff}_\mathrm{ex}|}{L_\mathrm{h} + L_\mathrm{J}\!/|\!\cos\!\phi^\mathrm{eff}_\mathrm{ex}|}\frac{\Phi}{\Phi_0}\right)^3 + \cdots \right].
\end{equation}
\end{widetext}
The effective external flux $\phi^\mathrm{eff}_\mathrm{ex}$ is defined as
\begin{equation}
\label{eq:phieffex}
    \phi^\mathrm{eff}_\mathrm{ex} = \frac{\Phi_\mathrm{ex}- L_\mathrm{loop}\dot{Q}_\mathrm{c}\!(0)}{2\Phi_0},
\end{equation}
where $\dot{Q}_\mathrm{c}\!(0)$ is the circulating current in the loop at $\Phi=0$, determined such that
\begin{equation}\label{eq:zero_ex}
\frac{L_\mathrm{loop}\dot{Q}_\mathrm{c}\!(0)}{2\Phi_0} = \pm \sin\!\left(\frac{\Phi_\mathrm{ex}- L_\mathrm{loop}\dot{Q}_\mathrm{c}\!(0)}{2\Phi_0}\right),
\end{equation}

\begin{widetext}
As discussed below for the derivation of a flux-driven JPA, the coefficient $\epsilon$ needs to be sufficiently small for efficient flux pumping and stable operation ($\epsilon \ll 1$). 
Thus, Eq.~(\ref{eq:QPhi_SQUID_Mac}) can be well approximated and simplified to
\begin{equation}
    \dot{Q} = \frac{2}{L_\mathrm{h} + L_\mathrm{J}\!/|\!\cos\!\phi^\mathrm{eff}_\mathrm{ex}|}\left[\Phi - \frac{1}{6}\left(\frac{L_\mathrm{J}\!/|\!\cos\!\phi^\mathrm{eff}_\mathrm{ex}|}{L_\mathrm{h} + L_\mathrm{J}\!/|\!\cos\!\phi^\mathrm{eff}_\mathrm{ex}|}\frac{\Phi}{\Phi_0}\right)^3 + \cdots \right].
\end{equation}
By considering only up to the third-order nonlinearity terms, this expression can be approximated as
\begin{equation}
\label{eq:QPhi_SQUID_sin}
\dot{Q} = \frac{\Phi_0}{L_\mathrm{h}/2 + L_\mathrm{J}\!/(2|\!\cos\!\phi^\mathrm{eff}_\mathrm{ex}|)}\left[\frac{L_\mathrm{h}}{L_\mathrm{h} + L_\mathrm{J}\!/|\!\cos\!\phi^\mathrm{eff}_\mathrm{ex}|}\frac{\Phi}{\Phi_0}+ \sin\!\left(\frac{L_\mathrm{J}\!/|\!\cos\!\phi^\mathrm{eff}_\mathrm{ex}|}{L_\mathrm{h} + L_\mathrm{J}\!/|\!\cos\!\phi^\mathrm{eff}_\mathrm{ex}|}\frac{\Phi}{\Phi_0}\right)\right].
\end{equation}
\end{widetext}
By comparing this with Eq.~(\ref{eq:QPhiLLJ_series}), we find that a SQUID with a loop inductance can be understood as a series circuit of a linear inductor with $L_\mathrm{loop}/4 \, (= L_\mathrm{h}/2)$ and a Josephson junction with an effectively flux-tunable Josephson inductance $L_\mathrm{J}^\mathrm{eff} = L_\mathrm{J}/(2|\!\cos\!\phi^\mathrm{eff}_\mathrm{ex}|)$, as schematically shown in Fig.~\ref{fig:app_circuit2}(b).
Using $L_\mathrm{J}^\mathrm{eff}$, Eq.~\eqref{eq:QPhi_SQUID_sin} is simplified as
\begin{equation}
\begin{split}
\dot{Q} = \frac{\Phi_0}{L_\mathrm{loop}/4+L_\mathrm{J}^\mathrm{eff}}\bigg[ &\bigg(\frac{L_\mathrm{loop}/4}{L_\mathrm{loop}/4+L_\mathrm{J}^\mathrm{eff}}\frac{\Phi}{\Phi_0}\bigg) \\ &+ \sin\!\left(\!\frac{L_\mathrm{J}^\mathrm{eff}}{L_\mathrm{loop}/4+L_\mathrm{J}^\mathrm{eff}}\frac{\Phi}{\Phi_0}\!\right)\!\bigg].
\end{split}
\end{equation}
This yields an explicit charge--flux relation in which the flux \(\Phi\) is written as a function of the charge \(\dot{Q}\), i.e.,
\begin{equation}
\Phi = \frac{L_\mathrm{loop}}{4} \dot{Q}  + \Phi_0 \arcsin\!\left(L_\mathrm{J}^\mathrm{eff}\frac{\dot{Q}}{\Phi_0}\right).
\end{equation}

Moreover, the participation ratio of the tunable Josephson inductance in the total inductance of the effective SQUID model, defined in the main text and denoted by \(p_\mathrm{SQ}\), provides a compact description of the correction coefficient for the self-Kerr nonlinearity in Eq.~\eqref{eq:epsilonKerr}, as
\begin{equation}
\epsilon = 3(1-p_\mathrm{SQ} ) \tan^2\!\phi^\mathrm{eff}_\mathrm{ex}.
\end{equation}


\subsubsection{SQUID array}

%
%
Incorporating a SQUID array into a JPA is key to engineering the nonlinearity, thereby enhancing both the gain and the dynamic range.
Here, we consider an inductive element consisting of an array of \(N\) identical symmetric SQUIDs, where each SQUID is characterized by a Josephson inductance \(L_\mathrm{J}\), a loop inductance \(L_\mathrm{loop}\), and an externally applied magnetic flux \(\Phi_\mathrm{ex}\) (see the circuit diagram highlighted in green in Fig.~\ref{fig:f1_1st}).
As discussed in the previous section, each SQUID can be considered as a series circuit of a linear inductor with $L_\mathrm{loop}/4$ and a Josephson junction with an flux-tunable Josephson inductance $L_\mathrm{J}^\mathrm{eff} = L_\mathrm{J}/(2\cos\!\phi^\mathrm{eff}_\mathrm{ex})$. 
In addition, we include a geometric inductance \(L_\mathrm{g}\) in series with the SQUID array. This inductance serves both to control the participation of the Josephson inductance in the total inductance and to account for stray inductance arising from the circuit geometry (see the effective circuit diagrams in Fig.~\ref{fig:f1-2_1st}).

By considering Kirchhoff’s current law and defining a shared current $\dot{Q}$ flowing through all the inductive elements in series, the total flux $\Phi$ across the SQUID array, including the geometric inductance, is given by
\begin{equation}
\begin{split}
\Phi = \left(\!\frac{N L_\mathrm{loop}}{4} + L_\mathrm{g}\!\right)\dot{Q} + N \Phi_0\arcsin\left(\!L_\mathrm{J}^\mathrm{eff} \frac{\dot{Q}}{\Phi_0}\!\right).
\end{split}
\end{equation}
Here, it is useful to rewrite the charge-flux relation in the form of $\dot{Q}$ as a function of $\Phi$ in order to calculate the potential energy of the SQUID array.
To this end, the expression is expanded around $\dot{Q}=0$ as
\begin{equation}\label{eq:phi_1}
\Phi = L_\mathrm{tot}\dot{Q} + \frac{N}{\Phi_0^2}\left(L_\mathrm{J}^\mathrm{eff} \right)^3 \frac{\dot{Q}^3}{6} + \cdots,
\end{equation}
where the total inductance of the SQUID array is defined as
\begin{equation}
\label{eq:Ltot}
L_\mathrm{tot} = N \left( \frac{L_{\mathrm{loop}}}{4} + L_\mathrm{J}^\mathrm{eff}\right) + L_\mathrm{g}.
\end{equation}

The Lagrange inversion theorems enable one to approximately derive the inverse function of Eq.~\eqref{eq:phi_1} as
\begin{equation}
\dot{Q} = \frac{\Phi_0}{L_\mathrm{tot}} \Bigg[\frac{\Phi}{\Phi_0}- \frac{N}{6}\left(\frac{L_\mathrm{J}^\mathrm{eff}}{L_\mathrm{tot}}\frac{\Phi}{\Phi_0}\right)^3 + \cdots \Bigg].
\end{equation}
By considering only up to the leading term of nonlinearity, the expression can be approximated as
\begin{widetext}
\begin{equation}
\label{eq:QPhi_SQUIDarr}
\dot{Q} = \frac{\Phi_0}{L_\mathrm{tot}}\Bigg[\frac{N (L_\mathrm{loop}/4) + L_\mathrm{g}}{L_\mathrm{tot}} \frac{\Phi}{\Phi_0} + N \sin\!\left(\frac{L_\mathrm{J}^\mathrm{eff}}{L_\mathrm{tot}}\frac{\Phi}{\Phi_0}\right)\Bigg],
\end{equation}
By integrating Eq.~(\ref{eq:QPhi_SQUIDarr}) with $\Phi$, we obtain the potential energy of the SQUID array as
\begin{equation}
U = \frac{{\Phi_0}^2}{2\left(N (L_\mathrm{loop}/4) + L_\mathrm{g}\right)} \!\left( \frac{N (L_\mathrm{loop}/4) + L_\mathrm{g}}{L_\mathrm{tot}} \frac{\Phi}{\Phi_0} \right)^2\\
+ N\frac{\:{\Phi_0}^2\:}{2L_\mathrm{J}^\mathrm{eff}} (-2)\!\cos\!\left(\frac{L_\mathrm{J}^\mathrm{eff}}{L_\mathrm{tot}}\frac{\Phi}{\Phi_0}\right).
\end{equation}
\end{widetext}
Thus, the total energy can be intuitively determined as the sum of the energies stored in the individual inductive elements. Each element’s energy is calculated by considering the flux weighted by the portion of the linear part of the corresponding inductance to the total inductance.
Although this statement is obvious for linear circuits, this conclusion would not be trivial for an inductive element including one or more Josephson junctions.

By introducing the participation ratio of the effective Josephson inductance relative to the total inductance, denoted by \(p_\mathrm{J}\), together with the inductive energies associated with one reduced flux quantum for the linear inductance and the individual flux-tunable Josephson junctions, denoted by \(E_{L_l}\) and \(E_{L_\mathrm{J}^\mathrm{eff}}\), respectively, we arrive at Eq.~\eqref{eq:squidenergy} in the main text, as schematically illustrated in Fig.~\ref{fig:f1-2_1st}(b).

Moreover, introducing the inductive energy of the total inductance, defined as
\begin{equation}
\label{eq:ELtotal}
E_{L_\mathrm{tot}} = \frac{{\Phi_0}^2}{2L_\mathrm{tot}}
\end{equation}
and using the participation ratio $p_\mathrm{J}$, the potential energy is rewritten as
\begin{equation}
U = E_{L_\mathrm{tot}}\left[(1-p_\mathrm{J}) \phi^2 + \frac{N^2}{p_\mathrm{J}}(-2)\!\cos\left(\frac{p_\mathrm{J}}{N}\phi\right)\right],
\end{equation}
where we use the dimensionless flux ($\phi =\Phi/\Phi_0$).
The potential energy can be, then, approximated by retaining only terms up to the leading-order nonlinearity, expressed as
\begin{equation}
\label{eq:potentialenergywithphi}
U = E_{L_\mathrm{tot}} \left(\phi^2 - \frac{p_\mathrm{J}^3}{12N^2}\phi^4 + \cdots\right),
\end{equation}
where a constant term is neglected. 
This explicitly shows that the number of SQUIDs (\(N\)) provides a means to control the nonlinear (self-Kerr) term independently of the total inductance.

\subsection{Flux pumping of SQUIDs} \label{app:fluxpumping}
\subsubsection{Equivalent circuit of a flux-pumped SQUID}

\begin{figure}[t]
\begin{center}
  \includegraphics[width=85.771mm]{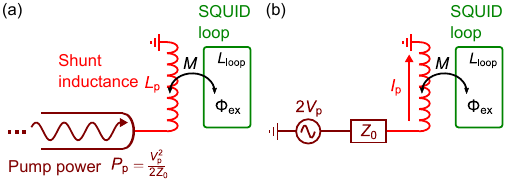} 
\caption{
Schematic of flux pumping of a SQUID loop.
(a)~Flux pumping through a waveguide.
(b)~Equivalent lumped-element circuit model.
} 
  \label{fig:effpumpmodel} 
\end{center}
\end{figure}

To operate the flux-driven JPA, the external magnetic flux through each SQUID loop needs to be modulated at a frequency twice that of the JPA. As shown in Fig.\ref{fig:effpumpmodel}(a), this modulation is typically achieved by applying a microwave pump through a waveguide that is terminated with a shunt inductance $L_\mathrm{p}$, which is mutually coupled to the loop inductance $L_\mathrm{loop}$ of the SQUID with a mutual inductance $M$.
In our design, \( L_\mathrm{p} \approx 150~\mathrm{pH}\), \(L_\mathrm{loop} \approx 20~\mathrm{pH}\), and \(M \approx 5~\mathrm{pH}\).

For a given pump power \(P_\mathrm{p}\), an incoming pump voltage signal is described as
\(
V_\mathrm{p}(t) = V_\mathrm{p}\cos(\omega_\mathrm{p} t + \theta_\mathrm{p}),
\)
where \(\omega_\mathrm{p}\) is the pump frequency, \(\theta_\mathrm{p}\) is the pump phase, and the peak pump voltage is given by
\(V_\mathrm{p}=\sqrt{2P_\mathrm{p} Z_0}\), with \(Z_0\) denoting the characteristic impedance of the waveguide.

Using the equivalent lumped-element circuit model with an alternating voltage source of amplitude \(2V_\mathrm{p}\) and input impedance \(Z_0\), as shown in Fig.~\ref{fig:effpumpmodel}(b), the resulting alternating current through the shunt inductance $L_\mathrm{loop}$ can be calculated as
$I_\mathrm{p}(t) = I_\mathrm{p}\cos(\omega_\mathrm{p}t + \theta_\mathrm{p})$, where the peak current amplitude is given by
\begin{equation}
I_\mathrm{p} = \left|\frac{2V_\mathrm{p}}{Z_0 + i \omega_\mathrm{p} L_\mathrm{p}} \right|
= \frac{2\sqrt{2}\sqrt{P_\mathrm{p}Z_0}}{\sqrt{{Z_0}^2 + (\omega_\mathrm{p} L_\mathrm{p})^2}}.
\end{equation}
Note that the pump phase $\theta_\mathrm{p}$ is updated accordingly.
Using this expression, the  alternating external flux at frequency $\omega_\mathrm{p}$ is obtained as
$\delta\Phi_\mathrm{ex}(t)  = \delta\Phi_\mathrm{ex}\cos(\omega_\mathrm{p}t+ \theta_\mathrm{p})$, where 
\begin{equation}
\delta\Phi_\mathrm{ex} = M I_\mathrm{p} =\frac{2\sqrt{2}M\sqrt{P_\mathrm{p}Z_0}}{\sqrt{{Z_0}^2 + (\omega_\mathrm{p} L_\mathrm{p})^2}}.
\end{equation}
The dimensionless flux pump amplitude is defined as
\begin{equation}
\phi_\mathrm{p} = \frac{\delta\Phi_\mathrm{ex}}{\Phi_0}= \frac{2\sqrt{2}M\sqrt{P_\mathrm{p}Z_0}}{\Phi_0 \sqrt{{Z_0}^2 + (\omega_\mathrm{p} L_\mathrm{p})^2}}.
\end{equation}
For an efficient pump, the condition $\omega_\mathrm{p} L_\mathrm{p} \ll Z_0$ would be satisfied, which is the case for our design ($\omega_\mathrm{p}/2\pi \approx 20~\mathrm{GHz}$, $L_\mathrm{p} \approx 150~\mathrm{pH}$, and $Z_0 = 50~\Omega$).
Consequently, the dimensionless flux pump amplitude is simplified as
\begin{equation}
\phi_\mathrm{p} = \frac{2\sqrt{2}M}{\Phi_0}\sqrt{\frac{P_\mathrm{p}}{Z_0}}.
\end{equation}

%

\subsubsection{Hamiltonian of a flux-pumped SQUID array}
To derive the pump Hamiltonian, we calculate the derivative of the total potential energy with respect to the external flux $\Phi_\mathrm{ex}$.
Using the chain rules, the derivative of Eq.~\eqref{eq:ELtotal} is described as
\begin{equation}
\label{eq:partialEL}
\frac{\partial E_{L_\mathrm{tot}}}{\partial \Phi_\mathrm{ex}} =  E_{L_\mathrm{tot}} \frac{ p_\mathrm{J}}{|\cos\!\phi^\mathrm{eff}_\mathrm{ex}|}\frac{\partial |\cos\!\phi^\mathrm{eff}_\mathrm{ex}|}{\partial \Phi_\mathrm{ex}}.
\end{equation}
Using Eq.~\eqref{eq:phieffex}, the derivative of the cosine term can be evaluated explicitly as
\begin{equation}
\frac{\partial|\cos\!\phi^\mathrm{eff}_\mathrm{ex}|}{\partial \Phi_\mathrm{ex}} = \mp\frac{\sin\!\phi^\mathrm{eff}_\mathrm{ex}}{2\Phi_0} \left( 1 - L_\mathrm{h}\frac{\partial (2\dot{Q}_\mathrm{c}\!(0))}{\partial \Phi_\mathrm{ex}}\right).
\end{equation}
The derivative of Eq.~\eqref{eq:zero_ex} can be also explicitly described as
\begin{equation}
\frac{\partial (2\dot{Q}_\mathrm{c}\!(0))}{\partial \Phi_\mathrm{ex}} = \frac{|\cos\!\phi^\mathrm{eff}_\mathrm{ex}|}{L_\mathrm{J}}\left( 1 - L_\mathrm{h}\frac{\partial (2\dot{Q}_\mathrm{c}\!(0))}{\partial \Phi_\mathrm{ex}}\right).
\end{equation}
Combining these two equations and using the participation ratio $p_\mathrm{SQ} $, we have
\begin{equation}
    \frac{\partial |\cos\!\phi^\mathrm{eff}_\mathrm{ex}|}{\partial \Phi_\mathrm{ex}}  =  \mp\frac{p_\mathrm{SQ} \sin\!\phi^\mathrm{eff}_\mathrm{ex}}{2\Phi_0} .
\end{equation}
By substituting this to Eq.~(\ref{eq:partialEL}), we have
\begin{equation}
\label{eq:ELtotPhiex}
\frac{\partial E_{L_\mathrm{tot}}}{\partial \Phi_\mathrm{ex}} = -E_{L_\mathrm{tot}}\frac{p_\mathrm{J} p_\mathrm{SQ}  \tan\!\phi^\mathrm{eff}_\mathrm{ex}}{2\Phi_0}.
\end{equation}

Using a time-dependent pump flux defined in the previous section as $\delta\Phi_\mathrm{ex}\!(t) = -\Phi_0 \phi_\mathrm{p} \cos(\omega_\mathrm{p}t)$ (\(\theta_\mathrm{p}=\pi\) for convenience), combined with Eq.~\eqref{eq:ELtotPhiex} and the quadratic potential energy term in Eq.~\eqref{eq:potentialenergywithphi}, the pump Hamiltonian is obtained as
\begin{equation}
\label{eq:pumphamiltonian}
\begin{split}
H_\mathrm{p}(t) &=  \frac{\partial E_{L_\mathrm{tot}}}{\partial \Phi_\mathrm{ex}} \:\delta\Phi_\mathrm{ex}\!(t) \: \phi^2\\
\quad &= E_{L_\mathrm{tot}} \frac{p_\mathrm{J} \, p_\mathrm{SQ} \tan\!\phi^\mathrm{eff}_\mathrm{ex}}{2}\phi_\mathrm{p} \phi^2 \cos(\omega_\mathrm{p}t).
\end{split}
\end{equation}

\subsection{Hamiltonian of JPA with SQUID array}~\label{app:jpasquidham}
\subsubsection{Canonical form}
Here, we consider a flux-driven lumped-element JPA, formed by a parallel circuit of a capacitor with a total capacitance and a SQUID array, as shown in Fig.~\ref{fig:f1_1st}.
The Hamiltonian consists of the kinetic (capacitive) energy in Eq.~\eqref{eq:capenergy}, the potential (inductive) energy in Eq.~\eqref{eq:potentialenergywithphi}, and the pump term in Eq.~\eqref{eq:pumphamiltonian}.
By describing the Hamiltonian with the dimensionless conjugate variables and replacing them with the corresponding operators ($\phi \rightarrow \hat{\phi}$ and $n \rightarrow \hat{n}$, where $[\hat{\phi},\:\hat{n}]=i$), we obtain the Hamiltonian of the JPA as
\begin{equation}
\begin{split}
\hat{H} &= E_\mathrm{C_\mathrm{tot}}{\hat{n}}^2 + E_{L_\mathrm{tot}}{\hat{\phi}}^2 - E_{L_\mathrm{tot}} \frac{p_\mathrm{J}^3}{N^2}\frac{{\hat{\phi}}^4}{12}\\
\quad & + E_{L_\mathrm{tot}}  \frac{ p_\mathrm{J} \, p_\mathrm{SQ}  \tan\!\phi^\mathrm{eff}_\mathrm{ex}}{2} \phi_\mathrm{p} {\hat{\phi}}^2 \cos(\omega_\mathrm{p}t).
\end{split}
\end{equation}
By introducing an annihilation operator $\hat{a}$ such that $\left[\hat{a},\:\hat{a}^\dag\right]=1$, the dimensionless conjugate operators are described as
\begin{equation}
\begin{split}
\hat{\phi} = \left(\frac{E_{C_\mathrm{tot}}}{E_{L_\mathrm{tot}}}\right)^{\frac{1}{4}}\frac{\hat{a}+\hat{a}^\dag}{\sqrt{2}},\\
\hat{n} = \left(\frac{E_{L_\mathrm{tot}}}{E_{C_\mathrm{tot}}}\right)^{\frac{1}{4}}\frac{\hat{a}-\hat{a}^\dag}{\sqrt{2}i}.
\end{split}
\end{equation}
Using these relations, the Hamiltonian is rewritten as
\begin{equation}
\begin{split}
\hat{H} &= \hbar\omega_\mathrm{a} \: \hat{a}^\dag \hat{a} + \frac{\hbar K}{2}\hat{a}^{\dag^2}{\hat{a}}^2 \\ &+ \frac{\hbar \Omega_\mathrm{p}}{4}\!\left(\hat{a}^{\dag^2}e^{-i\omega_\mathrm{p}t} + {\hat{a}}^2e^{i\omega_\mathrm{p}t}\right),
\end{split}
\end{equation}
where the resonance frequency is given by
\begin{equation}
\hbar\omega_\mathrm{a} = 2 \sqrt{E_{L_\mathrm{tot}}E_{C_\mathrm{tot}}},
\end{equation}
the self-Kerr coefficient is given by
\begin{equation}
\hbar K = -\frac{p_\mathrm{J}^3 E_{C_\mathrm{tot}}}{4N^2},
\end{equation}
and the two-photon drive amplitude is given by
\begin{equation}
\hbar \Omega_\mathrm{p} = \frac{\hbar \omega_\mathrm{a}\: p_\mathrm{J} \, p_\mathrm{SQ}  \tan\!\phi^\mathrm{eff}_\mathrm{ex}}{4}  \phi_\mathrm{p}.
\end{equation}
Note that the rotating wave approximation is applied in the Hamiltonian.
The two-photon drive amplitude is defined such that amplification will appear when $\Omega_\mathrm{p} \approx \kappa$.
Furthermore, the external coupling rate of the JPA to the waveguide is given by
\begin{equation}
\kappa = \omega_\mathrm{a} \frac{C_\kappa^2}{C_\mathrm{tot}^2}\frac{Z_0}{Z},
\end{equation}
where $Z = \sqrt{L_\mathrm{tot}/C_\mathrm{tot}}$ and $Z_0$ are the characteristic impedances of the JPA and the waveguide, respectively.

\section{Gain saturation} \label{app:gainsat}
One of the metrics of JPAs is the dynamic range, i.e., the saturation signal input power, or the compression point. The dominant cause of saturation for flux-driven JPA is determined by the self-Kerr effect, which shifts the JPA resonance frequency as the number of signal and idler photons in the JPA increase. As a result, the gain of the signal is decreased since the JPA frequency shifts away from the optimal point~\cite{bib:sqarray2, bib:frattini2018optimizing}. Here, we will discuss the scaling of the compression point concerning the system parameters of the JPA.

By the definition of gain, we can relate the input and output photon flux as
\begin{equation}
    G = \frac{n_\mathrm{out}}{n_\mathrm{in}},
\end{equation}
where $G$ is the JPA gain and $n_\mathrm{out}$ ($n_\mathrm{in}$) is the output (input) photon flux from (into) the JPA.
Moreover, from the input--output relation, we can relate the output photon flux to the photon number $n_\mathrm{a}$ in the JPA as
\begin{equation}
    n_\mathrm{out} = \kappa n_\mathrm{a},
\end{equation}
where $\kappa$ is the external-coupling rate of the JPA. We can safely neglect the contribution of the input photon flux and its interference because the gain is large for the photon flux.

To understand the scaling of the saturation of the gain for the input photon power, we can define the saturation condition so that the self-Kerr effect induced by the photons in the JPA becomes comparable with the JPA bandwidth. This can be expressed as
\begin{equation}
    K n_\mathrm{a} \sim \kappa,
\end{equation}
where $K$ is the self-Kerr coefficient. Here, we assume the internal loss of the JPA is negligible compared to the external loss.

By using these equations, we show that the saturation input photon flux ($n_\mathrm{in}^\mathrm{sat.}$) scales as
\begin{equation}
    n_\mathrm{in}^\mathrm{sat.} \propto \frac{\kappa^2}{GK}.
\end{equation}

\section{input--output theory}\label{app:iojpafp}

\subsection{Model description}

To understand the effect of an imperfect environment, for example, a JPA connected to a circulator that might have an impedance mismatch, we consider the case of a quantum system with annihilation operator $\hat{a}$, emitting into a waveguide in which a mirror with transmittance $\eext$ is located at a distance $v\tau$ (where $v$ is the microwave velocity in the waveguide, and $\tau$ is the time for propagation from the system to the mirror) from the quantum system. 
To take losses in the waveguide between the quantum system and this mirror into account, we also insert a beam splitter with transmission $\eint$ for both propagation directions.
The position of the mirror simulating the waveguide propagation loss does not matter here and we set it at the same virtual position as the quantum system. 

The Heisenberg equation of the field of the quantum system ($\hat{a}$), and its input--output relation to the fields of the Fabry--P\'erot cavity ($\ain$ and $\aout$), as well as its relation to the intrinsic loss channel ($\din$), are found in the main text. 
Moreover, the relations of the fields of the Fabry--P\'erot cavity ($\bin$ and $\bout$) to the fields directly interacting with the quantum system ($\ain$ and $\aout$), the external fields ($\cin$ and $\cout$), and those of the intrinsic loss channels ($\uin$ and $\vin$) are also defined in the main text.

\subsection{Effective Heisenberg equation}
In this section, we derive an effective Heisenberg equation for the combined system of the JPA and the Fabry--P\'erot cavity. 



By substituting the first row of Eq.~\eqref{eq:beamsplitter_eta} into Eq.~\eqref{eq:beamsplitter_eta0}, we have
\begin{widetext}
\begin{equation}
    \ain\!(t) = T^* \!\sqrt{\eint}\:\cin\!(t) + R\sqrt{\eint} \:\bout\!(t-2\tau) +\sqrt{1-\eint} \:\uin\!(t).
\end{equation}
The field $\bout$ can be eliminated using Eq.~(\ref{eq:beamsplitter_eta0-2}), leading to 
\begin{equation}
    \ain\!(t) = T^* \! \sqrt{\eint}\:\cin\!(t) + R \eint \: \aout\!(t-2\tau) +\sqrt{1-\eint} \:\uin\!(t)  + R \sqrt{\eint(1-\eint)}\:\vin\!(t-2\tau).
\end{equation}
Using Eq.~(\ref{eq:aoutaina}), the auxiliary Fabry--P\'erot cavity degrees of freedom can be eliminated, resulting in
\begin{equation}
\label{eq:ain_ain}
    \ain\!(t) = R\eint \: \ain\!(t-2\tau) - iR \eint  \sqrt{\kext} \: \hat{a}(t-2\tau) + T^* \! \sqrt{\eint} \: \cin\!(t) +\sqrt{1-\eint} \: \uin(t)  + R\sqrt{\eint(1-\eint)} \: \vin\!(t-2\tau).
\end{equation}
By recursively substituting Eq.~(\ref{eq:ain_ain}) into itself, we obtain
\begin{equation}
\begin{split}
\label{eq:ain}
    \ain\!(t) = &- i \sqrt{\kappa} \sum_{n=1}^{\infty} \left(R\eint\right)^{n} \:\hat{a}(t-2n\tau) + T^*\!\sqrt{\eint}\sum_{n=0}^{\infty}  \left(R\eint\right)^{n}\:\cin\!(t-2n\tau) \\
    & -i \sqrt{(1-\eint)} \sum_{n=0}^{\infty}  \left(R\eint\right)^{n} \:\uin\!(t-2n\tau) + R\sqrt{\eint(1-\eint)} \sum_{n=1}^{\infty}  \left(R\eint\right)^{n-1} \:\vin\!(t-2n\tau).
\end{split}
\end{equation}
Using the input field, the Heisenberg equation for the internal field of the quantum system, Eq.~(\ref{eq:Heisenbergeq_a}), is obtained as
\begin{equation}
\label{eq:Heisenberg_FP}
\begin{split}
    \frac{\di\hat{a}(t)}{\di t} = &  \frac{i}{\hbar} \left[\hat{H}, \:\hat{a}(t) \right]  - \frac{\kext + \kint}{2}\hat{a}(t)  - \kappa \sum_{n=1}^{\infty} \left(R\eint\right)^{n} \:\hat{a}(t-2n\tau) -i T^*\!\sqrt{\eint\kext}\sum_{n=0}^{\infty}  \left(R\eint\right)^{n}\:\cin\!(t-2n\tau)\\
    & -i \sqrt{(1-\eint)\kext} \sum_{n=0}^{\infty}  \left(R\eint\right)^{n} \:\uin\!(t-2n\tau) -i R\sqrt{\eint(1-\eint)\kext} \sum_{n=1}^{\infty}  \left(R\eint\right)^{n-1} \:\vin\!(t-2n\tau)  - i \sqrt{\kint}\:\din\!(t).    
\end{split}
\end{equation}
This expression implies that the quantum system is driven by the external fields at multiple discrete times, with delays introduced by successive reflections within the Fabry--P\'erot cavity.

Similarly, the output field, $\cout$, can be obtained by eliminating the auxiliary degrees of freedom of the Fabry--P\'erot cavity.
By substituting Eq.~(\ref{eq:ain}) into the input-relation of the quantum field, i.e., Eq.~(\ref{eq:aoutaina}), we have
\begin{equation}
\begin{split}    
    \aout\!(t) = &-i \sqrt{\kappa} \sum_{n=0}^{\infty} (R\eint)^n \hat{a}(t - 2\tau n) + T^* \sqrt{\eint} \sum_{n=0}^{\infty} (R\eint)^n \cin\!(t - 2n\tau)\\
    &+ \sqrt{1 - \eint} \sum_{n=0}^{\infty} (R\eint)^n \uin\!(t - 2n\tau ) + R \sqrt{\eint(1-\eint)} \sum_{n=1}^{\infty} (R\eint)^{n-1} \vin\!(t - 2n\tau).
\end{split}
\end{equation}
By substituting the above equation into Eq.~\ref{eq:beamsplitter_eta0-2}, we have
\begin{equation}
\label{eq:beamsplitter_eta0-2_delay}
\begin{split}    
    \bout\!(t) = &-i \sqrt{\eint \kappa} \sum_{n=0}^{\infty} (R\eint)^n \hat{a}(t - 2n\tau) + T^* \eint \sum_{n=0}^{\infty} (R\eint)^n \cin\!(t - 2n\tau)\\
    &+ \sqrt{\eint(1 - \eint)} \sum_{n=0}^{\infty} (R\eint)^n \uin\!(t - 2n\tau) + \sqrt{1-\eint} \sum_{n=0}^{\infty} (R\eint)^{n} \vin\!(t - 2n\tau).
\end{split}
\end{equation}
Finally, by substituting Eq.~(\ref{eq:beamsplitter_eta0-2_delay}) into the second row of Eq.~(\ref{eq:beamsplitter_eta}), we obtain the input--output relation for the incoming and outgoing fields to the Fabry--P\'erot cavity, which is given by
\begin{equation}
\label{eq:input--output_FP}
\begin{split}    
    \cout\!(t) = &  -i T\sqrt{\eint \kappa} \sum_{n=0}^{\infty} (R\eint)^n a(t - 2\tau n) - R^*\cin(t+2\tau) + |T|^2 \eint \sum_{n=0}^{\infty} (R\eint)^n \cin\!(t - 2n\tau)\\
    &+ T\sqrt{\eint(1 - \eint)} \sum_{n=0}^{\infty} (R\eint)^n \uin\!(t - 2n\tau) + T\sqrt{1-\eint} \sum_{n=0}^{\infty} (R\eint)^{n} \vin\!(t - 2n\tau),
\end{split}
\end{equation}
where the time argument is shifted as $t \rightarrow t + \tau$. In summary, Eqs.~(\ref{eq:Heisenberg_FP}) and (\ref{eq:input--output_FP}) represent the generalized Heisenberg equation and the input--output relation for the quantum system under the Fabry--P\'erot interference.

When the quantum system is characterized by a frequency \(\omega\) (such as the resonance frequency of the free-evolving system or the drive frequency of an input field), the time delay introduced by the Fabry--P\'erot cavity can be approximated as a phase shift, for example, 
\(
\hat{a}(t-2\tau n) \approx \hat{a}(t)\, e^{i 2\omega \tau n}.
\)
Therefore, the Heisenberg equation in Eq.~(\ref{eq:Heisenberg_FP}) and its input-output relation in Eq.~\eqref{eq:input--output_FP} can be approximated as
\begin{equation}
\begin{split}
    \frac{\di\hat{a}(t)}{\di t} &= \frac{i}{\hbar} \left[ \hat{H}, \:\hat{a}(t) \right]  - \frac{\kext + \kint}{2}\hat{a}(t)  - \kappa \sum_{n=1}^{\infty} \left(R\eint e^{i 2\omega \tau}\right)^{n} \:\hat{a}(t) -i T^*\!\sqrt{\eint\kext}\sum_{n=0}^{\infty}  \left(R\eint \: e^{i 2 \omega \tau}\right)^{n}\:\cin\!(t) 
     \\ & - i \sqrt{(1-\eint)\kext} \sum_{n=0}^{\infty}  \left(R\eint \: e^{i 2 \omega \tau}\right)^{n} \:\uin\!(t) -i R\sqrt{\eint(1-\eint)\kext} \cdot \: e^{i 2 \omega\tau} \sum_{n=1}^{\infty}  \left(R\eint \: e^{i 2 \omega \tau}\right)^{n-1} \:\vin\!(t) - i \sqrt{\kint}\:\din\!(t)
\end{split} 
\end{equation}
and
\begin{equation}
\begin{split}    
    \cout\!(t) = &  -i T\sqrt{\eint \kappa} \sum_{n=0}^{\infty} (R\eint e^{i 2\omega \tau})^n a(t) - R^*e^{-i 2\omega \tau}\cin(t) + |T|^2 \eint \sum_{n=0}^{\infty} (R\eint e^{i 2\omega \tau})^n \cin\!(t)\\
    &+ T\sqrt{\eint(1 - \eint)} \sum_{n=0}^{\infty} (R\eint e^{i 2\omega \tau})^n \uin\!(t) + T\sqrt{1-\eint} \sum_{n=0}^{\infty} (R\eint e^{i 2\omega \tau})^{n} \vin\!(t),
\end{split}
\end{equation}

Provided that \(|R\,\eint\,e^{i2\omega\tau}|<1\), the corresponding infinite geometric series converges. This allows us to derive the effective Heisenberg equation and input--output relation in the presence of the Fabry--P\'erot interference, which are given by Eqs.~\eqref{eq:heisenberg_main} and \eqref{eq:inout_main} in the main text.


\end{widetext}

\subsection{Effective input--output relation} \label{app:effectiveJPAinput}
In the following, we consider a flux-driven JPA as the quantum system described in the previous section.
The Hamiltonian of the flux-driven JPA is given by
\begin{equation}
    \hat{H} = \hbar \omega_\mathrm{a} \: \hat{a}^\dag \hat{a} + \frac{\hbar \Omega_\mathrm{p}}{4} \left( \hat{a}^{\dag 2} e^{-i \omega_\mathrm{p} t} + \hat{a}^2 e^{i \omega_\mathrm{p} t} \right),
\end{equation}
where we neglect the self-Kerr energy term for simplicity. In this analysis, we are primarily interested in the gain spectrum, or equivalently, the reflection spectrum of the JPA. To this end, a Fourier-domain treatment proves particularly useful. The Heisenberg equation (\ref{eq:heisenberg_main}) for the signal field at frequency $\omega_\mathrm{s}$ can be written in the Fourier domain as
\begin{widetext}
\begin{equation}
\label{eq:Heisenberg_FP_sig}
\begin{split}
    -i\omega_\mathrm{s} \hat{a}(\omega_\mathrm{s}) = & -i\omega_\mathrm{a} \hat{a}(\omega_\mathrm{s}) + i\frac{\Omega_\mathrm{p}}{2} \hat{a}^\dag\!(\omega_\mathrm{i}) - \frac{\kext + \kint}{2}\hat{a}(\omega_\mathrm{s})  -\kext 
 \frac{\mathcal{R}_{(\omega_\mathrm{s})}}{1 - \mathcal{R}_{(\omega_\mathrm{s})}}\hat{a}(\omega_\mathrm{s}) -i \frac{T^*\!\sqrt{\eint\kext}}{1 - \mathcal{R}_{(\omega_\mathrm{s})}}\cin\!(\omega_\mathrm{s})\\
    & -i \frac{\sqrt{(1-\eint)\kext}}{1 - \mathcal{R}_{(\omega_\mathrm{s})}}\uin\!(\omega_\mathrm{s}) -i \sqrt{\frac{(1-\eint)\kext}{\eint}}\frac{\mathcal{R}_{(\omega_\mathrm{s})}}{1 - \mathcal{R}_{(\omega_\mathrm{s})}}\vin\!(\omega_\mathrm{s})  - i \sqrt{\kint}\:\din\!(\omega_\mathrm{s}),   
\end{split}
\end{equation}
where the round-trip coefficient $\mathcal{R}_{(\omega)}$ is defined in Eq.~\eqref{eq:r} of the main text.
All field operators are now expressed in the Fourier domain, with their arguments given in terms of the corresponding frequencies.
In this expression, the signal frequency component is coupled to the idler frequency component with $\omega_\mathrm{i} = \omega_\mathrm{p} - \omega_\mathrm{s}$ via the two-photon drive term in the Hamiltonian.
Therefore, to obtain the solution for the signal frequency component, we must also consider the Heisenberg equation for the idler frequency component, where it is important to note that the Fabry--P\'erot interference must be evaluated at the idler frequency $\omega_\mathrm{i}$.
This can be seen more clearly by directly applying the Fourier transform to Eqs.~(\ref{eq:Heisenberg_FP}) and \eqref{eq:input--output_FP}.

These two coupled equations can be solved in matrix form as
\begin{equation}
\label{eq:a_solutions}
\begin{aligned}
\begin{pmatrix}
\hat{a}(\omega_\mathrm{s}) \\
\hat{a}^\dagger(\omega_\mathrm{i})
\end{pmatrix}
=
i \begin{pmatrix}
\chi^{-1}_{(\omega_\mathrm{s})} & i \frac{\Omega_\mathrm{p}}{2} \\
- i \frac{\Omega_\mathrm{p}^*}{2} & \chi^{-1*}_{(\omega_\mathrm{i})} 
\end{pmatrix}^{-1}
\Bigg[
\sqrt{\eint \kext}
\begin{pmatrix}
\displaystyle \frac{T^*}{1 - \mathcal{R}_{(\omega_\mathrm{s})}}\cin\!(\omega_\mathrm{s}) \\
\displaystyle -\frac{T}{1 - \mathcal{R}^*_{(\omega_\mathrm{i})}}\cin^\dagger\!(\omega_\mathrm{i})
\end{pmatrix}
+ \sqrt{(1 - \eint) \kext}
\begin{pmatrix}
\displaystyle \frac{1}{1 - \mathcal{R}_{(\omega_\mathrm{s})}}\uin\!(\omega_\mathrm{s}) \\
\displaystyle -\frac{1}{1 - \mathcal{R}^*_{(\omega_\mathrm{i})}}\uin^\dagger\!(\omega_\mathrm{i})
\end{pmatrix} \\[6pt]
+ \sqrt{\frac{(1 - \eint)\kext}{\eint}}
\begin{pmatrix}
\displaystyle \frac{\mathcal{R}_{(\omega_\mathrm{s})}}{1 - \mathcal{R}_{(\omega_\mathrm{s})}} \vin\!(\omega_\mathrm{s}) \\
\displaystyle -\frac{\mathcal{R}^*_{(\omega_\mathrm{i})}}{1 - \mathcal{R}^*_{(\omega_\mathrm{i})}}\vin^\dagger\!(\omega_\mathrm{i})
\end{pmatrix}
+ \sqrt{\kint}
\begin{pmatrix}
\din\!(\omega_\mathrm{s}) \\
-\din^\dagger\!(\omega_\mathrm{i}),
\end{pmatrix}
\Bigg],
\end{aligned}
\end{equation}
where the linear response function $\chi^{-1}_{(\omega)}$ is defined in Eq.~\eqref{eq:chi-1omega} of the main text.

To obtain the input--output relation of the JPA at the signal frequency of $\omega_\mathrm{s}$ in the Fourier domain, Eq.~\eqref{eq:inout_main}) is transformed into
\begin{equation}
\label{eq:input--output_FP_Fourier} 
    \cout\!(\omega_\mathrm{s}) =   -i  \frac{T\sqrt{\eint \kappa}}{1 - \mathcal{R}_{(\omega_\mathrm{s})}} \hat{a}(\omega_\mathrm{s})
    + \left( - \frac{\mathcal{R}^*_{(\omega_\mathrm{s})}}{\eint} 
    + \frac{|T|^2 \eint}{1 - \mathcal{R}_{(\omega_\mathrm{s})}} \right) \cin\!(\omega_\mathrm{s})
    + \frac{T\sqrt{\eint(1 - \eint)}}{1 - \mathcal{R}_{(\omega_\mathrm{s})}} \uin\!(\omega_\mathrm{s}) 
    + \frac{T\sqrt{1-\eint} }{1 - \mathcal{R}_{(\omega_\mathrm{s})}} \vin\!(\omega_\mathrm{s}).
\end{equation}
By substituting the solution for $\hat{a}(\omega_\mathrm{s})$ into Eq.~(\ref{eq:input--output_FP_Fourier}), we obtain the input--output relation of the JPA as
\begin{equation}
\label{eq:input--output_JPA-FP}
\begin{split}
\cout\!(\omega_\mathrm{s}) = &\left[ - \frac{\mathcal{R}^*_{(\omega_\mathrm{s})}}{\eint} 
    + \frac{\eext \eint}{1 - \mathcal{R}_{(\omega_\mathrm{s})}} \left\{1 + \frac{\kext \chi^{-1*}_{(\omega_\mathrm{i})}}{\left(1 - \mathcal{R}_{(\omega_\mathrm{s})}\right)\left(\chi^{-1}_{(\omega_\mathrm{s})}\chi^{-1*}_{(\omega_\mathrm{i})} - \frac{|\Omega_\mathrm{p}|^2}{4}\right)} \right\}\right] \cin\!(\omega_\mathrm{s}) \\
    & + \frac{\sqrt{\eext \eint(1-\eint)}}{1 - \mathcal{R}_{(\omega_\mathrm{s})}}\left[1 + \frac{\kext \chi^{-1*}_{(\omega_\mathrm{i})}}{\left(1 - \mathcal{R}_{(\omega_\mathrm{s})}\right)\left(\chi^{-1}_{(\omega_\mathrm{s})}\chi^{-1*}_{(\omega_\mathrm{i})} - \frac{|\Omega_\mathrm{p}|^2}{4}\right)} \right] \uin\!(\omega_\mathrm{s}) \\
    & + \frac{\eint\!\sqrt{\eext(1-\eint)}}{1 - \mathcal{R}_{(\omega_\mathrm{s})}}\left[1 + \frac{\kext \chi^{-1*}_{(\omega_\mathrm{i})} \: \mathcal{R}_{(\omega_\mathrm{s})}}{\left(1 - \mathcal{R}_{(\omega_\mathrm{s})}\right)\left(\chi^{-1}_{(\omega_\mathrm{s})}\chi^{-1*}_{(\omega_\mathrm{i})} - \frac{|\Omega_\mathrm{p}|^2}{4}\right)} \right] \vin\!(\omega_\mathrm{s}) \\
     &+ \frac{\sqrt{\eext \eint \kext \kint}\:\chi^{-1*}_{(\omega_\mathrm{i})}}{\left(1 - \mathcal{R}_{(\omega_\mathrm{s})}\right)\left(\chi^{-1}_{(\omega_\mathrm{s})}\chi^{-1*}_{(\omega_\mathrm{i})} - \frac{|\Omega_\mathrm{p}|^2}{4}\right)} \din\!(\omega_\mathrm{s}) \\
    & + i \: \frac{\Omega_\mathrm{p}}{2}\frac{\kext \left[\eext\eint\: \cin^\dag\!(\omega_\mathrm{i}) + \sqrt{\eext\eint(1-\eint)}\: \uin^\dag\!(\omega_\mathrm{i}) +\eint\sqrt{\eext(1-\eint)} \: \mathcal{R}^*_{(\omega_\mathrm{i})} \:\vin^\dag\!(\omega_\mathrm{i})\right]}{\left(1 - \mathcal{R}_{(\omega_\mathrm{s})}\right)\left(1 + \mathcal{R}^*_{(\omega_\mathrm{i})}\right)\left(\chi^{-1}_{(\omega_\mathrm{s})}\chi^{-1*}_{(\omega_\mathrm{i})} - \frac{|\Omega_\mathrm{p}|^2}{4}\right)} \\
    & + i \: \frac{\Omega_\mathrm{p}}{2} \frac{\sqrt{\eext \eint \kext \kint} }{\left(1 - \mathcal{R}_{(\omega_\mathrm{s})}\right)\left(\chi^{-1}_{(\omega_\mathrm{s})}\chi^{-1*}_{(\omega_\mathrm{i})} - \frac{|\Omega_\mathrm{p}|^2}{4}\right)}\:\din^\dag\!(\omega_\mathrm{i}),
\end{split}
\end{equation}
where we use the relation $\sqrt{\eext}= |T|$ to eliminate the explicit dependence on $T$. Note that the argument of the complex transmission coefficient is absorbed into the definition of the input noise fields through corresponding phase shifts, as these phase factors do not affect the input--output relation for the signal. The coefficient of $\cin\!(\omega_\mathrm{s})$ corresponds to the reflection spectrum of the combined system of the JPA and Fabry--P\'erot cavity, yielding Eq.~\eqref{eq:ana_gain_spe} of the main text,
\end{widetext}

\subsection{Effective amplifier model}\label{app:noise_eff_model}
\begin{figure}[t]
\begin{center}
  \includegraphics[width=83.595mm]{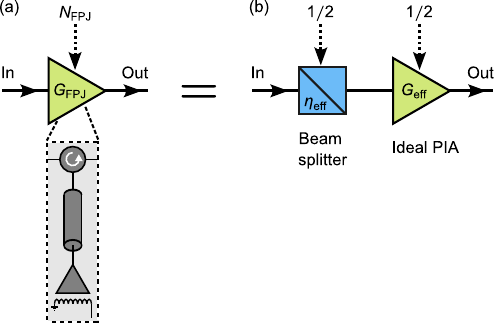} 
\caption{
Effective amplifier model.
(a)~Amplifier model of the combined JPA–Fabry--P\'erot cavity system, characterized by a gain \(G_\mathrm{FPJ}\) and an added photon noise \(N_\mathrm{FPJ}\).
(b)~Equivalent effective model consisting of a beam splitter with transmittance \(\eta_\mathrm{eff}\) followed by an ideal phase-insensitive amplifier (PIA) with gain \(G_\mathrm{eff}\). Both inputs are assumed to be in the vacuum state, contributing a noise of \(1/2\).
} 
  \label{fig:effampmodel} 
\end{center}
\end{figure}

As described in Eq.~\eqref{eq:input--output_JPA-FP}, we obtain the input--output relation of the signal field for the combined JPA–Fabry--P\'erot cavity system, which can be characterized by a gain \(G_\mathrm{FPJ}\) and an added photon noise \(N_\mathrm{FPJ}\) as shown in Fig.~\ref{fig:effampmodel}(a). 
The added noise is defined as the sum of the inevitable vacuum noise and the excess noise arising from imperfections of the phase-insensitive amplifier (PIA).
Here, we discuss the effective amplifier model for the input--output relation.

The coefficients of the input fields in Eq.~\eqref{eq:input--output_JPA-FP} can be defined so that 
\begin{equation}
\begin{split}
    \cout(\omega_\mathrm{s}) &= C \cin\!(\omega_\mathrm{s}) + C'^*\cin^\dag\!(\omega_\mathrm{i}) \\
    &+ U\uin\!(\omega_\mathrm{s}) + U'^*\uin^\dag\!(\omega_\mathrm{i})\\
    & + V\vin\!(\omega_\mathrm{s}) + V'^*\vin^\dag\!(\omega_\mathrm{i}) \\
    &+ D\din\!(\omega_\mathrm{s}) + D'^*\din^\dag\!(\omega_\mathrm{i}).
\end{split}
\end{equation}

%

The effective input noise field at signal frequency is defined as
\begin{equation}
    \hat{e}_\mathrm{in}\!(\omega_\mathrm{s}) = \frac{U\uin\!(\omega_\mathrm{s}) + V\vin\!(\omega_\mathrm{s}) + D\din\!(\omega_\mathrm{s})}{\sqrt{|U|^2 + |V|^2 + |D|^2}}.
\end{equation}
Similarly, the effective input noise field at the idler frequency is defined as 
\begin{equation}
\begin{split}
    \hat{f}_\mathrm{in}\!(\omega_\mathrm{i}) =\, &\frac{C' \cin\!(\omega_\mathrm{i}) + U'\uin\!(\omega_\mathrm{i}) + V'\vin\!(\omega_\mathrm{i})}{\sqrt{|C'|^2 + |U'|^2 + |V'|^2 + |D'|^2}} \\ &+ \frac{D'\din\!(\omega_\mathrm{i})}{\sqrt{|C'|^2 + |U'|^2 + |V'|^2 + |D'|^2}}.
\end{split}
\end{equation}
These effective noise fields maintain the commutation relations of the input noise fields.
Moreover, we define the effective gain and transmittance as
\begin{equation}
\begin{split}
    G_\mathrm{eff} = |C|^2 + |U|^2 + |V|^2 + |D|^2, \\
\eta_\mathrm{eff} = \frac{|C|^2}{|C|^2 + |U|^2 + |V|^2 + |D|^2}.
\end{split}
\end{equation}
To ensure that the input--output relation satisfies the unitary condition, the following constraint must hold:
\begin{equation}
\begin{split}
    &|C|^2 + |U|^2 + |V|^2 + |D|^2 \\&- |C'|^2 - |U'|^2 - |V'|^2 - |D'|^2 = 1,
\end{split}
\end{equation}
which we have verified numerically. By using these parameters, the input--output relation for the signal field can be rewritten as
\begin{equation}
\begin{split}
    \cout(\omega_\mathrm{s}) = &\sqrt{G_\mathrm{eff}}\Big[\sqrt{\eta_\mathrm{eff}}\:\cin(\omega_\mathrm{s}) \\ 
    &\qquad\quad+ \sqrt{1- \eta_\mathrm{eff}}\:\hat{e}_\mathrm{in}\!(\omega_\mathrm{s})\Big] \\ 
    &+ \sqrt{G_\mathrm{eff}-1}\:\hat{f}^\dag_\mathrm{in}\!(\omega_\mathrm{i}),
\end{split}
\end{equation}
where the input signal field $\cin\!(\omega_\mathrm{s})$ is redefined to absorb the additional phase introduced by the convention of treating the effective transmittance coefficient as real-valued for simplicity.
As shown in Fig.~\ref{fig:effampmodel}(b), this simplified expression suggests that the input--output relation can be interpreted as a transmission process through a beam splitter with effective transmittance $\eta_\mathrm{eff}$, which couples the signal to vacuum noise, followed by an ideal phase-insensitive amplification (PIA) process with effective gain $G_\mathrm{eff}$.

The effective model can be used to obtain the analytical expressions of the added noise and gain of the combined FPJPA system as follows.
By assuming that both effective noise fields are in vacuum, the gain can be expressed as
\begin{equation}
G_\mathrm{FPJ} = G_\mathrm{eff} \: \eta_\mathrm{eff}
\end{equation}
and the added noise can be expressed as
\begin{equation}
    N_\mathrm{FPJ} = \frac{1}{\eta_\mathrm{eff}}\left(\frac{1-\eta_\mathrm{eff}}{2} + \frac{1}{2}\right),
\end{equation}
where $G_\mathrm{eff} \gg 1$ is assumed.
The analytical expressions are used to compute theoretical predictions for the added noise using the model parameters extracted from fits to the measured gain spectra (see Fig.~\ref{fig:f6}).

\bibliography{bibtemp.bib}

\end{document}